\documentclass[
    reprint,
    preprintnumbers,
    superscriptaddress,
    nofootinbib,
    amsmath,amssymb,
    aps,
    prd,
    floatfix,
    longbibliography]{revtex4-2}
    
    \usepackage{graphicx}
    \usepackage{subfigure}
    \usepackage{multirow}
    \usepackage[usenames,dvipsnames]{xcolor} 
    \usepackage{mathrsfs} 

    \usepackage{pgfplots}
    \usepackage[utf8]{inputenc}
    \usepackage{color}
    \usepackage[normalem]{ulem} 
    \usepackage{physics}
    \usepackage{enumitem}
    \usepackage{comment}
    \usepackage{bm}
    \usepackage{aas_macros}
    \usepackage{multirow}
    \usepackage{xspace}
    \usepackage{hyperref}

\hypersetup{
  breaklinks = true,
  colorlinks   = true, 
  urlcolor     = Blue, 
  linkcolor    = Blue, 
  citecolor   = Blue 
}

\usepackage{float}

\newcommand{\lcdm}{$\Lambda$CDM\xspace}

\newcommand{\oxford}{Astrophysics, University of Oxford, DWB, Keble Road, Oxford OX1 3RH, United Kingdom}

\newcommand{\llr}{{\tt LLR}\xspace}
\newcommand{\euclid}{{\tt Euclid}\xspace}
\newcommand{\rubin}{{\tt Rubin}\xspace}
\newcommand{\stiskalek}{{\tt Stiskalek25}\xspace}
\newcommand{\tmpp}{{\tt 2M++}\xspace}
\newcommand{\srd}{{\tt DESC SRD}\xspace}
\newcommand{\desi}{{\tt DESI}\xspace}
\newcommand{\sdss}{{\tt SDSS}\xspace}
\newcommand{\boss}{{\tt BOSS}\xspace}
\newcommand{\desipv}{{\tt DESI PV+BGS}\xspace}
\newcommand{\planck}{{\tt Planck}\xspace}
\newcommand{\wmap}{{\tt WMAP}\xspace}
\newcommand{\act}{{\tt ACT}\xspace}
\newcommand{\spt}{{\tt SPT}\xspace}
\newcommand{\planckact}{{\tt Planck}+\act}
\newcommand{\des}{{\tt DES}\xspace}
\newcommand{\desyf}{{\tt DES-Y5}\xspace}
\newcommand{\desdov}{{\tt DES-Dovekie}\xspace}
\newcommand{\pantheonp}{{\tt Pantheon+}\xspace}
\newcommand{\uniont}{{\tt Union3}\xspace}
\newcommand{\stolzner}{{\tt St\"olzner+18}\xspace}
\newcommand{\krolewski}{{\tt Krolewski+22}\xspace}
\newcommand{\reeves}{{\tt Reeves+25}\xspace}

\newcommand{\nvss}{{\tt NVSS}\xspace}
\newcommand{\tmpz}{{\tt 2MPZ}\xspace}
\newcommand{\wisc}{{\tt WIxSC}\xspace}
\newcommand{\unwise}{{\tt unWISE}\xspace}

\newcommand{\desisnecmb}{{\tt DESI}+SNe+CMB\xspace}
\newcommand{\desidesdovcmb}{{\tt DESI}+{\tt DES-Dovekie}+CMB\xspace}

\begin{document}

\title{The Status of Single Scalar Field Dark Energy}

\author{Carlos Garc\'ia-Garc\'ia}
\email{carlos.garcia2@ciemat.es}
\affiliation{\oxford}
\affiliation{Waterloo Centre for Astrophysics, University of Waterloo, Waterloo, ON N2L 3G1, Canada}
\affiliation{Department of Physics and Astronomy, University of Waterloo, Waterloo, ON N2L 3G1, Canada}
\affiliation{CIEMAT, Avenida Complutense 40, E-28040 Madrid, Spain}
\author{Pedro G. Ferreira}
\email{pedro.ferreira@physics.ox.ac.uk}
\affiliation{\oxford}
\author{William J. Wolf}
\email{william.wolf@physics.ox.ac.uk}
\affiliation{\oxford}

\begin{abstract}
We present an assessment of the current observational status of single scalar field models of dark energy. Motivated by recent cosmological measurements -- including baryon acoustic oscillations, Type Ia supernovae, and CMB data -- we examine whether a dynamical scalar field offers a viable explanation for the accelerated expansion of the Universe. Working within an effective field theory (EFT) framework, we argue that cosmological observations are fundamentally limited and can at most constrain a small number of parameters that govern scalar field dynamics. We show that quintessence remains only marginally distinguishable from a cosmological constant, $\Lambda$, and that more general EFT extensions exhibit modest statistical preference, though such evidence is sensitive to data set selection and prior assumptions. These extended models generically predict fifth forces and modifications to the growth of structure, raising challenges from astrophysical constraints.  We compare their predictions with current growth rate measurements, Integrated Sachs-Wolfe (ISW) effect and Solar System constraints. We emphasize that viable screening mechanisms remain theoretically non-trivial and observationally testable. On the other hand, we find that current ISW and growth data remain largely in agreement. Looking ahead to Stage IV surveys we forecast improvements in constraints on the dark energy behaviour; although there will be some tightening of bounds, we argue that the problem of underdetermination will persist. We conclude that while single scalar field dark energy remains a natural and flexible framework, its ultimate viability will hinge on improved low-redshift growth measurements and a clearer understanding of gravitational screening.
\end{abstract}

\maketitle

\section{Introduction}\label{sec:intro}

For almost three decades, the $\Lambda$ Cold Dark Matter model, or $\Lambda$CDM for short, has been the overwhelming front-runner model of the Universe. A key ingredient is the cosmological constant, $\Lambda$, which drives the accelerated expansion of space at late times and has an energy density which is constant in time. For almost as long, it has been conjectured that it is actually driven by a more general model, in which $\Lambda$ is replaced by a form of {\it dark energy} with a time-evolving energy density. Recent cosmological observations have provided tentative evidence for the presence of such dynamical dark energy \cite{DESI:2024mwx, DESI:2025zgx}.

From the moment dark energy was first proposed, a huge number of candidate models have been formulated, involving fundamental fields and extensions to general relativity. Building on an idea first proposed by Ratra and Peebles \cite{Ratra:1987rm, Peebles:1987ek} (and subsequently resurrected in \cite{Ferreira:1997au,Ferreira:1997hj, Caldwell:1997ii, Perrotta:1999am, Amendola:1999er, Brax:1999gp, Barreiro:1999zs, Sahni:1999gb}) many of them involve a fundamental ``rolling'' (or time evolving) scalar field. While it may seem that this involves just one narrow subset of proposals, it turns out that many models of dark energy can be reformulated in terms of such a fundamental scalar \cite{Clifton:2011jh}. Hence, arguably, scalar field dark energy, constitutes an overwhelming majority of candidate theories.

While one might hope that ever more precise and abundant cosmological measurements might allow us to sift through the menagerie of dark energy theories, the  limited scope of such data means that the nature of dark energy is effectively underdetermined. This means that it is practically impossible to distinguish between a large number of microphysical models and, at best, one will be able to pin down a few parameters in a suitably constructed effective field theory of scalar field dark energy \cite{Wolf:2023uno}. In a number of publications, we have undertaken a programme of mapping what those parameters are and how well we can constrain them with current data \cite{Wolf:2024eph, Wolf:2024stt, Wolf:2025jed, Wolf:2025acj}.

In this paper we build on our work and make a comprehensive assessment of single scalar field dark energy and what can be said with cosmological data as well as constraints on ancillary gravitational effects. Furthermore, we look ahead, and speculate on what might be determined with up and coming Stage IV survey data. While we will show that prospects for scalar field dark energy are riven with underdetermination, it will allow us to establish how one should approach such a situation and how to assess whether scalar fields are a viable model.

We structure our paper as follows. Section \ref{sec:scalarfields} reviews the single scalar field paradigm as applied to the problem of dark energy, and in particular argues that viewing dark energy through the lens of effective field theory restricts us to three main classes of scalar field constructs. Section \ref{sec:data} reviews the cosmological data and evidence for dynamical dark energy. Section \ref{sec:prior_procedure} briefly recaps our approach for projecting theories onto the widely used CPL parametrization of the dark energy equation of state. Section \ref{sec:cosmoconstraints} assesses scalar field dark energy in light of data mainly constraining the cosmological expansion history. Section \ref{sec:ancillary} explores the ancillary gravitational consequences that follow from these scalar field models. Section \ref{sec:growth} assesses the important role that measurements of the growth of structure will have on constraining scalar field couplings to gravity, and in particular argues that low-$z$ growth measurements of $f\sigma_8$ will be especially significant. In Section \ref{sec:isw} we study the compatibility of these extended models with a particular probe of dark energy -- the Integrated Sachs-Wolfe effect. 
Section \ref{sec:future} discusses the prospects of breaking the undetermination with forthcoming Stage-IV data. Finally, we conclude in Section \ref{sec:conclusions}.

\section{Scalar Fields}
\label{sec:scalarfields}

By far the most widely favoured class of dark energy theories involve a cosmological scalar field, $\varphi$, whose energy density dominates at late times. The simplest model one might consider is {\it quintessence}: 
 \begin{equation}\label{eq:fullaction}
S=\int d^4 x\sqrt{-g}\left[\frac{M^2_{\rm Pl}}{2}R+X 
-V(\varphi)\right] +\mathcal{S}_{\mathrm{m}},
\end{equation}
 where $R$ is the Ricci scalar, $M_{\mathrm{Pl}}$ is the Planck mass, $X=-\frac{1}{2}\partial_\mu\varphi\partial^\mu\varphi$ 
and $\mathcal{S}_{\mathrm{m}}$ is the action for matter fields. If $V(\varphi)=0$ we have a free scalar field; its energy density, $\rho_\varphi$ (and pressure, $P_\varphi$) have an equation of state, $w_\varphi$, such that $w_\varphi\equiv P_\varphi/\rho_\varphi=1$. In that case $\rho_\varphi\propto a^{-6}$ (where $a$ is the scale factor of the Universe) and thus decays too fast to dominate at late times. The potential, $V(\varphi)$, changes the dynamics such that, if  at late times $X\ll V$, one has $w_\varphi\sim -1$, and the scalar field can drive accelerated expansion. The art is then to pick a potential that has the desired behaviour. 
 
Effective field theory (EFT) supplies a systematic approach to go beyond quintessence. By adding all  operators which are allowed by locality and symmetry, one can construct a hierarchy of terms which are gradually suppressed by inverse powers of the energy scale, $M$,  which signals when the theory breaks down. The lowest terms one might add will be 
  \cite{Park:2010cw}:
\begin{equation}\label{eq:EFTlowest}
\Delta S_1=\int d^4 x\sqrt{-g}\left[\frac{M^2_{\rm Pl}}{2}f(\varphi)R+k(\varphi)X\right].
\end{equation}
Through a field redefinition, it is generally possible to set $k(\varphi)$ to a constant. If one wants to go to ${\cal O}(M^{-3})$, the situation becomes more complicated. For a start one must include the cubic Galileon term
\begin{eqnarray}\label{eq:EFTSS}
\Delta S_2 = \int d^4 x\sqrt{-g} &\frac{h_1(\varphi)}{M^3}X\Box \varphi
\end{eqnarray}
But also one needs to include a term in $\Box \varphi/M^3$ in $f$. This term inevitably leads to higher order equations of motion and the inevitable Ostrogradsky instabilities \cite{Woodard:2015zca}. While there is a well established approach to dealing with these instabilities \cite{Burgess:2014lwa}, we have opted not to include this complication in the analysis in this paper and we leave it for future work.

\begin{figure}[t]
   \centering
   \includegraphics[width=\columnwidth]{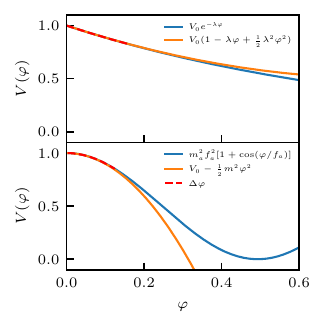}
   \vskip -0.1in
   \caption{(Top) The exponential potential for a scalar field with a non-minimal coupling to gravity,the Taylor expansion of this potential, and the total field excursion over all of cosmic history. (Bottom) The axion potential in minimally coupled quintessence, its Taylor expansion (with $V_0=2m^2_af^2_a$ and $m^2=m^2_a$), and the total field excursion. In both cases, over the relevant range of field excursions, the potentials are essentially indistinguishable from their Taylor expansion out to quadratic order.}
   \label{fig:potentials_quad_expansion}
\end{figure}

Following the EFT approach, we have ended up with a number of free functions which, in principle, can take any shape or form. However, in practice, observations only probe a very narrow range of the scalar field evolution. For example, in the case of {\it quintessence} where we only have a potential, observations impose $V(\varphi)\gg X$ to agree with the preferred $w\sim -1$.  In terms of the field variation, $\Delta \varphi$, this means that $\Delta \varphi/M_{\rm Pl}\simeq {\dot \varphi}/(M_{\rm Pl}H_0)\ll 1$ over the range that is effectively probed. This is not specific to quintessence and is in fact general to scalar field dark energy (although one should take care in the case of shift symmetric theories, such as Galileons). As a consequence, the EFT functions $f$ and $V$, $h_i$ can, without loss of generality, be Taylor expanded in $\varphi/M_{\rm Pl}$ to, at most, quadratic order. Note that, in the case in which we include the cubic Galileon term, we should, for consistency, expand to cubic order in the $f$ and $V$. In practice, it makes no difference in the observables, compared to truncating them at quadratic order \cite{Wolf:2025acj}.

To illustrate this point, let us consider two examples: a minimally coupled axion, 
\begin{equation}
    V(\varphi) = m^2_af^2_a\left(1+\cos\left[\frac{\varphi}{f_a}\right]\right) ;\quad f(\varphi) = 1,
\end{equation}
and the non-minimally coupled exponential potential
\begin{equation}
    V(\varphi) = V_0e^{-\lambda\varphi} ;\quad f(\varphi) = 1-\xi\left(\frac{\varphi}{M_{\rm Pl}}\right)^2,
    \label{eq:nonmimexp}
\end{equation}
and let us approximate their potentials by their Taylor expansions up to quadratic order,
\begin{equation}
    V(\varphi) \simeq V_0 + \beta \varphi + \frac{1}{2}m^2\varphi^2.
\end{equation}
If we evolve the scalar field and Friedmann equations (with dark energy density today $\Omega_{\varphi0} = 0.69$ and Hubble parameter $H_0 = 67~\mathrm{km\,s^{-1}\,Mpc^{-1}}$), we find that $\Delta \varphi/M_{\rm Pl} \simeq \mathcal{O}(0.1)$ and that the models and their Taylor expansions are essentially indistinguishable over this range of field values (see Figure~\ref{fig:potentials_quad_expansion}). For the axion example, $w_\varphi(z=0) \simeq -0.80$ and for this non-minimally coupled exponential $w_\varphi (z=0) \simeq -0.90$; in both cases the resulting equation of state determined from the Taylor expansion matches these values to sub-percent levels of precision. As is clear from examining Figure \ref{fig:potentials_quad_expansion}, this approximation would hold to a high degree of accuracy even with significantly more field evolution, indicating a substantial observational degeneracy among microphysical dark energy proposals. Furthermore, this approach is valid in any dark energy theory where the potential is given by an analytic function admitting such an expansion. 
Consequently, one can comprehensively explore all single scalar field models of dark energy with ease, as in \cite{Wolf:2023uno, Wolf:2024eph, Wolf:2025jed, Wolf:2025acj} where various corners of this theory space have been comprehensively explored. 

\begin{table*}[t]
\centering
\renewcommand{\arraystretch}{1.3}
\setlength{\tabcolsep}{12pt}
\begin{tabular}{lcccc}
\hline\hline
Scalar field EFT operators 
& $V(\varphi)$ 
& $k(\varphi)$ 
& $f(\varphi)$ 
& $h_1(\varphi)$ \\
\hline
Minimal quintessence
& $V_0 + \frac{1}{2}m^2\varphi^2$
& $1$
& $1$
& $0$
\\

Non-minimal quintessence (NM)
& $V_0 + \beta\varphi + \frac{1}{2}m^2\varphi^2$
& $1$
& $1-\xi \varphi^2$
& $0$
\\

Massive Galileon (mG)
& $V_0 + \frac{1}{2}m^2\varphi^2$
& $\alpha$
& $1$
& $\gamma$
\\
\hline\hline
\end{tabular}
\caption{
Schematic summary of the effective field theory ingredients for the scalar--tensor models considered in this work, in the conventions from Eqs.~\ref{eq:EFTlowest} and \ref{eq:EFTSS}.
}
\label{tab:eft_schematic}
\end{table*}

The EFT presented here is in the Jordan frame, where the non-minimal coupling, $f(\varphi)$, is manifest. It is often convenient to work in the Einstein frame where one has a canonical Einstein-Hilbert term but in which matter now couples directly to the scalar field.\footnote{See \cite{Perrotta:1999am, Baccigalupi:2000je, Baccigalupi:2001aa, Boisseau:2000pr, Gannouji:2006jm, Amendola:1999er, Amendola:1999qq} for some of the original literature on scalar field dark energy where the non-minimal coupling is present in either the gravitational or matter sectors.} Indeed, it is in this frame that one can see most clearly, for non-minimally coupled theories,  that there is a universal fifth force (in the sense that the scalar field couples equally to all components of the energy momentum tensor of the Universe) -- such a fifth force satisfies the Weak Equivalence Principle. In terms of the Einstein frame metric, $g^{\rm E}_{\mu\nu}$, we  have that, schematically, the coupling to the matter is explicit in the matter Lagrangian which takes the form ${\cal L}_{\rm M}(f^{-1}g^{\rm E}_{\mu\nu},\psi)$, where $\psi$ represents all the matter fields in the Universe. If we consider, as argued above, a Taylor expansion for $f(\varphi)$ as in Equation \ref{eq:nonmimexp}, we have that\footnote{We can, of course, redefine the field so that it has a canonical kinetic energy in the Einstein frame but one can easily show that for small $\sqrt{\xi} \varphi$ the Einstein and Jordan frame fields are essentially equivalent.} the coupling to the Einstein metric is of the form 
\begin{eqnarray}
f^{-1}g^{\rm E}_{\mu\nu}\simeq \left[1+\xi\left(\frac{\varphi}{M_{\rm Pl}}\right)^2\right]g^{\rm E}_{\mu\nu}.
\end{eqnarray}

In summary, we are considering the case of dark energy taking the form of a single scalar field which, when non-minimally coupled will do so universally, satisfying the WEP. The case of a WEP violating scalar fields, such as in the case of interacting dark matter/dark energy models, or the case of multiple scalar fields will briefly be discussed in Appendix \ref{sec:appendixB}.

The full EFT is encapsulated in the parametrizations shown in Table \ref{tab:eft_schematic}, covering all possible single scalar field models at low energies which are relevant for the cosmological energy scales being probed by the late time expansion of the Universe. For convenience, when looking beyond the standard, minimally coupled, scalar field theory -- quintessence -- we have divided the space of {\it extended models} into two: the non-minimally coupled scalar field (NM) -- note that minimal thawing quintessence is a subset of these -- and a massive Galileon (mG), which is a cubic Galileon where the shift symmetry is broken by a massive potential. Ideally, we would have explored the full space given by the EFT in one go, varying all parameters simultaneously. It turns out that is numerically unfeasible (at least at the moment) and the results will be no different, in practice, from slicing the space up the way we do here.

One could consider a combination of the two, but our results would be unaltered. One could also consider additional higher order terms but their effects will be suppressed relative to these. Finally, if one were to choose higher order terms {\it without} including the lowest order terms we are discussing here, one would have to invoke some new principle for suppressing the latter. This is an added complication which severely limits the generality of what we are trying to achieve here. Thus, we will look at single scalar field dark energy in its full generality, from the perspective of EFT.

It is useful to quickly mention one particular case where symmetry excludes the presence of certain lower order terms: shift symmetric theories, like Galileons. There one can argue that $f=V=0$. But, as was shown in \cite{Traykova:2021hbr, Wolf:2025acj} and proven in \cite{Tsujikawa:2025wca}, such theories lead to an equation of state $w<-1$, i.e.~which is phantom at all times and which is severely counter to current observational evidence. Thus it is not worthwhile pursuing such theories further.

As discussed in our prior work \cite{Wolf:2024eph, Wolf:2024stt, Wolf:2025acj, Wolf:2025jed, Wolf:2025jlc}, the NM and mG models have been implemented in \texttt{hi\_class} \cite{hi_class1, hi_class2, CLASS} in order to compute the background quantities and linear cosmological perturbations, and we have used \texttt{Cobaya} \cite{Cobaya, Cobaya2} to sample and perform Bayesian inference for cosmological parameters using Markov Chain Monte Carlo (MCMC) methods \cite{metropolismc, hastingsmc, Lewis:2013hha}. For thawing quintessence and the non-minimally coupled scalar field, $V_0$ is in units of $M^2_{\mathrm{Pl}} M^2_{\mathrm{H}}$ (where $M_{\mathrm{H}} = H_0/h$), $\beta$ is in units of $M_{\mathrm{Pl}} M^2_{\mathrm{H}}$, $m^2$ is in units of $M^2_{\mathrm{H}}$ and $\xi$ is dimensionless. For the massive Galileon model, $V_0$ is in units of $M^2_{\mathrm{Pl}} \tilde{H}_0^2$, $m^2$ is in units of $\tilde{H}_0^2$, and $\alpha$ is dimensionless. Here, $\tilde{H}_0 = 67.5 \, \mathrm{km\,s^{-1}\,Mpc^{-1}}$ is set to a fiducial value along with setting $\gamma = -1/\tilde{H}_0^{2}$ to fix the normalization of the field as was done in \cite{Wolf:2025acj, Traykova:2021hbr}.

\section{The evidence for evolving dark energy}
\label{sec:data}

It is useful to examine in some detail the current evidence for evolving dark energy. At late times, one assumes a Friedmann--Robertson--Walker cosmology described by
\begin{eqnarray}
H^2(a)\equiv\left(\frac{\dot a}{a}\right)^2
=H_0^2\left[\Omega_{\rm m}a^{-3}+(1-\Omega_{\rm m})S(a)\right],
\end{eqnarray}
where radiation and massive neutrinos have been neglected for simplicity in this argument. Here $a$ is the scale factor and $S(a)$ encodes the time evolution of the dark-energy density,
\begin{eqnarray}
S(a) = \frac{\rho_{\rm DE}(a)}{\rho_{\rm DE, 0}}=\exp\!\left[-3\int_a^1\!\frac{1+w(a')}{a'}\,{\rm d}a'\right].
\end{eqnarray}
The dark energy equation of state is commonly parametrized using the Chevallier-Polarski-Linder (CPL) form \cite{Chevallier:2000qy, Linder:2002et},
\begin{equation}\label{eq:CPL}
w(a)=w_0+w_a(1-a),
\end{equation}
where $w_0$ denotes the present-day value and $w_a$ captures its time variation \cite{Linder:2002et,Chevallier:2000qy}.\footnote{Other two-parameter forms have been explored extensively, but current data are largely insensitive to the precise functional choice \cite{DESI:2025fii, Wolf:2025jlc, Shlivko:2025fgv, Montefalcone:2026iga}.}  
The $\Lambda$CDM limit corresponds to $(w_0,w_a)=(-1,0)$ and evidence for evolving dark energy arises when $w_a\neq0$ is preferred.

The original evidence for $w_a\neq0$, at up to $4.2\sigma$, came from combining Dark Energy Spectroscopic Instrument (\desi) Data Release (DR) 2 measurements of Baryon Acoustic Oscillations (BAO) with  Dark Energy Survey Year 5 (\desyf) Type Ia supernovae (SNe) and Cosmic Microwave Background (CMB) data (specifically \planck 2018 primary anisotropies together with \planck+ Atacama Cosmology Telescope (\act) DR6 lensing). Choosing other SNe samples, the significance ranges from $2.8\sigma$ (\pantheonp) to $3.8\sigma$ (\uniont). In all cases, the preferred equation of state is phantom ($w<-1$) at earlier times and crosses $w=-1$ at $a\simeq0.75$ ($z\simeq0.33$). A reanalysis of the \desyf data in which the photometric redshifts were recalibrated (the ``Dovekie'' analysis (\desdov) \cite{DES:2025sig}) has reduced the significance to $3.2\sigma$, while a similar re-analysis of \uniont (``Union3.1'') has slightly reduced its significance to $3.4\sigma$ \cite{Hoyt:2026fve}. Below we briefly summarize the role of the main data sets.

\noindent{\it DESI DR2 BAO:}
\desi is a Stage-IV spectroscopic survey targeting a variety of galaxy populations -- Bright Galaxies with the Bright Galaxy Survey (BGS), Emission Line Galaxies (ELG), Luminous Red Galaxies (LRG) and Quasars (QSO) -- to measure BAO across a wide redshift range. BAO constrain the angular-diameter distance $D_A(z)$ and the Hubble distance $D_H(z)=H^{-1}(z)$, sometimes summarized by the isotropic distance
$D_V(z)\equiv\!\left(zD_M^2D_H\right)^{1/3}$.
Relative to the \planck 2018 $\Lambda$CDM best fit \cite{Planck:2018vyg}, \desi DR2 prefers a smaller late-time volume and a matter density lower by $\sim1.8\sigma$, with a $\sim2\sigma$ discrepancy in the $(\Omega_{\rm m},hr_{\rm d})$ plane \cite{DESI:2025zgx}. Notably, the \desi best fit lies along the CMB degeneracy direction $\Omega_{\rm m}h^3$, and full-shape analyses yield consistent conclusions \cite{DES:2025sig}. It is also worth noting that, independent of any SNe data, the combination of \desi BAO and CMB data is in $\sim 2\sigma$ tension with $\Lambda$.

\noindent{\it Type~Ia supernovae:}
The \desyf sample comprises 1635 high-quality \des supernovae in $0.10<z<1.13$, supplemented by lower-redshift samples \cite{DES:2024jxu}. While $\Lambda$CDM provides an acceptable fit, \desyf shows a $\sim2\sigma$ preference for evolving dark energy with $w_0>-1$ and $w_a<0$. As mentioned above, a recent recalibration of DES photometric redshifts (the ``Dovekie'' analysis) reduces this preference \cite{DES:2025sig} from $4.2\sigma$ to $3.2 \sigma$.  

\pantheonp combines 1550 SNe spanning $0.001<z<2.26$ from multiple surveys \cite{Brout:2022vxf}. When combined with CMB data, it yields constraints consistent with \desyf along the same $(w_0,w_a)$ degeneracy direction. However, without external data, \pantheonp alone is compatible with $w_0=-1$, and the degeneracy is primarily fixed by the CMB. Differences between \pantheonp and \desyf methodologies, particularly at low redshift, have been discussed extensively \cite{Efstathiou_2024, DES:2025tir}, with the latter arguing that \desyf better controls systematics.

\noindent{\it CMB experiments:}
The combined BAO+SN+CMB analyses use \planck 2018 primary CMB data \cite{Planck:2018vyg, Planck:2019nip} together with \act DR6, as well as \planckact lensing. When combined with \desi BAO, a preference for dynamical dark energy emerges largely independently of the specific CMB dataset used (\planck, \planckact, or \wmap+\act) \cite{AtacamaCosmologyTelescope:2025nti}. Recent South Pole Telescope  (\spt) results strengthen this picture: combining \spt, \act, \planck, \desi BAO and \pantheonp SNe yields a $2.9\sigma$ deviation from $\Lambda$ \cite{SPT-3G:2025bzu}. Within $\Lambda$CDM, there is also a $2$--$3.7\sigma$ inconsistency between CMB and \desi in the $(\Omega_{\rm m},hr_{\rm d})$ plane, which is alleviated in models with evolving dark energy.

A comprehensive analysis of various combinations of data sets has been undertaken in \cite{Ong:2026tta} where the tensions between data sets and dependence of evidence on priors is explored in detail. The authors conclude that there is no evidence for evolving dark energy. 
Following the last official \planck PR3 release, there has also been subsequent re-analyses of the data which have resulted in the publication of new PR4 likelihoods \cite{Planck:2020olo, PR4_hilli_lolli}. The change on likelihood impacts the so-called \textit{$A_{\mathrm{lens}}$} anomaly, where \textit{$A_{\mathrm{lens}}$} is a phenomenological rescaling parameter that artificially re-weights the contribution of the CMB weak lensing correction to the CMB primary anisotropies ($A_{\text{lens}}=1$ is the fiducial value for $\Lambda$CDM). The PR3 releases favour a $A_{\text{lens}} > 1$ with greater than $2\sigma$ significance \cite{Planck:2019nip}, which could impact inferences regarding the presence of signals for extended models \cite{Planck:2018vyg}. The PR4 release brings $A_{\text{lens}} \rightarrow 1$, effectively eliminating the anomaly.

Despite the care taken in constructing these datasets, residual systematics remain possible. Discrepancies between \desyf and \pantheonp supernova magnitudes have been partially traced to differences in selection functions, intrinsic scatter models and host-mass corrections, leaving a small unexplained offset but that should have no impact in the cosmological constraints \cite{DES:2025tir}. Combining \desi with \sdss BAO reduces the significance of evolving dark energy to $2$--$3.5\sigma$ depending on the SNe sample \cite{DESI:2024mwx}, although \desi DR2 has substantially greater statistical power. Detailed consistency tests find \desi DR2 and \sdss to be statistically compatible \cite{DESI:2025zgx}.

Additional tensions and degeneracies complicate the picture. Within $\Lambda$CDM, \desi BAO+CMB prefer very low (or even formally negative\footnote{Negative neutrino masses are unphysical and their preference should be interpreted as encoding another physical effect or systematic that has the same effect. We refer to \cite{Elbers:2025vlz} for a detailed discussion. In short, lowering neutrino masses allows for a larger inferred $H_0$ that brings the \lcdm CMB prediction for $D_V/r_d$ into closer alignment with \desi data, which translates into agreement on $H_0r_{\rm d}$. It also allows to fit the oscillatory pattern of low-$\ell$ CMB temperature power spectrum, and for low values of the reionization optical depth $\tau$ parameter, in agreement with Planck's polarization data, which, in turn, increases $H_0 r_{\rm d}$ in better agreement with \desi.}) neutrino masses, in tension with laboratory constraints which set a lower bound on their masses of $0.06$eV, assuming normal ordering, that is the case least in tension \cite{Esteban:2024eli}.  However, opening the parameter space to allow for dynamical dark energy, for instance in terms of $w_0$--$w_a$, restores the compatibility with terrestrial constraints \cite{Elbers:2025vlz, Elbers:2024sha, Green:2024xbb, Kibris:2026cqq}. If spatial curvature is considered, the neutrino mass constraints are relaxed, reducing the tension. In addition, one finds a mild ($\sim2\sigma$) preference for $\Omega_K>0$ ($<0$ in DR1 \cite{DESI:2024mwx}) that goes away once evolving dark energy and SNe are included \cite{Chen:2025mlf, DESI:2025zgx}. Finally, a possible underestimation of the optical depth $\tau$ from low-$\ell$ CMB polarization could reduce the \desi--CMB tension (as well as a few other cosmological tensions) within $\Lambda$CDM if $\tau$ were higher by $\sim2$--$3\sigma$   \cite{Allali:2025yvp, Sailer:2025lxj, Jhaveri:2025neg}. However, replacing the low-$\ell$ polarization data with constraints from the latest measurements of the redshift evolution of the neutral hydrogen fraction from the Lyman-$\alpha$ forest and from damping wing measurements bring the constraints on $\tau$ firmly down to the canonical value obtained with the full \planck CMB \cite{2508.21069, Kageura:2026ryq}.

A key assumption in deriving constraints on the expansion rate is that the Universe is perfectly homogeneous and isotropic. This has been, until now, an excellent approximation. But as measurements have become increasingly precise, it may make sense to revisit these assumptions and see how crucial they are in the scientific interpretation of current observations. The authors of \cite{Ginat:2026fpo} have argued that different aspects of large scale inhomogeneities, while small, may bias constraints towards evolving dark energy, even though the underlying cosmology may be $\Lambda$ dominated.

\begin{figure}[t]
   \centering
    \includegraphics[width=\columnwidth]{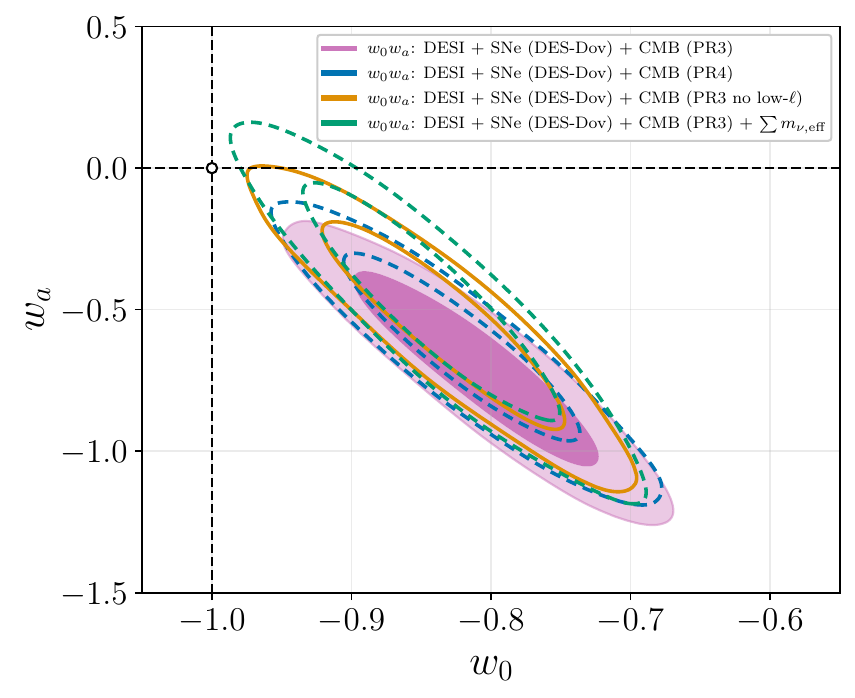}
    
   \vskip -0.1in
   \caption{Constraints on the $w_0w_a$ CPL parameters (68\% and 95\% C.L.) under a few different data combinations and assumptions. All scenarios depicted above display a preference for dynamical dark energy over $\Lambda$, but the strength of this preference varies.}
   \label{fig:cpl_data}
\end{figure}

In summary, currently there is no clear evidence for $\Lambda$ versus evolving dark energy or vice versa. Any preference for $w_a\neq0$ is marginal, dependent on prior or choices of data sets \cite{Ong:2025utx, Ong:2026tta}. Figure \ref{fig:cpl_data} depicts constraints on the CPL parameters Eq.~\eqref{eq:CPL} under the various scenarios discussed in this section. Thus we are firmly in a regime where prior assumptions on what model one wishes to explore dictates the way forward. In this paper, as advertised, we will explore what the data have to say about single scalar field dark energy.

\section{$(w_0, w_a)$ as data compression}\label{sec:prior_procedure}

Before embarking onto the cosmological analysis of scalar field dark energy, it is worth explaining carefully how we interpret $(w_0, w_a)$ constraints in this and the preceding papers. First, it is important to remark that $(w_0, w_a)$ are not the parameters of a Taylor expansion of the equation of state of a dark energy model. Instead, they must be understood as a way of compressing the phenomenology of cosmological models into two variables. In some sense it is a projection of the full Lagrangian onto a reduced functional space that is able to capture the whole variety of observational changes that a dark energy model can produce.

The idea dates back to our goal of finding theoretical priors in a reduced parameter space as we did for quintessence \cite{Garcia-Garcia:2019cvr} and shift-symmetric theories \cite{Traykova:2021hbr}, finding that the CPL parametrization did an excellent job on compressing the whole background-level phenomenology\footnote{Note that one can also compress, in a similar way, the linear perturbations into a reduced basis \cite{Traykova:2021hbr}.} of these models when fitting cosmological observables like $H(z)$, $D_A(z)$, $f\sigma_8(z)$, etc.\ to within 1\%. However, this is not enough to compare with the constraints obtained with actual data as observations do not come at all redshifts and with equal errors. In \cite{Wolf:2023uno, Wolf:2024eph}, we refined the procedure to use the observables at the redshifts probed by a given data set, and with their associated relative covariance.

The method can be summarized as follows:
\begin{enumerate}
    \item Define a Lagrangian $\mathcal{L}$ with fields $(\varphi,\psi,\cdots)$, parameters $(\alpha,\beta,\cdots)$, and initial conditions $(\varphi_i,\dot{\varphi}_i,\cdots)$. 
    \item Generate a Latin hypercube for all the parameters in your model, with the prior ranges that are physically sound.
    \item For each sample  (i.e.\ point in the hypercube), generate the expansion history $H^2_{\varphi}$ predicted by the microphysical model. From this predicted expansion history, calculate the observables probed by the data sets you want to compare the $(w_0, w_a)$ constraints against, at the redshifts they were observed. For example, in the case of \desi BAO, these would be $D_{\rm A}$, $D_{\rm M}$ and $D_{\rm V}$ at the \desi redshifts. Or, with \planck, it would be the background observables that can be used to effectively compress the power spectra to remarkable precision \cite{Wolf:2024eph, Chen:2018dbv}.
    \item Now, using a $w_0w_a$CDM cosmology, compute $H^2_{(w_0,w_a)}$ and the equivalent set of observables, and determine the $(w_0, w_a)$ parameters that best fit the observables predicted by the model Lagrangian $\mathcal{L}$, weighted by the associated relative covariance of the chosen observables.
\end{enumerate}
Following these steps, one can populate the $(w_0, w_a)$ plane and find a distribution that we called ``theory prior''. As these theory priors were determined by using the same covariance as a given set of data, one can then make an ``apples to apples'' comparison between the distribution of $(w_0, w_a)$ points determined as a compression of the microphysical theory's predicted expansion history and observables and the $(w_0, w_a)$ distribution determined directly from the data measurements. Additionally, it is worth emphasizing that there is no unique representation of a theory's compression into this parameter space, as the best fitting $(w_0, w_a)$ parameters will depend on the specific observables and redshifts probed, in addition to the respective uncertainties on these quantities \cite{Wolf:2023uno, Wolf:2024eph}. 
Furthermore, we have demonstrated that the $(w_0, w_a)$ distribution fitted through this procedure recovers the predicted observables for the microphysical model to an excellent degree of accuracy \cite{Wolf:2025jlc}. In this paper, when compressing a microphysical model's expansion history into $(w_0, w_a)$, we always utilize the observables and uncertainties given by our baseline data combination which consists of {\tt plik} \planck PR3 CMB power spectra, \planckact CMB lensing, \desdov SNe, and \desi DR2 BAO \cite{Planck:2018vyg, ACT:2023kun, DESI:2025zgx, DES:2025sig} (\desidesdovcmb, in short), where we have used both exact background observables or compressed versions, where necessary, as described in more detail in \cite{Wolf:2024eph, Wolf:2025jlc}.

Throughout this paper we will follow this philosophy -- that $(w_0, w_a)$ is a form of data compression -- and we can use it as an efficient way of {\it visualizing}, in a compact form, the impact of each scalar field theory on cosmic evolution. Unless stated otherwise, we always use the same data combination for the projection, \desidesdovcmb. We should emphasize, however, that, we {\it do not} use this data compression (the $(w_0, w_a)$ parameters) when undertaking a full Bayesian inference -- we do that by comparing the model predictions of full scalar field dynamics directly with the data. That is, for the goodness of fit $\Delta \chi^2$, evidences, scalar field parameter constraints, etc.\ we run MCMC, solving the field equation with \texttt{hi\_class} \cite{hi_class1, hi_class2} and sampling with \texttt{Cobaya} \cite{Cobaya, Cobaya2}. 

\section{Cosmological constraints on scalar field dark energy}
\label{sec:cosmoconstraints}

\begin{figure*}[t]
   \centering
   \includegraphics[width=0.48\textwidth]{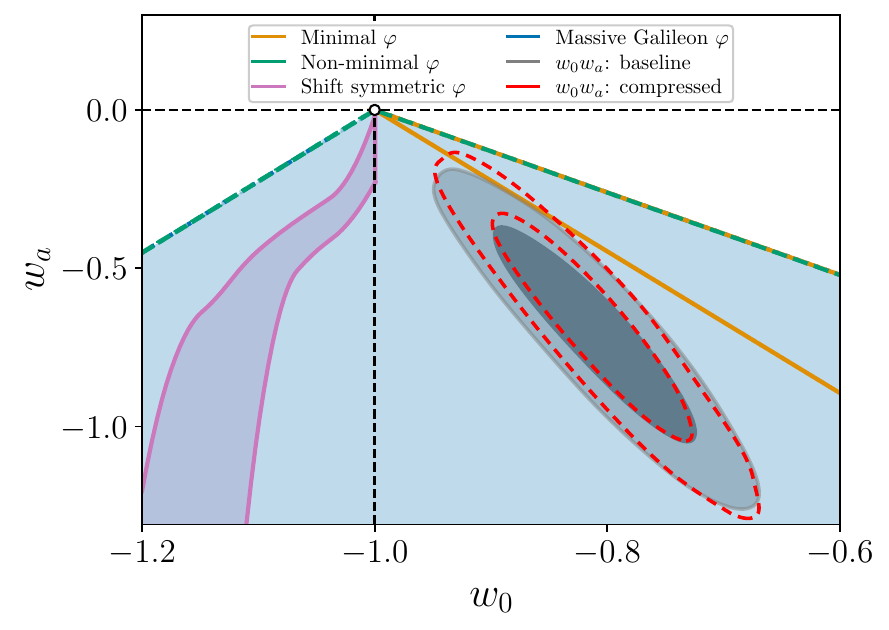}
   \hfill
   \includegraphics[width=0.48\textwidth]{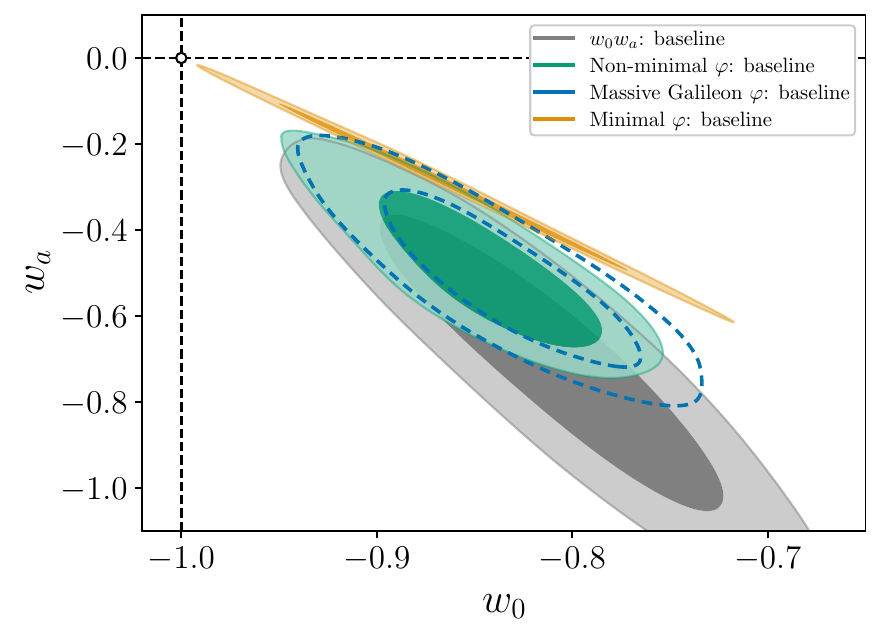}
   \vskip -0.1in
   \caption{On the left, we depict the area spanned by the theory priors for each dark energy model considered in this work against both the constraints from the baseline data combination (\planck PR3 CMB power spectra, \planckact CMB Lensing, \desdov SNe, and \desi DR2 BAO) and their compressed version used in the projection (see Section~\ref{sec:prior_procedure}), interpreted in terms of the CPL $(w_0, w_a)$ parameterization. On the right, after obtaining constraints on the dark energy models, we then sample from the converged chains and project these models' posteriors into the $(w_0, w_a)$ parameter space following Section~\ref{sec:prior_procedure}. Here one sees that the phantom-crossing dark energy models project right into the heart of the $(w_0, w_a)$ data constraints, whereas quintessence projects into a tiny sliver right on the edge of the data posteriors.}
   \label{fig:theory_priors_post}
\end{figure*}

Having examined the cosmological evidence for evolving dark energy, we now turn to what the data can tell us about its underlying nature, assuming the observations are correct. Much of the interpretation of this phenomenon has relied on phenomenological descriptions in which dark energy is characterized solely by an equation of state, $w(a)$. While such parameterizations are useful, our goal is more ambitious: to identify a microphysical explanation for dark energy that can be embedded within a broader framework of fundamental physics.

Cosmological data are often interpreted using simple parametric models of the equation of state, such as the CPL form given in Eq.~\eqref{eq:CPL}. However, the equation of state predicted by a microphysical dark energy model does not generally follow an arbitrary phenomenological parameterization. Instead, it depends on the dynamics of an underlying field, typically a scalar field, and its time derivatives. In this case, the equation of state is given by $w_{\varphi} \equiv P_{\varphi}/\rho_{\varphi}$ and can exhibit complicated and highly non-linear behavior. While minimally coupled canonical quintessence models satisfy $w_\varphi \geq -1$, more general theories with non-minimal couplings to gravity or non-trivial kinetic terms can violate this bound, allowing for phantom behavior and crossings of the phantom divide \cite{Boisseau:2000pr, Gannouji:2006jm}.

As a first pass at assessing how well the classes of scalar-field models discussed in Section~\ref{sec:scalarfields} describe the expansion history favoured by current constraints coming from \planck PR3 CMB power spectra, \planckact CMB lensing, \desdov SNe, and \desi DR2 BAO \cite{Planck:2018vyg, ACT:2023kun, DESI:2025zgx, DES:2025sig}, we apply the procedure described in Section \ref{sec:prior_procedure} to sample from broad prior ranges for the scalar field models free parameters and determine the theory priors. Figure~\ref{fig:theory_priors_post} (left) shows the resulting theory priors for quintessence, NM, mG, and shift-symmetric scalar fields (using the general shift-symmetric model of \cite{Traykova:2021hbr}), consistent with our previous work \cite{Wolf:2024eph, Wolf:2025jed, Wolf:2025acj}. While quintessence provides a slightly better fit than $\Lambda$, it fails to describe the full expansion history data to high accuracy across all redshifts. In contrast, both NM and mG models provide an excellent fit, fully spanning the allowed $(w_0, w_a)$ parameter space. This arises from the additional structure in their inferred pressure and energy density, which allows the equation of state to enter the phantom regime before thawing at late times.

While the former allows us to find the ``theory priors''; i.e. the area of the $(w_0, w_a)$ plane that the models span; we can use it directly on the cosmological parameters posterior distributions to find the corresponding $(w_0, w_a)$ posterior. In this case, we use directly the steps of the converged MCMC chains (to be discussed shortly) rather than sampling from broad theory priors, with corresponding cosmological and dark energy parameters constraints shown in Table~\ref{tab:cosmo_constraints}. The resulting mapping, shown in Figure~\ref{fig:theory_priors_post} (right), demonstrates that the theory priors collapse in the expected manner: the extended models cluster around the observational $(w_0, w_a)$ posteriors, while minimally coupled quintessence occupies a narrow region at the edge of the $95\%$ confidence level of the CPL data constraints, as suggested by the theory priors. Let us highlight here that the resulting $(w_0, w_a)$ distributions represent a powerful compression of the microphysical model's constrained expansion histories as they reproduce the observables they were fit to with excellent accuracy (see also \cite{Wolf:2025jlc} for more details). Between the BAO, compressed SNe, and compressed CMB observables, we are fitting the CPL model to $\simeq20$ data points (along with the corresponding relative uncertainties associated to each data set) for each realization of a microphysical dark energy expansion history. The fits are typically excellent, with $\chi^2 \lesssim 1$ and a reduced $\chi^2_{\rm r}$ of $< 0.1$. In other words, the $(w_0, w_a)$ projection depicted on the right side of Figure \ref{fig:theory_priors_post} faithfully represents the posterior distributions and constrained expansion histories of the respective dark energy models.

\begin{table*}[t]
\centering
\renewcommand{\arraystretch}{1.25}
\setlength{\tabcolsep}{4pt}
\footnotesize

\begin{tabular}{lccccccccc}
\hline\hline
Model & $\Omega_{\rm m}$ & $h$ & $w_0$ & $w_a$ & $V_0$ & $\beta$ & $m^2$ & $\xi$ & $\alpha$ \\
\hline

$\Lambda$CDM
& $0.303\pm 0.004$ & $0.683\pm 0.003$ & $-1$ & 0
& -- & -- & -- & -- & -- \\

CPL ($w_0w_a$)
& $0.314\pm 0.006$ & $0.674\pm 0.006$ & $-0.804^{+0.054}_{-0.062}$ & $-0.72^{+0.24}_{-0.21}$
& -- & -- & -- & -- & -- \\

Minimal $\varphi$
& $0.317^{+0.006}_{-0.008}$ & $0.667\pm 0.007$ & -- & --
& $0.991\pm 0.016$ & 0 & $-56^{+22}_{-34}$ & 0 & 1 \\

Non-minimal $\varphi$
& $0.312\pm 0.005$ & $0.674^{+0.005}_{-0.006}$ & -- & --
& $ 0.71^{+0.10}_{-0.12}$ & $2.45^{+0.54}_{-0.36}$ & $-2.4^{+1.5}_{-1.2}$ & $2.21^{+0.85}_{-0.35}$ & 1 \\

M. Galileon $\varphi$
& $0.313^{+0.005}_{-0.006}$ & $0.672\pm 0.006$ & -- & --
& $2.23^{+0.08}_{-0.51}$ & 0 & $-21.4^{+14}_{-8.8}$ & 0 & $-2.22^{+0.29}_{-0.42}$ \\

\hline\hline
\end{tabular}

\caption{
Constraints (mean and 68\% C.L. errors) on the cosmological and dark energy parameters with our baseline data combination (\desidesdovcmb). Note that the \lcdm and, specially, the CPL constraints reproduce very closely the constraints from \cite{DES:2025sig}.
}
\label{tab:cosmo_constraints}
\end{table*}

From these $(w_0, w_a)$ posterior distribution representations, we can estimate the tension of each model with \lcdm through (e.g. \cite{Motloch:2018pyt, Raveri:2018wln})
\begin{equation}\label{eq:suspicion}
\chi^2=\left(\mu_A-\mu_B\right)^T\left(\operatorname{Cov}_{\mathrm{A}}+\operatorname{Cov}_{\mathrm{B}}\right)^{-1}\left(\mu_A-\mu_B\right),
\end{equation}
where we have approximated the posterior distributions as Gaussians, with mean $\mu_i$ and covariance ${\rm Cov}_i$, for the case $i$. We determine the $(w_0, w_a)$ representation for \lcdm using the same procedure  (sampling the converged chains and then fitting the CPL model to the resulting expansion history using the uncertainties of the relevant data). While there is a very small spread in $w_0$ and $w_a$ values, in practice it is too small to see in Figure \ref{fig:theory_priors_post} as the distribution is essentially a delta function at $w_0=-1$ and $w_a=0$ with negligible covariance. In this case, Eq.~\ref{eq:suspicion} becomes the squared Mahalanobis distance. For each $\chi^2$, we can compute the corresponding $p$-value
\begin{equation}\label{eq:p-value}
p
= 
\int_{|\chi^2|}^{\infty}
f_{\chi^2_{k}}(x)\,dx,
\end{equation}
where $f_{\chi^2_{k}}$ is the PDF of a $\chi^2$ distribution with $k$ degrees of freedom (in this case $k =2$). This can then be converted to an equivalent number of Gaussian sigmas through
$n_\sigma
=
\sqrt{2}\,
\operatorname{erf}^{-1}(1-p).$
The result is that we see $n_\sigma\simeq 2.27$ (quintessence), $n_\sigma\simeq 3.62$ (mG), and $n_\sigma\simeq 3.77$ (NM) discrepancies with respect to \lcdm when analyzing their inferred $(w_0, w_a)$ distributions.

\begin{table*}[t]
\centering
\renewcommand{\arraystretch}{1.15}
\setlength{\tabcolsep}{6pt}

\begin{tabular}{lcccc}
\hline\hline
 & Baseline & \planck PR4 & $\sum m_{\nu,\mathrm{eff}}$ & High-$\tau$/no low-$\ell$ EE \\
\hline
\multicolumn{5}{c}{$\Delta\chi^2_{\varphi\Lambda}$ ($\Delta \mathrm{AIC}_{\varphi\Lambda}$) [$\sigma$]} \\
\hline
Minimal $\varphi$   & -7.5 (-3.5) [2.26$\sigma$] & -6.9 (-2.9) [2.15$\sigma$] & -9.3 (-5.3) [2.59$\sigma$] & -8.7 (-4.7) [2.49$\sigma$] \\
$w_0w_a$    & -13.4 (-9.4) [3.23$\sigma$] & -11.7 (-7.7) [2.98$\sigma$] & -8.7 (-4.7) [2.49$\sigma$] & -8.1 (-4.1) [2.38$\sigma$] \\
Non-minimal $\varphi$   & -15.7 (-9.7) [3.21$\sigma$] & -12.0 (-6.0) [2.68$\sigma$] & -11.8 (-5.8) [2.65$\sigma$] & -11.6 (-5.6) [2.62$\sigma$] \\
Massive Galileon $\varphi$  & -15.4 (-9.4) [3.17$\sigma$] & -10.6 (-4.6) [2.45$\sigma$] & -12.1 (-6.1) [2.69$\sigma$] &  -11.8 (-5.8) [2.65$\sigma$]\\
\hline\hline
\end{tabular}
\caption{
Best-fit $\Delta\chi^2_{\mathrm{MAP}}$ (relative to $\Lambda$CDM) and $\Delta \mathrm{AIC}$ for different dark-energy models under different data-combination assumptions, along with their statistical significance inferred from Wilk's theorem. Our baseline combination consists of CMB data (\planck PR3 polarization power spectra and \planckact lensing), BAO data (\desi DR2), and SNe data (the newly released \desdov). We also modify the baseline analysis in a number of ways. We replace the PR3 likelihoods with PR4 likelihoods and we remove the low-$\ell$ EE data, finding in both cases that the preference for dynamical dark energy is reduced. Finally, we perform the baseline analysis while allowing the sum of the effective neutrino mass to vary, also to negative values, finding that its effect is somewhat degenerated with dark energy and reduces the preference for dynamical dark energy over $\Lambda$ and reduces the statistical differences between the various scalar field models.}\label{table:stats}
\end{table*}

We turn now to a comparison of these models with the full complement of our baseline data (\desidesdovcmb). In \cite{Wolf:2025acj, Wolf:2025jed, Wolf:2024eph}, we have explored the respective parameter spaces using both MCMC methods \cite{metropolismc,hastingsmc,Lewis:2013hha} and nested sampling \cite{Handley:2015fda} through {\tt Cobaya} \cite{Torrado:2020dgo}, and derived the posterior constraints for all of these various microphysical dark energy proposals.\footnote{There has been a large volume of recent works analyzing various dark energy constructs in light of the latest \desi and SNe datasets. For just a small sample of the recent literature, see \cite{Gialamas:2025pwv, Shlivko:2024llw, Luu:2025fgw, DESI:2025fii, Das:2023rvg, RoyChoudhury:2024wri,RoyChoudhury:2025dhe, RoyChoudhury:2025iis, Bhattacharya:2024hep, Liu:2025bss, Alestas:2025syk, Bayat:2025xfr, Lin:2025gne, Park:2025fbl, Shlivko:2025krk, Berghaus:2024kra, Najafi:2026kxs, Toomey:2025xyo, Khoury:2025txd, Braglia:2025gdo, Bansal:2025ipo, Andriot:2025los,Ye:2024ywg, Gomez-Valent:2025mfl, Chakraborty:2025syu, Goldstein:2025epp, Tsujikawa:2025wca, Pan:2025psn, Yao:2025wlx, Adam:2025kve, SanchezLopez:2025uzw, Cataneo:2025vae, Nojiri:2026uvn, Shlivko:2026jxa, Bedroya:2025fwh, Mishra:2025goj, Wang:2026vqw, Ye:2026yqk, Gomez-Valent:2026ept, Kibris:2026cqq, Mishra:2026tzn, Baisri:2026pqe, Efstathiou_2024, Gialamas:2024lyw, Naidoo:2026umv, Rosatello:2026znt, Ren:2026jyw, Cai:2025mas, Silva:2025hxw, Delaunay:2026fse, Delaunay:2026jto, Gonzalez-Fuentes:2025lei, Gonzalez-Fuentes:2026rgu, Yang:2026vkc, Argudo-Panes:2026qkw, Lagos:2026pbr, Alestas:2024eic, Adam:2026ajg}.} Here, we update those constraints by using the \planckact lensing likelihood {\tt actplanck\_baseline} in place of the \act-only {\tt act\_baseline}, and by replacing the \desyf likelihood \cite{DES:2024jxu} with the newly recalibrated \desdov likelihood \cite{DES:2025sig}. Our baseline data combination then consists of {\tt plik} \planck PR3 CMB power spectra, \planckact CMB lensing, \desdov SNe, and \desi DR2 BAO \cite{Planck:2018vyg, ACT:2023kun, DESI:2025zgx, DES:2025sig}. In Table \ref{table:stats}, we report the best-fit $\Delta\chi^2_{\mathrm{MAP}}$ relative to $\Lambda$CDM as well as the Akaike information criterion which penalizes the introduction of new parameters and is defined at $\Delta$AIC $=\Delta\chi^2 + 2 \Delta k$ (where $k$ is the number of parameters in the model). Additionally, as $\Lambda$CDM is nested within all of the models considered here ($w_0 = -1, w_a = 0$ and $\beta=m^2=\xi=\alpha=\gamma =0$), we can follow Wilk's theorem \cite{Wilks:1938dza} to convert this to a frequentist significance by 
approximating $|\Delta\chi^2|$ as a chi-squared distributed variable with $\Delta k$ degrees of freedom. That is, we determine the resulting $p$-value and $n_\sigma$ as was done in Eq.~\ref{eq:p-value}, except now we use  $\Delta \chi^2_{\rm MAP}$ and $\Delta k$ with respect to \lcdm from the full analysis of each model (see Table~\ref{tab:priors} to see the varied parameters in each case). Note that the resulting $n_\sigma$ values in Table \ref{table:stats} are very similar to those reported from the $(w_0, w_a)$ compression, further validating the quality of that approximation.

We consistently find both the NM model and the mG model are on par with or even superior to the phenomenological $(w_0, w_a)$ model in terms of their ability to fit the data, while minimally coupled quintessence lags behind all three. As the new \desdov SNe sample decreases the preference for dynamical dark energy, the significance of the preference for these dynamical dark energy models is similarly reduced (see Table \ref{table:stats}). In comparison to when the \desyf original calibration is used, that yields $\Delta \chi^2 \simeq -24$ $(\simeq 4.2 \sigma)$ for both NM and mG models respect to $\Lambda$ \cite{Wolf:2025acj, Wolf:2025jed}, now $\Delta \chi^2 \simeq -16$ $(\simeq 3.2 \sigma)$ with \desdov. However, there still remains a mild preference for the more exotic dark energy proposals, even when one penalizes for the introduction of new parameters, while the minimally coupled scalar fields are essentially statistically indistinguishable from $\Lambda$.

\begin{figure}[t]
   \centering
    \includegraphics[width=\columnwidth]{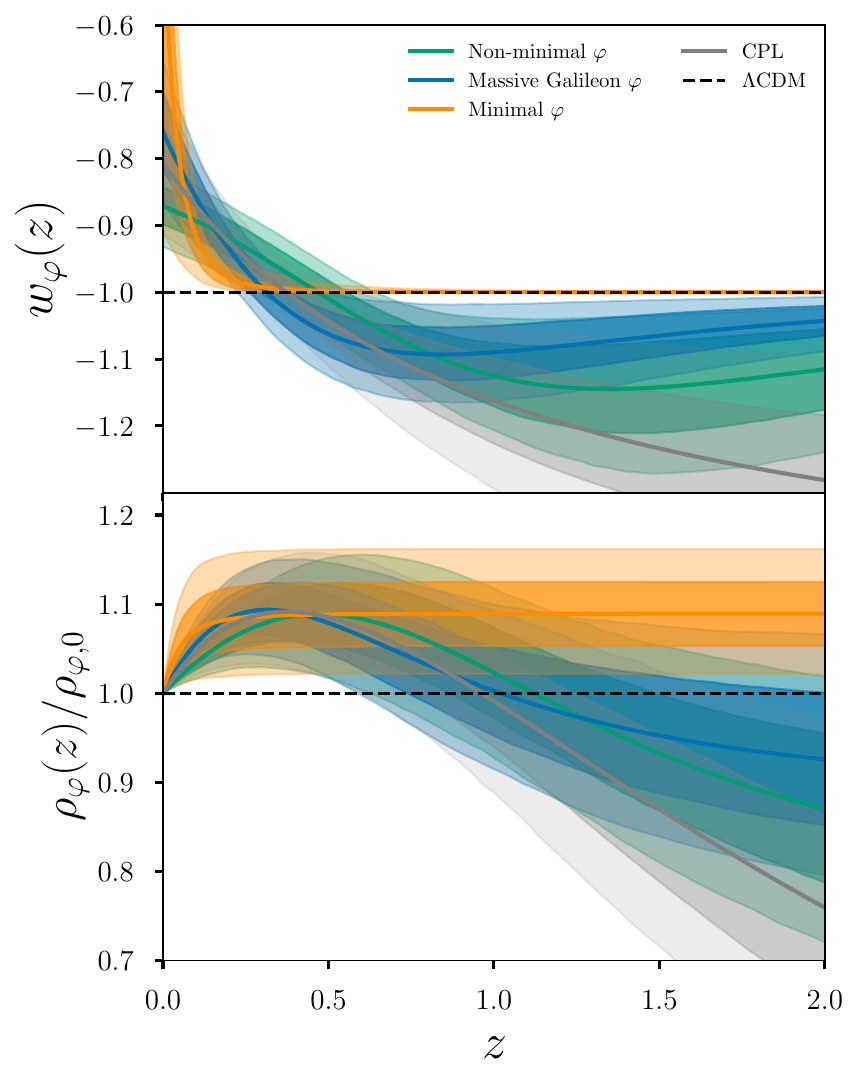}
    
   \vskip -0.1in
   \caption{Constraints on dark energy equation of state $w_{\varphi}(z)$ and energy density  $\rho_\varphi (z)/\rho_{\varphi,0}$ (68\% and 95\% C.L.) using our baseline data combination of \desi DR2 \cite{DESI:2025zgx}, \desdov SNe \cite{DES:2025sig}, \planck PR3 power spectra \cite{Planck:2019nip}, and \planckact CMB lensing \cite{ACT:2023kun}.}
   \label{fig:wz}
\end{figure}

\begin{figure}[t]
   \centering
    \includegraphics[width=\columnwidth]{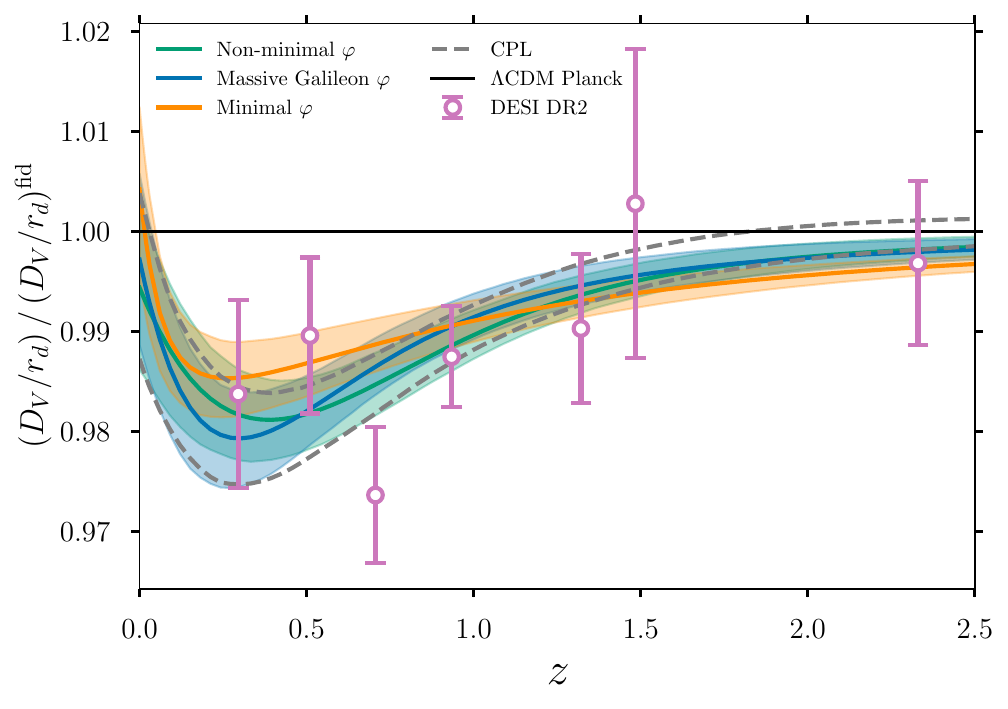}
    
   \vskip -0.1in
   \caption{Constraints on the isotropic distance ratio $D_V/r_d$ (68\% C.L.) resulting from our baseline cosmological data combination compared with the \desi DR2 measurements normalized to the predictions from \planck 2018 TTTEEE+lowE+lensing \lcdm cosmology (Table 2 of \cite{Planck:2018vyg}).}
   \label{fig:dv_constraints}
\end{figure}

The equation of state posterior distributions for these various microphysical proposals are depicted in Figure \ref{fig:wz} (top). For the minimally coupled field we have $w(z=0.00) = -0.45 \pm 0.30$; for NM, $w(z=0.00) = -0.87 \pm 0.03$; and for mG, $w(z=0.00) = -0.76\pm 0.06$. These reflect $\sim 2\sigma$ and $\sim 4\sigma$ tensions with the $\Lambda$CDM value $w=-1$. Furthermore, the exotic dark energy models favour a phantom crossing around $z\simeq0.5$, a curious fact that has been repeatedly found in the data when investigating dark energy from a number of different perspective (e.g.\ microphysical models, parameterizations, and model-agnostic reconstructions) \cite{Berti:2025phi, DESI:2025fii, Wolf:2025jed, Wolf:2025acj}. This can also be seen in the inferred constraints on $\rho_{\varphi}(z)$ (Figure~\ref{fig:wz}, bottom) where the dark energy density actually increases throughout the evolution of the phantom models, before reaching a peak at approximately $z\simeq0.5$ and beginning to thaw in the familiar quintessence manner. Figure \ref{fig:dv_constraints} depicts the constraints on $D_V(z)$ compared with the \desi DR2 measurements of this quantity, where one can see that the more exotic models follow the shape of the data more closely than quintessence which has a more restricted dynamical behavior.
Additionally posterior constraints for all dark energy models consistently favour a negative squared mass term ($m^2 < 0$), and for the more exotic models there is a slight preference for exotic physics, with a $\sim2.6\sigma$ preference for the  non-minimally coupled field ($\Delta \chi^2_{\rm NM, Min\, \varphi} = -8.2$) or for the massive Galileon ($\Delta \chi^2_{\rm mG, Min\, \varphi} = -7.9$) over a minimally coupled scalar field.

Finally, we quantify the preference of these dark energy models respect to $\Lambda$CDM in a Bayesian sense, using the Bayesian evidence
\begin{equation}
    \log \mathcal{Z} = \log \int \mathcal{L}(D |\theta, M) P(\theta | M) \, d\theta.
\end{equation}
Here, $\mathcal{L}(D | \theta, M)$ is the likelihood,  $P(\theta| M)$ is the prior, $\theta$ the sampled parameters and $M$ a given model. We compute the evidence using {\tt Cobaya} \cite{Torrado:2020dgo} and the nested sampler {\tt PolyChord} \cite{Handley:2015fda}. We report the Bayes factor given by the difference $\Delta \log \mathcal{Z}_{\varphi\Lambda} = \log \mathcal{Z}_{\varphi} - \log\mathcal{Z}_{\Lambda}$ with respect to $\Lambda$CDM, where the favoured model has a positive Bayes factor and less favoured model has a negative one. 

We compute the Bayesian evidence for our baseline data combination given by PR3 CMB power spectra, \planckact CMB Lensing, \desdov SNe, and \desi DR2 BAO and report the results in Table~\ref{tab:bayes}. Of particular note is that this combination of data is very similar to that used for all of the same dark energy models in our previous works \cite{Wolf:2025jed, Wolf:2025acj}, with the main difference being that the \desyf likelihood has been replaced with the re-calibrated \desdov likelihood. When referring to the Jeffrey's scale \cite{Jeffreys1939, Trotta:2005ar, Liddle:2007fy, Trotta:2008qt}, the statistical evidence for quintessence over $\Lambda$ goes from moderate to statistically indistinguishable, and the statistical evidence for the non-minimally coupled and massive Galileon models goes from strong to bordering between moderate and weak. These results align with changes seen in $\Delta \chi^2_{\varphi\Lambda}$, $\Delta \mathrm{AIC}_{\varphi\Lambda}$, and the Wilk's $\sigma$ results in Table~\ref{table:stats}, indicating that across all statistical measures \desdov induces a suggestive shift back towards $\Lambda$. We note that our findings do not contradict those of \cite{Ong:2026tta, Dinda:2025iaq, Dinda:2026ktu}, as the analysis greatly depends on the model one is considering and the associated priors.

\begin{table}[t]
\centering
\begin{tabular}{lc}
\hline
Cosmological Model & $\Delta \log \mathcal{Z}$ \\
\hline
$\Lambda$CDM & $0$ \\
Minimal $\varphi$ & $0.31 \pm 0.44$ \\
$w_0w_a$ [DESI Priors] & $0.78 \pm 0.46$ \\
$w_0w_a$ [Reduced Priors] & $2.10 \pm 0.45$ \\
Non-minimal $\varphi$ & $3.20 \pm 0.45$ \\
Massive Galileon $\varphi$ & $2.05 \pm 0.45$ \\
\hline
\end{tabular}
\caption{The values for the logarithm of the evidence ($\log \mathcal{Z}$) relative to $\Lambda$CDM under our baseline dataset combination (\planck PR3 CMB power spectra, \planckact CMB Lensing, \desdov SNe, and \desi DR2 BAO) for the different classes of dark energy model.}\label{tab:bayes}
\end{table}

While we have presented values of the Bayesian evidence for these models under our baseline data combination, we remain cautious about using it to make definitive statements about model preference (for an exhaustive analysis of model comparison with CPL, see \cite{Ong:2026tta}). For example, we present the Bayesian evidence for CPL $w_0w_a$ using both the \desi priors ($w_0 \in [-3.0, 1.0]$ and $w_a \in [-3.0, 2.0]$) and a more restricted set of priors ($w_0 \in [-1.5, 0.0]$ and $w_a \in [-2.0, 0.5]$). Under both sets of priors, $w_0$ and $w_a$ are both very well constrained as the prior edges for the restricted priors are still $\sim 12\sigma$ and $\sim 6 \sigma$ away from the posteriors on $w_0$ and $w_a$ respectively. Yet, this change in priors increases the Bayes evidence by a factor $\sim 2.7$, without having any effect on parameter inference or the model's ability to fit the data.

With the current iteration of the data, it is not possible to find strong evidence away from $\Lambda$CDM; the mild evidence we find for extended scalar field models will depend strongly on the assumptions about prior ranges and, crucially, on priors about the models themselves. Given our current (limited) understanding of fundamental physics on cosmological scales, it is not possible to establish these priors with any certainty. We report the priors used in this work in Appendix \ref{sec:appendixA}.

While the above is reflective of the baseline analysis performed by the \desi collaboration \cite{DESI:2025zgx} (with the exception that we have deployed the latest SNe likelihood which was released after the \desi DR2 paper), there are a number of other intriguing curiosities in the cosmological data that could suggest pursuing alternative routes. The results, in term of the $\Delta \chi^2_{\varphi \Lambda}$, $\Delta {\rm AIC}_{\varphi \Lambda}$ and the statistical significance from Wilk's theorem are shown in Table~\ref{table:stats} for each dark energy model.

{\it Planck PR4}:  As mentioned earlier, \planck's PR4 likelihoods notably reduce the \textit{$A_{\mathrm{lens}}$} anomaly and are more consistent with $\Lambda$CDM. Consequently, we would expect at least some of the preference for the extended scalar field models seen in our baseline data combination to be driven by PR3's preference for $A_{\text{lens}} > 1$. Thus, we replace PR3 low-$\ell$ and high-$\ell$ likelihoods with their PR4 counterparts {\tt LoLLiPoP} and {\tt HiLLiPoP}. Doing so slightly reduces the preferences for dynamical dark energy seen in the minimally coupled scalar field and $(w_0, w_a)$ (as was also seen in \cite{DESI:2025zgx}), but even more notably reduces the preference for the extended scalar field models. While they are still slightly favored over $\Lambda$CDM and the most vanilla scalar fields, the preference is not as striking, which is to be expected considering that the PR4 likelihoods are more consistent with $\Lambda$.

We also note that there is another possible choice of PR4 likelihood, the \texttt{CamSpec} NPIPE release \cite{Rosenberg:2022sdy, Efstathiou:2019mdh}. This likelihood, while reducing the $A_{\rm lens}$ anomaly relative to PR3, does not reduce it to the same extent as {\tt LoLLiPoP} and {\tt HiLLiPoP} and, as can be seen in Figure 18 of \cite{DESI:2025zgx}, it is slightly more favorable to dynamical dark energy than {\tt LoLLiPoP} and {\tt HiLLiPoP}. Thus, our choices of PR3 and PR4 likelihoods result in the most- and least-evidence cases for dynamical dark energy, respectively.

{\it Optical depth $\tau$}: The possibility of a larger optical depth $\tau$ can be achieved by removing the low-$\ell$ polarization data, which in turn can relax constraints on $\Omega_{\rm m}$ and $H_0$ from the CMB. This is particularly relevant for dark energy scalar fields with a minimal coupling, as the preference for thawing dark energy will push $\Omega_{\rm m}$ higher and $H_0$ lower relative to $\Lambda$ in these models. Consequently, relaxing these constraints allows thawing quintessence to drift closer to the $\Omega_{\rm m}$ values preferred by \desi BAO measurements and become on par with $w_0w_a$, while the overall preference for dynamical dark energy is reduced.

\begin{table}[t]
\centering
\renewcommand{\arraystretch}{1.2}
\begin{tabular}{l c}
\hline\hline
Cosmological Model & $\sum m_{\nu,\mathrm{eff}}$ [eV] \\
\hline
$\Lambda$CDM (CMB+DESI)             & $-0.099^{+0.049}_{-0.060}$ \\
$w_0w_a$ (baseline)           & $-0.052\pm 0.088$ \\
Minimal $\varphi$ (baseline)       & $-0.106^{+0.052}_{-0.062}$ \\
Non-minimal $\varphi$ (baseline)      & $-0.066^{+0.060}_{-0.087}$ \\
Massive Galileon $\varphi$ (baseline)      & $-0.068^{+0.068}_{-0.082}$ \\
\hline
\end{tabular}
\caption{Constraints on the summed neutrino masses under different dark energy models resulting from the baseline data combination of \desi DR2 \cite{DESI:2025zgx}, \desdov SNe \cite{DES:2025sig}, \planck PR3 power spectra \cite{Planck:2019nip}, and \planckact CMB lensing \cite{ACT:2023kun}.}
\label{tab:neutrino_mass_constraints}
\end{table}

\begin{figure}[t]
   \centering
    \includegraphics[width=\columnwidth]{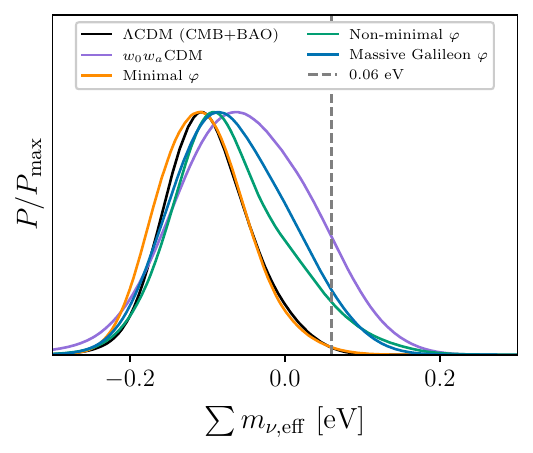}
    
   \vskip -0.1in
   \caption{Marginalized posterior distributions of the effective neutrino mass sum $\sum m_{\nu,\mathrm{eff}}$ for $\Lambda$CDM, $w_0w_a$CDM, and dark energy models (minimal $\varphi$, non-minimal $\varphi$, and massive Galileon $\varphi$), normalized to their maximum likelihood values. The dashed vertical line indicates the lowest bound imposed by terrestrial experiments $\sum m_\nu = 0.06\,\mathrm{eV}$ \cite{Esteban:2024eli}. Interestingly, only the dark energy models that allow for phantom crossing reduce the tension with the terrestrial experiments.}
   \label{fig:neutrinos}
\end{figure}

{\it Neutrinos}: Using the effective neutrino mass model of \cite{Elbers:2024sha}, we vary $\sum m_{\nu,\mathrm{eff}}$ and allow the effective mass to go negative in all of the microphysical dark energy models considered here. Here, we assume three degenerate neutrino mass states. In contrast, all other analyses in this paper assume one massive neutrino with a mass of $\sum m_{\nu} = 0.06$~eV (in line with terrestrial constrains \cite{Esteban:2024eli}). Interestingly, allowing the neutrino masses to vary can bring quintessence more into line with the CPL or extended scalar field models in terms of their ability to fit the data (as seen in the $\Delta \chi^2$ values of Table \ref{tab:neutrino_mass_constraints}); however, this comes with an attendant cost. Dynamical dark energy only alleviates the neutrino mass tension in the exotic models (Figure \ref{fig:neutrinos}). Minimally coupled quintessence makes the preference for negative neutrino masses worse than it is in $\Lambda$CDM alone, while extended scalar field models absorb some of this tension and make positive neutrino masses compatible with the cosmological data. 

Whatever the source of the tension that causes the unphysical preference for negative neutrino masses, it can only be absorbed by dynamical dark energy if we assume some kind of exotic structure or coupling in the dark energy sector. We report the constraints on the sum of the effective neutrino mass in Table \ref{tab:neutrino_mass_constraints}. We note that our $\Lambda$CDM constraint with the CMB+\desi BAO data agrees with that reported in \cite{Elbers:2025vlz}, while we also recover the constraint the sum of the effective neutrino mass reported there with the CPL model when using CMB+\desi+\desyf ($\sum m_{\nu,\mathrm{eff}}$ =$-0.042\pm 0.085$~eV). When using \desdov, the constraints shift slightly to $\sum m_{\nu,\mathrm{eff}}$ =$ -0.052\pm 0.088$~eV and the extended scalar field model constraints produce similar values.

\begin{table}[t]
\centering
\renewcommand{\arraystretch}{1.2}
\begin{tabular}{l c}
\hline\hline
Cosmological Model & $\Delta \varphi$ \\
\hline
Minimal $\varphi$        & $0.112^{+0.025}_{-0.028}$ \\
Non-minimal $\varphi$       & $0.335^{+0.068}_{-0.094}$ \\
Massive Galileon $\varphi$       & $0.652^{+0.097}_{-0.058}$ \\
\hline
\end{tabular}
\caption{Excursion of the scalar field, $\Delta \varphi \equiv |\varphi_0-\varphi_i|$, posterior distributions for the different scalar field scenarios and the baseline data combination of \desi DR2 \cite{DESI:2025zgx}, \desdov SNe \cite{DES:2025sig}, \planck PR3 power spectra \cite{Planck:2019nip}, and \planckact CMB lensing \cite{ACT:2023kun}.}
\label{tab:delta_phi}
\end{table}

\begin{figure}[t]
   \centering
    \includegraphics[width=\columnwidth]{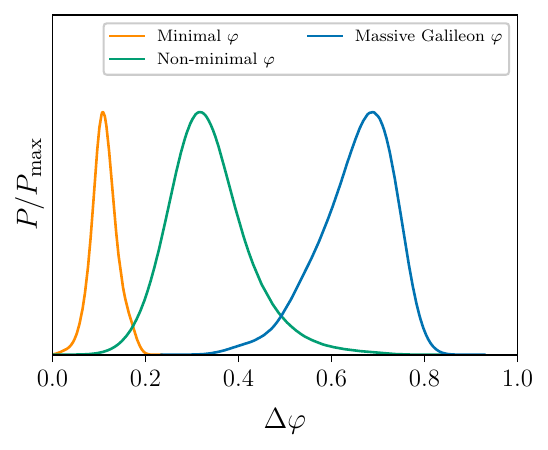}
    
   \vskip -0.1in
   \caption{Marginalized posterior distributions of the scalar field excursion $\Delta \varphi \equiv |\varphi_0-\varphi_i|$ for the different scalar field dark energy models, normalized to their maximum.
   }
   \label{fig:delta_phi}
\end{figure}

Finally, we briefly comment on the cosmologically favoured evolution of the scalar field $\varphi$ for the three branches of our scalar field EFT as it has direct bearing on the utility of the EFT philosophy we have adopted for dark energy. From our baseline data combination, we report the cosmological constraints on field excursion $\Delta \varphi \equiv |\varphi_0-\varphi_i|$ from its initial value, $\varphi_i$, to today, $\varphi_0$ in Figure \ref{fig:delta_phi} and Table \ref{tab:delta_phi}. For the minimal and non-minimal models, the mean values of $\Delta \varphi$ are $\simeq 0.1$ and $\simeq 0.3$ respectively, with both being very comfortably within the range where a Taylor expansion of the scalar field potential is justified as an excellent approximation. As stated earlier, the massive Galileon excursion mean $\simeq 0.7$ approaches values where the Taylor expansion becomes a more questionable approximation. 

Consider again as an illustrative example the exponential potential $V(\varphi) = V_0e^{-\lambda \varphi} \simeq V_0 (1 - \lambda\varphi+\frac{\lambda^2}{2}\varphi^2 \cdots)$ approximated by its Taylor expansion about $\varphi =0$. Furthermore, let us assume that $\lambda = 1$ for simplicity (current cosmological data for minimally coupled quintessence \cite{Akrami:2025zlb} and for non-minimally coupled quintessence  \cite{Ye:2024ywg, Wolf:2024stt} favours $\lambda\simeq 1$). Under this assumption, the field excursions of $\Delta \varphi \simeq 0.1$ and $\Delta \varphi \simeq 0.3$ lead to fractional errors of $\simeq 0.02 \%$ and $\simeq 0.5 \%$ respectively when comparing numerical values of the exponential potential and its Taylor expansion at the final field value. For $\Delta \varphi \simeq 0.7$, the fractional error jumps to $\simeq 9\%$. Furthermore, it is worth emphasizing that in this brief example we are considering the \textit{entire} field excursion, whereas in practice cosmological data is only meaningfully sensitive to a fraction of this field excursion. Given the uncertainties in the current (and future) cosmological data, which can only constrain a limited number of parameters in the expansion, such an approximation is well-justified. Additionally, the results of this simple example are a separate question from whether the cosmological constraints on the EFT parameters fall within a regime in which the expansion is valid. Examining the relative sizes of the constrained dark energy EFT terms in the case of the minimal and non-minimal models indicates that the expansion remains well-controlled in the cosmological regimes of interest. Conversely, for the massive Galileon model, the relative sizes of the constrained coefficients indicates that the expansion may break down. In future work, we will further elaborate and quantify the validity and limits of this approximation for a variety of potentials within our EFT framework.

\section{Ancillary Gravitational Effects and Gravitational Screening}
\label{sec:ancillary}

Both of the extensions to canonical quintessence -- NM and mG -- discussed in this paper lead to new, fifth forces, mediated by the scalar field. A schematic way of understanding this is as follows. If we expand the scalar field into a background component, $\varphi_0$ and a fluctuating term, $\delta\varphi$ so that $\varphi=\varphi_0+\delta\varphi$, we can see that in the quasistatic limit (akin to the Newtonian limit in standard General Relativity), we have that (in Fourier space) the evolution equation for the fluctuating term in the scalar field is given by 
\begin{eqnarray}
-k^2\delta\varphi\propto 4\pi Ga^2 \delta\rho_{\rm M}
\end{eqnarray}
where $\delta\rho_{\rm M}$ is the fluctuation in the matter density above the cosmological background and $k$ is the wave number. Note that $G$ is the "bare" Newton's constant which appears in the action and that the constant of proportionality may depend on both time and $k$. The effective gravitational potentials -- $\Phi$ in the case of massive bodies and $\Phi_+$ in the case of light rays --  will then receive a contribution from $\delta\varphi$ so that their modified Newton-Poisson equations are now\footnote{We are assuming the perturbed metric in Newtonian gauge
\begin{equation}
    ds^2 = a^2(\tau) [-(1 + 2\Psi) d\tau^2 + (1 + 2\Phi)\gamma_{ij}dx^idx^j] \nonumber
\end{equation}}
\begin{eqnarray}
    -k^2\Phi&=&4\pi \mu (\eta, k)Ga^2\delta\rho \nonumber \\
    -k^2\Phi_+&=&4\pi \Sigma(\eta, k)G a^2\delta\rho
\end{eqnarray}

Following \cite{DeFelice:2011hq}, we have that, for the model with non-minimal coupling,
\begin{align*}
&\mu
= \frac{M_{\rm Pl}^2}{M_*^2}\;
\frac{\big[M_*^2+8\xi^2\varphi^2\,\big]\,(k/a)^2 + M_*^2\left[m^2 + \xi R\right]}
{\big[\,M_*^2+6\xi^2\varphi^2\,\big]\,(k/a)^2 + M_*^2\left[m^2 + \xi R\right]},\\
& \Sigma = \frac{M^2_{\rm Pl}}{M^2_*}
\end{align*}
where $M^2_{*} \equiv M^2_{\rm Pl}-\xi\varphi^2$ is the effective Planck mass and $R=6(\dot{H}+2H^2)$
For the massive Galileon model, we have,
\begin{align*}
&\mu=\Sigma
= \frac{m^2 - \left(2\gamma\,\ddot\varphi + 4\gamma H\dot\varphi -\alpha \right)\,(k/a)^2}{m^2 - \Big(2\gamma\,\ddot\varphi + 4\gamma H\dot\varphi -\alpha + \dfrac{\gamma^2 \dot{\varphi}^4}{2M_{\rm Pl}^2}\Big)\,(k/a)^2}.
\end{align*}
On sufficiently sub-horizon scales, $k/(aH)\gg 1$, one finds that these expressions agree with those that have previously been derived (see \cite{Bellini:2014fua}).

\begin{figure}[t]
   \centering
   \includegraphics[width=\columnwidth]{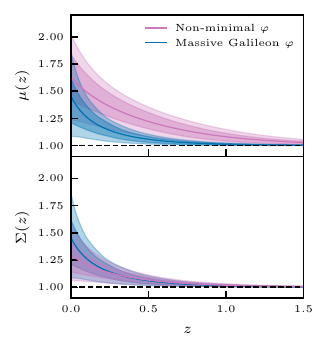}
   \vskip -0.1in
   \caption{Constraints (68\% and 95\%) on $\mu(z)$ and $\Sigma(z)$ in the quasi-static approximation resulting from the baseline data combination of DESI DR2 \cite{DESI:2025zgx}, DES-Dovekie SNe \cite{DES:2025sig}, Planck PR3 power spectra \cite{Planck:2019nip}, and ACT+Planck CMB lensing \cite{ACT:2023kun}.}
   \label{fig:mu_sigma}
\end{figure}

In Figure \ref{fig:mu_sigma} we plot the expected time evolution of $\mu$ and $\Sigma$ given the constraints we found in Section \ref{sec:cosmoconstraints}. We see that once dark energy begins to play a dominant role in the expansion of the Universe, from $z\simeq1$ until today, these parameters will deviate substantially from their general relativistic values. Such deviations will have striking consequences. On smaller scales -- for example on the scale of the Solar System or near binary pulsars -- it will affect local dynamics and all the ancillary effects that will arise.  There, one finds extremely tight constraints on such modifications, with  $\mu-1<10^{-5}-10^{-2}$ depending on the scale on which one is probing \cite{adelberger2003tests}. In fact, such constraints would immediately invalidate the scalar field models arising within the EFT we have considered, as an explanation for the observed accelerated expansion. 

It may be possible to circumvent such constraints by including gravitational screening \cite{Brax:2021wcv, Karam:2026sqg,Sirera:2026klo}. Here there are two view points one may have. The first, most economical view is that the EFT already contains all the necessary ingredients required for screening either in the terms already being considered in the action or at the next order. The second view point is to be more conservative and assume that the EFT we are considering on cosmological scales is not valid on astrophysical scales. Let us first consider the case that the EFT already contains screening. This can take three forms: Kinetic\cite{Vainshtein:1972sx,Brax:2014wla, Babichev:2009ee}, Chameleon \cite{Khoury:2003rn, Burrage:2017qrf} or Symmetron screening \cite{Hinterbichler:2010es}. We now look at these mechanisms in turn.

{\it Kinetic Screening:} Vainshtein screening is already built into the massive Galileon theories \cite{Deffayet:2011gz} through the $X\Box\varphi$ term, and arises in the next order terms in the EFT for non-minimally coupled theories. Specifically, in the case of the cubic Galileon, it can be shown that the fifth force, $F_5$ takes the form \cite{Babichev:2013usa,Koyama:2015vza} 
\begin{eqnarray}
\frac{F_5}{F_N}\propto \left(\frac{r}{r_V}\right)^{3/2} \nonumber
\end{eqnarray}
for $r<r_V$, where we have that the Vainshtein radius $r_V\simeq$ few Mpc. We emphasize that the scale is set by the parameters which are consistent with the observed expansion rate of the Universe and that we are not free to change them. In some sense, this is conceptually attractive as it is completely predictive.

In the case of the non-minimal theory, one can include one of the next terms in the EFT expansion,
\begin{eqnarray}
    \Delta S_3 = \int d^4 x\sqrt{-g} \frac{\varepsilon}{M^4}X^2,
\end{eqnarray}
which would be an example of K-mouflage screening \cite{Babichev:2009ee, Brax:2014wla}.
Note that now, unlike in the case of the mG model, both $\varepsilon$ and $M$ are unconstrained by the background expansion. The guiding principles give us a reasonable guess for these parameters but one can tune them so that $r_V$ should be able to mask the modifications on scales in which stringent tests to fifth forces have been found. The resulting fifth force, $F_5$ takes the form
\begin{eqnarray}
\frac{F_5}{F_N}\propto \left(\frac{r}{r_V}\right)^{4/3} \nonumber
\end{eqnarray}
for $r<r_V$.

A significant problem arises with Vainshtein screening if we take into account that they emerge as a valid EFT of some UV complete theory. As was argued in \cite{Burrage_2021}, if one considers, at higher energy scales, a heavy scalar field whose exchange generates the Galileon interactions -- the mass term -- one finds that the Vainshtein mechanism fails completely. Numerical solutions of the full UV theory show no suppression of the scalar force near the source; instead, the force remains unscreened and comparable to gravity.
The failure of screening can be  traced to the appearance of an infinite tower of higher-derivative operators generated when the heavy field is integrated out. Although each operator is individually suppressed by powers of the heavy mass, collectively they become important at macroscopic distances precisely where screening is supposed to occur. As a result, the usual EFT practice of truncating higher-order operators is invalid in the Vainshtein regime.
Furthermore,  improving the theoretical consistency of the model by raising its cutoff (i.e. making the UV completion explicit) eliminates the very nonlinear effects responsible for screening. The scalar field behaves effectively linearly all the way down to the source radius. It is conceivable that something might happen in the case of k-mouflage screening.

Finally, even if one could find a way of, somehow, protecting kinetic screening from what seems like an insurmountable problem, it still may not be enough to save the non-minimally coupled theory from constraints in the Solar System. In that case we have that screening will help circumvent the constraints on fifth forces which arise from the fluctuating part of the scalar field, $\delta\varphi$. But we also know that the background, $\varphi$, evolves with time and will directly affect the effective Newton's constant \cite{Boisseau:2000pr, Damour:1993id},
\begin{equation}
G_{\rm eff}
=
\frac{G}{f(\varphi)}\left[1+
\frac{f_{\varphi}^2}
     {2f(\varphi)+3f_{\varphi}^2}\right],
\nonumber
\end{equation}
where $f_\varphi \equiv df/d\varphi$. $G/f(\varphi)$ represents the purely cosmological evolution of $G_{\rm eff}$ while the additional terms represent a scalar mediated correction between matter sources relevant to Cavendish-type experiments and local tests of gravity.
For the non-minimally coupled model considered in this paper, this leads to a time variation in the effective gravitational constant given by,
\begin{equation}
\frac{\dot G_{\rm eff}}{G_{\rm eff}}
=
\frac{2\xi\varphi\dot{\varphi}}{f(\varphi)}
\left[
1+
\frac{
8\xi f(\varphi)
}{
\left[2f(\varphi)+16\xi^2\varphi^2\right]
\left[2f(\varphi)+12\xi^2\varphi^2\right]
}
\right].
\nonumber
\end{equation}
Non-minimal models consistent with the current expansion rate will lead to substantial evolution in ${\dot G}_{\rm eff}/G_{\rm eff}$. We can probe this time evolution directly  with, for example, the Lunar Laser Ranging (\llr) experiment \cite{adelberger2003tests} (see e.g. \cite{Adam:2025kve, March:2025uqt}); these constraints will be impervious to Vainshtein screening as they are directly affected by $\varphi_0$ \cite{Babichev:2011iz}. There, using our baseline data combination, one finds constraints on Newton's constant of $\dot G/G = (6.3^{+2.9}_{-2.4})\times 10^{-11}\,\mathrm{yr^{-1}}$, which is of order $\frac{\dot G}{G} \sim 10^{-1} H_0 $ (see also \cite{Wolf:2025jed, Adam:2025kve}). This is at odds with a number of constraints that have been found in the literature, such as \cite{Williams:2004qba, Uzan:2024ded, Hofmann_LLR}, which give constraints such as $\dot G/G = (4.0\pm9.0)\times 10^{-13}\,\mathrm{yr^{-1}}$, which is of order $\frac{\dot G}{G} \sim 10^{-3} H_0 $. However, given the large error bars of the background-inferred value, the tension with local constraints is only at $2.6\sigma$.

\begin{figure}[t]
   \centering
    \includegraphics[width=\columnwidth]{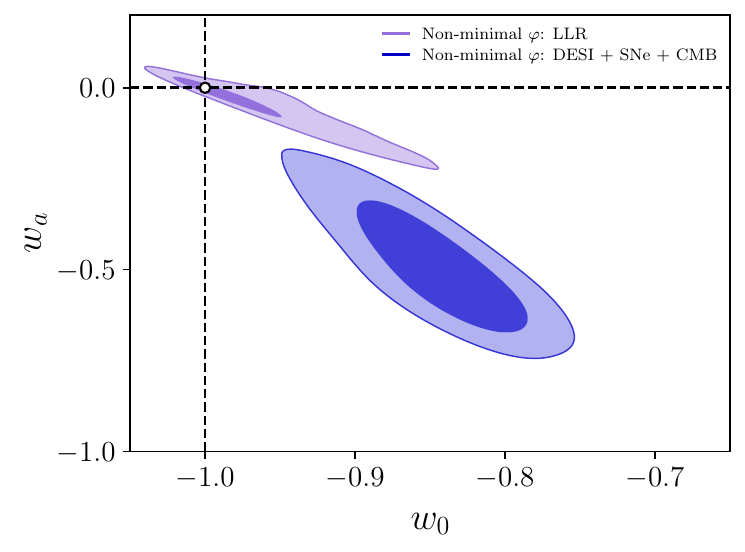}
   \vskip -0.1in
   \caption{Inferred ($w_0, w_a$) parameters obtained  using the baseline data combination (\desidesdovcmb) compared with those obtained when using the Lunar Laser Ranging data from \cite{Williams:2004qba}. Incorporating the \llr constraints immediately pulls the non-minimally coupled model back to $\Lambda$. In order to make a like for like comparison between the expansion histories favoured by the different data combinations, we follow the procedure explained in Section~\ref{sec:prior_procedure} and use the \desidesdovcmb data properties to project the model constraints into $(w_0, w_a)$ parameters.}
   \label{fig:LLR}
\end{figure}

It is interesting to think of this constraint as an additional {\it prior} on the model parameters and see what it implies in terms of allowed values of $w_0$ and $w_a$ (again, we are using these parameters as a useful way to visualize the impact of a scalar field model on the cosmological expansion as described in Section \ref{sec:prior_procedure}). In Figure \ref{fig:LLR} we do just that by running an MCMC with the constraints on $\dot G/G $ obtained in \cite{Williams:2004qba}, sampling the obtained parameter distributions, and then fitting the ($w_0, w_a$) model to the \desisnecmb observables predicted for this distribution\footnote{It is important to emphasize here that, while the parameters of the dark energy model were constrained with the \llr data, the projection into $(w_0, w_a)$ parameters was determined by fitting $H^2_{(w_0, w_a)}$ to the expansion history $H^2_\varphi$ predicted by the \llr constrained dark energy model \textit{using the redshift bins and uncertainties of the compressed background expansion data for our baseline data combination} (\desisnecmb), as explained in Section~\ref{sec:prior_procedure}, to provide a meaningful comparison with the expansion history inferred from the cosmological data.}.

When translating this distribution into ($w_0, w_a$) values, one can see that the large prior range, which previously encompassed the current cosmological constraints, is now greatly reduced and concentrated around what one expects for \lcdm. Furthermore, we can clearly see a tension between the resulting $(w_0, w_a)$ distributions inferred from the \llr data and those inferred through the expansion history data alone. In this case, when using the metric given by Eq.~\ref{eq:suspicion}, we find a tension of  $n_\sigma\simeq3.1$ and $n_\sigma\sim 2.9$ in the $(\xi,\beta, m^2)$ parameter space.

This does not quite tell the full story as the NM model has quintessence as a limit, meaning that this part of the parameter space can still come close to accessing the regions of the $(w_0, w_a)$ plane favoured by the expansion history data while still remaining consistent with the \llr data. However, it is worth noticing the \llr data pulls $\beta$ towards $0$ and $m^2 > 0$, effectively freezing the field in much of the constrained parameter space. This leaves the non-minimal coupling parameter loosely constrained, as the phenomenology remains insensitive to its value, with the field mimicking a cosmological constant or standard quintessence. For a scalar field theory with a non-minimal coupling to gravity, outside of delicately fine-tuning the initial conditions or functional forms of the model to engineer $G_{\rm eff}=G_N$ today (see e.g.~\cite{ Argudo-Panes:2026qkw, Banihashemi:2026twb}), screening mechanisms which can screen the background evolution of $G_{\rm eff}$ will be needed in order to realize the exotic phenomenology suggested by the cosmological data.

{\it Chameleon Screening:} Chameleon screening will only appear in theories with explicit non-minimal coupling, represented by the NM theory here. In practice, such a mechanism is usually studied in the Einstein frame where the effective potential is
\begin{eqnarray}
V_{\rm eff}(\varphi)\simeq V_0+\beta\varphi+\frac{1}{2}\left(m^2+\xi\frac{\rho_{\rm M}}{M_{\rm Pl}^2}\right)\varphi^2
\end{eqnarray}
which has a minimum and effective mass which both depend on the environment through $\rho_{\rm M}$:
\begin{eqnarray}
    \varphi_{\rm min}&=&-\frac{\beta}{m^2_{\rm eff}} \nonumber \\
    m^2_{\rm eff}&=&m^2+\xi\frac{\rho_{\rm M}}{M_{\rm Pl}^2}
\end{eqnarray}
Thus, in principle it should be possible to have Yukawa screening with a range that depends on the environment (see e.g. \cite{Nojiri:2026uvn}). 

While these are essential ingredients for the Chameleon mechanism, they are not sufficient. One needs to satisfy the ``thin-shell'' conditions and it is not clear that, for this model, it can be satisfied. In particular, the usual Chameleon example has a linear (or exponential) conformal coupling while here it is quadratic. In the case of NM considered here,  the thin-shell condition may only be satisfied if $m^2$ is very large or if $\xi\gg1$, both of which are outside the desired values for cosmological constraints. One could consider a purely linear (or exponential) field dependence for the conformal coupling to bring it in line with standard examples of successful Chameleon screening but that would defeat the goal of being general, which is the path we have taken here. 

{\it Symmetron Screening:} It is possible to have Symmetron screening in the NM case where all the correct ingredients are present {\it except} for the  presence of a linear term, $\beta\varphi$, in the potential, which breaks $Z_2$ symmetry, and the fact that we truncate the expansion of the potential at quadratic order. Now, the quadratic trunction arises because, as we have shown, this is all we need on cosmological scales (from the point of view of cosmological evolution). But screening manifests itself on much smaller scales where one would expect to, either, have to go to higher order in the expansion or assume an altogether EFT. So it might have been possible to accomodate a Symmetron screening potential, apart from the fact that the posteriors on $\beta$ favour a substantially non-negligable value. There are other issues. The range of parameters, and structure of the potential are such that Symmetron screening would also suffer from the same thin-shell problem. Furthermore, given the values, in the posterior, for the coupling constants in the potential, it is unclear when the symmetry breaking phase might occur, relative to the period of dark energy domination. 

In summary, if we consider our EFT expansion has all the elements for screening, we find clear problems that can, at best, introduce strong tensions between the cosmological background-level\footnote{Note that we are loosely calling \desisnecmb constraints "background" as most information comes from it.} and Solar System constraints, or, in the worst case scenario, completely rule out single scalar field dark energy.

However, there is no reason to believe that the EFT at cosmological scales will be valid on astrophysical and Solar System scales. The philosophy of EFT tells us that it {\it should} only be applicable within a well defined, relatively narrow, range of length and energy scales. Extrapolating it from $\sim$ Gpc scales down to A.U. scales is, arguably, a vast range of validity which is unlikely to be correct. If that is the case then one can consider completely alternative formulations of the theory at Solar System scales where, for example, other forms of screening may be valid. In particular, theories with screening may be able circumvent the \llr constraints. If that is the case, little can be said about the cosmological EFT as a result from constraints coming on astrophysical scales.

There are, thus, two ways to search for gravitational signatures for scalar field dark energy which is consistent with current cosmological observations. One is to, loosely, look for signs of screening in, for example, the morphology and dynamics of galaxies. Screening can lead to a number of observational phenomena that can be constrained with current and future high resolution, multi-wavelength maps of galaxies. Chameleon screening will lead to specific signatures in galactic warps, in offsets between the light, gas and the dark matter halos and in the velocity profiles \cite{Baker:2019gxo}. Stringent constraints have already been obtained \cite{Desmond:2020gzn,Landim:2024wzi} but it is hoped that, with the new wave of data, the situation will improve even further. Vainshtein screening will also lead to observational signatures, most notably the offset between the central black hole and galactic structure \cite{Bartlett:2020tjd}. Again, the situation is expected to improve with the current generation of surveys.

The other approach is to look on sufficiently large scales where gravitational collapse is still linear and screening is not effective. In particular, by measuring the growth rate of structure in different ways, one should be able to probe directly for the effect of the fifth forces. 
We turn to this approach in the next two sections.

\section{The Growth Rate of Structure}\label{sec:growth}

On larger scales, greater than tens of Megaparsecs, one might be able to probe the novel gravitational effects by measuring the growth rate of structure, $f$, defined by $f\equiv{d\ln\delta_{\rm m}}/{d\ln a}$. We have that the evolution equation for $f$ is
\begin{eqnarray}
    \frac{df}{d\ln a}+f^2+\left(1+\frac{d\ln{\cal H}}{d\ln a}\right)f=\frac{3}{2}\mu\Omega_{\rm m}(a). \label{eq:fevol}
\end{eqnarray}

In the case of Einstein-de Sitter Universe, where $\Omega_{\rm m}=1$ we have that $f=1$ and in \lcdm we have that $f$ can be approximated by $f\simeq\Omega_{\rm m}^{{6}/{11}}<1$.  One can clearly see in Eq.~\eqref{eq:fevol} that if $\mu>1$ (as is the case of the range of models being considered here), the driving term on the right hand side will enhance $f$. Nevertheless, as can be seen in Fig \ref{fig:mu_sigma}, $\mu$ only departs significantly from unity at $z<1$ which means that $(f-f_\Lambda)/f_\Lambda\simeq \ {\rm few} \times 10 \%$ at very late times (with $f_\Lambda$ being the $\Lambda$CDM value of the growth rate). Thus one still has $f<1$ for these theories.

\begin{figure}[t]
   \centering
\includegraphics[width=\columnwidth]{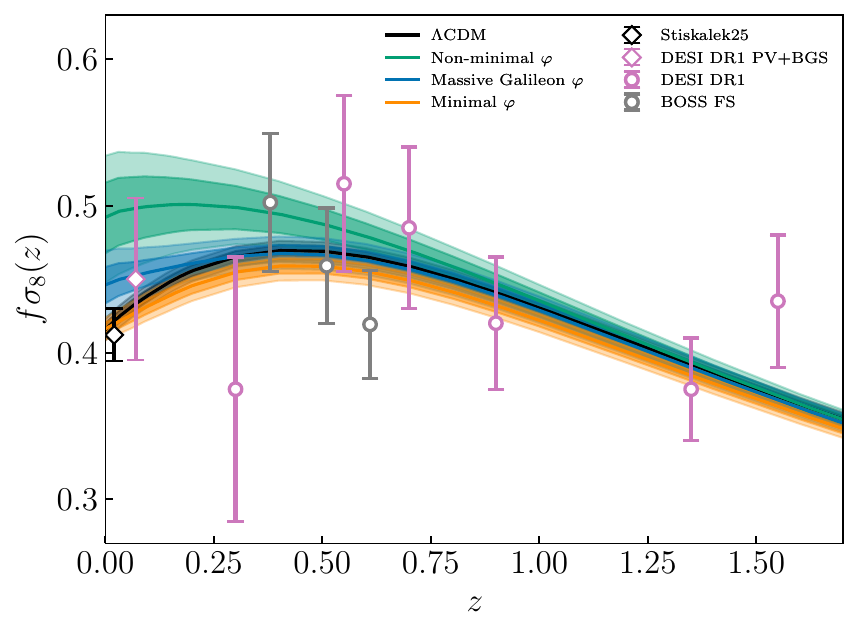}
   \vskip -0.1in
   \caption{The density-weighted growth rate of structure for the $\Lambda$CDM, the NM and mG models, as determined from the constraints coming from the expansion history data (i.e. \desidesdovcmb), compared to current measurements from \boss redshift space distortions \cite{BOSS:2016wmc}, \desi \texttt{ShapeFit} (taken from Fig 14 in \cite{DESI:2024jxi}) ($z>0.3$) and peculiar velocities \cite{Lai:2025xkf, Stiskalek:2025oht} (PV; $z\simeq 0.05$). As it can be seen, the largest deviations from $\Lambda$CDM occur at low redshift, where the \stiskalek point is.}
   \label{fig:fsigma8data}
\end{figure}

Measuring the growth rate, or more pragmatically, the {\it observable} growth rate, $f\sigma_8$ (where $\sigma_8$ is the mass variance on scales of $8h^{-1}$ Mpc radius) is one of the main objectives of current and future surveys\footnote{Note that \desi is actually using the {\tt Full Shape} clustering of its tracers \cite{DESI:2024jxi} or the {\tt ShapeFit} parameters \cite{Brieden:2021edu}, which include an effective $f\sigma_{s8}$ quantity that accounts for cosmology changes, instead of compressing the growth measurements into $f\sigma_8$. This is because $f\sigma_8$ measurements are less robust \cite{Brieden:2021edu, DESI:2024jxi}}. Stage IV surveys, through measurements of the Full Shape, Redshift Space Distortions (RSDs) \cite{1987MNRAS.227....1K} or the novel \texttt{ShapeFit} method \cite{Brieden:2021edu, DESI:2024jxi}, will be able to constrain the observable growth rate at the percent level \cite{Font-Ribera:2013rwa}, on scales large enough that screening mechanisms will not be effective. Most of these surveys are targeting the growth rate at intermediate to high redshifts where we do not expect substantial deviations from  $f_\Lambda$. Thus we do not expect them to be particularly effective at detecting or constraining the deviations we expect to arise in both the NM and mG models. This can be clearly seen in Figure \ref{fig:fsigma8data} where the data uncertainties, for $z>0.3$ swamp the differences between different models.

\begin{figure}[t]
   \centering
\includegraphics[width=\columnwidth]{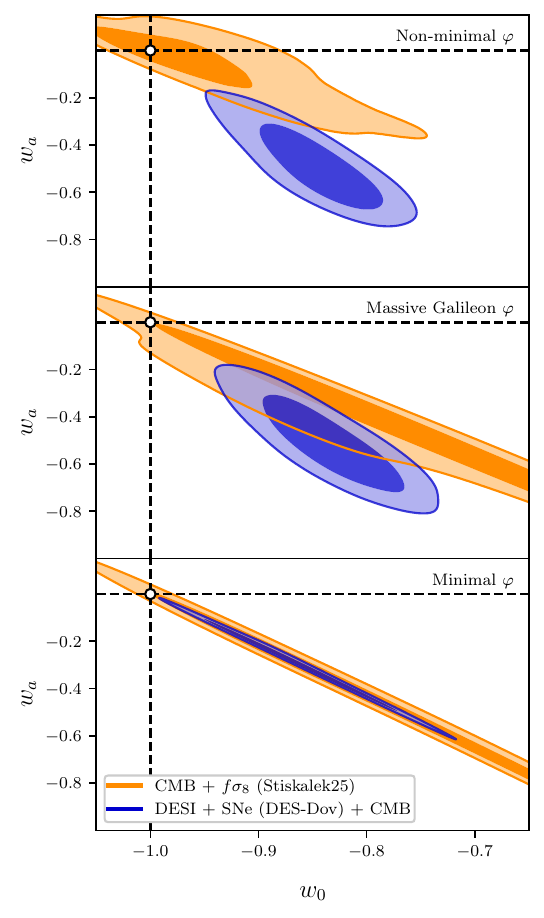}
   \vskip -0.1in
   \caption{Inferred ($w_0, w_a$) parameters obtained  using the baseline data combination (\desidesdovcmb) compared with those obtained when using CMB and the \stiskalek $f\sigma_8$ measurement \cite{Stiskalek:2025oht}. Including \stiskalek significantly pulls the extended models back towards $\Lambda$. In order to make a like for like comparison between the expansion histories favoured by the different data combinations, we follow the procedure explained in Section~\ref{sec:prior_procedure} to project both constraints onto the $(w_0, w_a)$ plane, using the \desidesdovcmb data properties.}
   \label{fig:nm_fsigma8}
\end{figure}

\begin{figure*}[t]
   \centering
   \includegraphics[width=0.45\textwidth]{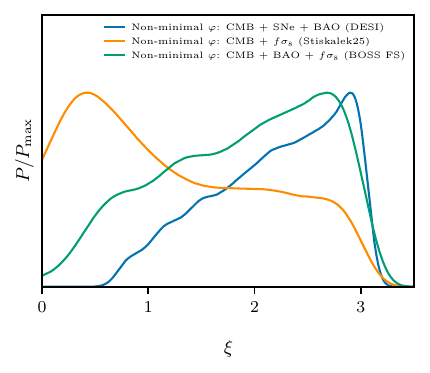}
   \hfill
   \includegraphics[width=0.45\textwidth]{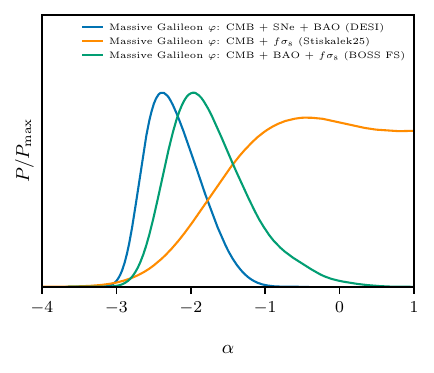}
   \vskip -0.1in
   \caption{Marginalized posterior distributions of the coupling parameters $\xi$ (left) and $\alpha$ (right) for the non-minimal and massive Galileon models. Results are shown for three dataset combinations: \desidesdovcmb (baseline), CMB+BAO+$f\sigma_8$ (\boss FS), and CMB+$f\sigma_8$ (\stiskalek). The inclusion of growth data significantly shifts and tightens the constraints relative to the baseline expansion-history combination, highlighting the sensitivity of structure formation observables to the parameters.
}
   \label{fig:fsigma8_constraints}
\end{figure*}

As pointed out above, we {\it do} expect larger deviations from the \lcdm at very low redshifts and these are much harder to measure accurately and precisely. There are statistical limitations -- one has much more limited samples in terms of volume -- and systematic limitations -- we are much deeper in the non-linear regime of structure growth as well as subject to the role of non-gravitational physics arising through baryon feedback. Yet, there are attempts at pinning down $f\sigma_8$ at very late times through, for example, peculiar velocities. The most precise attempt (\stiskalek) can be found in \cite{Stiskalek:2025oht}, where the author infers \(f\sigma_8\) by fitting a linear-theory peculiar--velocity field from the \tmpp reconstruction to distances from Tully--Fisher and SNe Ia in order to constrain the velocity scaling parameter $\beta_\star = f/b$. It assumes a linear, scale-independent galaxy bias $b$ relating the galaxy and matter density fields via $\delta_g = b\,\delta$, treats $b$ as constant and degenerate with $f$, and combines the fitted $\beta_\star$ with the measured \tmpp clustering amplitude, $\sigma_8^g$, to obtain stringent constraints on $f\sigma_8^{\rm NL} = \beta_\star \sigma_8^g $, which they relate to $f\sigma_8 = 0.412 \pm0.018$\footnote{\cite{Stiskalek:2025oht} provides also a more conservative value for $S_8$ based on \cite{Hollinger:2023mrp} sample variance. In this case, and assuming the same relative error change for both $S_8$ and $f\sigma_8$, the uncertainty goes from $\sim 4.5\%$ to $\sim 6.5\%$. We study the impact of this conservative error in Appendix~\ref{sec:appendix_growth}, which only adds to the underdetermination of dark energy.}
 at $z\simeq 0.02$ by using an emulator, trained in \lcdm, that maps the measured variance of the non-linear field, $\sigma_8^{\rm NL}$, to the usual $\sigma_8$. Figure \ref{fig:fsigma8data} shows how this data point can already play an important role in discriminating between the different models. Additionally, we also note that the \desi collaboration has also used peculiar velocities to measure low redshift growth, reporting $f\sigma_8 = 0.45 \pm0.055$
 at $z\simeq 0.07$ \cite{Lai:2025xkf}. As is apparent in Figure \ref{fig:fsigma8data}, this data point is more compatible with the enhanced growth seen in the extended models in addition to having larger error bars. Consequently, we focus most of our attention on \stiskalek as this will be more illustrative of the potential constraining role that low redshift growth measurements will play in the future. However, for comparison, we also provide the constraints using \desi's peculiar velocity measurement, and a forecast using its expected future accuracy (similar to \stiskalek), in Appendix~\ref{sec:appendix_growth}.
 
 Given the assumptions that go into \stiskalek, we have checked they remain valid for our scalar field models. In the case of the scale-independent galaxy bias, the approximation still holds. The NM and mG scale dependence is weak at the scales and redshifts of interest, with changes on the linear growth factor of $\sim 0.2\%$ for NM and $\lesssim 0.8\%$ for mG within 95\% C.L. between $k=0.1$/Mpc and $k=2$/Mpc, much smaller than the 5\% error in $f\sigma_8^{\rm NL}$. In contrast, the impact of mapping $\sigma_8^{\rm NL} \rightarrow \sigma_8$ assuming \lcdm is less clear. On the one hand, the NM and mG might enhance the non-linear matter power spectrum respect to \lcdm, biasing the estimated $\sigma_8$. On the other hand, for models close enough to \lcdm, the mismatch will most likely be small. We cannot quantify the impact of this approximation easily, as it would require the calibration of the non-linear regime for these dark energy models, which is beyond the scope of this paper. In any case, and as it will become apparent, a small bias will not change the qualitative picture. As we will show, the \stiskalek does not have enough discriminatory power and the model underdetermination persists.
 
A particularly useful way to visualise the role of such a low redshift constraints on $f\sigma_8$ (vis-\`a-vis the background cosmological constraints) is to combine it with the CMB and infer the allowed ($w_0$, $w_a$) parameters -- as shown in \cite{Ruiz-Zapatero:2022xbv}, the CMB plays a crucial role in breaking the degeneracy between $\Omega_{\rm m}$ and $\mu$). As in Section \ref{sec:ancillary}, we use different datasets to constrain the dark energy model (in this case growth vs. expansion data), but we use the same observables and redshifts (\desisnecmb) in the $(w_0, w_a)$ compression as detailed in Section \ref{sec:prior_procedure} so as to make a meaningful comparison between the distributions of dark energy parameters. 

In Figure \ref{fig:nm_fsigma8} we show the constraints from the expansion history (in blue) versus constraints from the \stiskalek low-$z$ growth rate measurement (and CMB) in orange. While for thawing quintessence there is substantial overlap, that is not the case for the NM model, where the growth rate constraint is pulling the model back towards $\Lambda$CDM (or the quintessence limit). Similarly, with the mG model, the growth data notably pulls the model back towards $\Lambda$CDM and quintessence, although in this case, since its growth departs less from that of $\Lambda$CDM, both posterior distributions overlap at $<2\sigma$. Qualitatively, assuming the distributions are Gaussian and using Eq.~\ref{eq:suspicion}, we find no tension ($n_\sigma\simeq0.4$) for quintessence, a mild tension ($n_\sigma\simeq1.6$) for mG, and a more pronounced ($n_\sigma\simeq2.7$), albeit not too strong, tension for NM. 

In addition to viewing this from the perspective of the inferred ($w_0, w_a$) parameters, we can also examine the resulting tension in terms of the EFT parameters $\xi$ and $\alpha$, which respectively quantify the magnitude of the non-minimal coupling to gravity for the NM model, and of the non-canonical kinetic coefficient for mG. Their constraints can be found in Table \ref{tab:xi_alpha_constraints} and Figure \ref{fig:fsigma8_constraints}. When using the baseline expansion history data, these parameters are constrained to be significantly away from 0; i.e. the $\Lambda$CDM limit. Yet, when one determines the constraints using the combination of CMB data and \stiskalek single low-$z$ point, one immediately obtains upper (and lower) bounds on $\xi$ (and $\alpha$), and the posteriors shift right back to the values expected in a $\Lambda$ or quintessence driven dark energy scenario. That is $\xi \rightarrow 0$ and $\alpha$ is pulled back to canonical positive values and becomes unconstrained in this direction as this corresponds to the quintessence limit of the theory where the ratio  $\alpha/\gamma \rightarrow -\infty$ (see Table~\ref{tab:eft_schematic}).

\begin{table}[t]
\centering
\renewcommand{\arraystretch}{1.2}
\begin{tabular}{l cc}
\hline\hline
Data & $\xi$ & $\alpha$ \\
\hline
Baseline
    & $2.21^{+0.85}_{-0.35}$
    & $-2.22^{+0.29}_{-0.42}$ \\

CMB+$f\sigma_8$ (\boss FS)
    & $1.89^{+1.10}_{-0.56}$
    & $-1.73^{+0.37}_{-0.64}$ \\

CMB+$f\sigma_8$ (\stiskalek)
    & $< 1.77$
    & $> -0.56$ \\
\hline
\end{tabular}
\caption{Constraints on the coupling parameters $\xi$ (non-minimal model) and $\alpha$ (massive Galileon model) for different dataset combinations. The errors  are 68\% C.L. and the limits the 95\% C.L. The very precise measurement of \stiskalek, at very low redshift, when the deviations between models are the largest, pulls the constraints back into agreement with \lcdm, although still in moderate agreement with the baseline \desidesdovcmb constraints.}
\label{tab:xi_alpha_constraints}
\end{table}

Interestingly, one finds that for these $\xi$ and $\alpha$ parameters, this single low-$z$ $f\sigma_8$ \stiskalek data point offers more constraining power than the final \sdss~ \boss full shape (FS) analysis that provides three intermediate-$z$ $f\sigma_8$ data points (depicted in gray in Figure \ref{fig:fsigma8data}) and $D_M$ and $H$ (see Table 7 of \cite{BOSS:2016wmc}) \footnote{We use the likelihood in \texttt{Cobaya} called \texttt{bao.sdss\_dr12\_consensus\_full\_shape}.}. When one uses \boss FS measurements, the constraints on $\xi$ and $\alpha$ are in good agreement with those coming from \desi BAO. While it will be interesting to see if future data releases from \desi can improve upon this situation, at present the \desi DR1 has similar constraining power as \sdss \cite{DESI:2024jxi}. This highlights the importance of obtaining high quality, low-$z$ growth measurements -- they have the potential to play a significant role in ascertaining the viability of extended models. We will return to this in Section \ref{sec:future} 

\section{Integrated Sachs-Wolfe Effect}\label{sec:isw}

The integrated Sachs-Wolfe (ISW) effect in the cosmic microwave background (CMB) relates measurements of the anisotropy of the relic radiation, $\Delta T/T$, with time variations of the sum of the gravitational potentials, $\Phi_+$, on large scales. It is given by
\begin{eqnarray}
\frac{\Delta T}{T}^{\rm ISW}({\bf {\hat n}})=2\int_{\eta_*}^{\eta_0}  d\eta\, e^{-\tau(\eta)}\partial_\eta\Phi_+[\eta,\chi(\eta){\bf {\hat n}}]
\end{eqnarray}
where $\eta$ is conformal time, ${\bf {\hat n}}$ is the unit direction vector in the sky, $\tau(\eta)$ is the optical depth, $\chi(\eta)$ is the comoving distance from a light ray emitted at $\eta_*$ and arriving at the origin at $\eta_0$ and where subscripts $*$ ($0$) are last-scattering (today). In minimally coupled theories, which obey the Einstein field equations, $\Phi_+=\Phi$.
In a landmark paper \cite{Crittenden:1995ak}, the authors showed that the ISW effect could be used to tease out the effect of the cosmological constant $\Lambda$ on the growth rate of structure on large scales through the Rees-Sciama effect \cite{Rees:1968zza}. 

The argument is as follows. Relativistic corrections aside one can use the Newton Poisson equation to show that $\partial_\eta\Phi=(f-1){\cal H}\Phi$. As discussed in the previous section,  in an Einstein-de Sitter universe, $f=1$, resulting in $\partial_\eta\Phi=0$ and so $\Delta T/T^{\rm ISW}=0$.  In a Universe with a cosmological constant, $f<1$ and so $\partial_\eta\Phi>0$ (note that $\Phi<0$) and $\Delta T/T^{\rm ISW}\neq 0$. 

It is not possible to isolate the ISW with measurements of the CMB alone as it competes against other primordial signals sourced at the surface of last scattering. But, if one correlates it with other probes of large scale structure at late times, it is possible to isolate the effect. In particular, a galaxy survey, with number density fluctuations $\delta_g({\bf {\hat n}})$, can be directly related to the matter density contrast through an expression which is similar to that of the ISW: $\delta_{\rm g}({\bf {\hat n}})=\int_{\eta_*}^{\eta_0}b_{\rm g}(z)p(z)\delta[z,\chi(z){\bf {\hat n}}]dz$ where we have assumed a linear, scale independent, bias factor, $b_{\rm g}(z)$, and an isotropic galaxy selection function, $p(z)$. Cross-correlating a galaxy survey with the map of the cosmic microwave background, it is possible to pick out the ISW effect through the cross spectrum

\begin{eqnarray}\nonumber
    C^{gT}_\ell= \frac{2}{\pi}\int dk\, k^2 dz\, dz'\, (W^{\rm ISW}_\ell (k, z) W^{\rm g}_\ell (k, z') \times \\ 
    P_{\rm mm}(k, z, z')) ,
\end{eqnarray}
where the two kernels are given by 
\begin{eqnarray}
    W^{\rm ISW}_\ell (k, z) &=& \frac{3\Omega_{\rm m} H^2_0}{k^2} e^{\tau(\eta)} [1-f(z)]j_\ell[k\chi(z)],\nonumber \\
    W^{\rm g}_\ell (k, z) &=&  b_{\rm g}(z) p(z) j_\ell[k\chi(z)], \nonumber 
\end{eqnarray}
$P_{\rm mm}$ is the matter power spectrum and $j_\ell(x)$ is a spherical Bessell function. We can already see, from what we have shown, that this cross spectrum will be positive for $\Lambda$CDM. 
We compute this quantity with \texttt{hi\_class}, with the precision parameters in Appendix~\ref{sec:ISW-prec}. 

Since the method was first proposed in \cite{Crittenden:1995ak} there have been a number of attempts at measuring $C^{gT}_\ell$ \cite{Boughn:1997vs,Boughn:2003yz,PhysRevD.77.123520,2016A&A...594A..21P,2018PhRvD..97f3506S,Krolewski:2021znk, Seraille:2024beb}. Limits on how well we should be able to measure $\Lambda$ from this method alone were shown in \cite{Hu:2004yd} and has been argued that it should be possible to distinguish between a $\Lambda$ dominated universe and a Universe with dark energy with a time varying equation of state \cite{Pogosian:2005ez}. 

The ISW effect can be used, in principle, to constrain the behaviour of dark energy. So, for example, cubic Galileons have been ruled out as a possible theory of dark energy at high significance due to their anomalous ISW \cite{Renk:2016olm,Renk:2017rzu}. More generally,
the presence of $\Sigma$ has an additional effect on the ISW, which can be inferred from the modified Newton-Poisson equation for $\Phi_+$. Now we have $ \partial_\eta\Phi_+=(\alpha_\Sigma+f-1){\cal H}\Phi_+ $ where $\alpha_\Sigma\equiv d\ln \Sigma/d\ln a$.  Thus if $\Sigma>1$ and grows (i.e.\ $\alpha_\Sigma>0$), the presence of fifth forces may have a dramatic effect on $\partial_\eta\Phi_+$. Indeed, just a moderate growth in $\Sigma$ will change the sign of $\partial_\eta\Phi_+$ and may lead to {\it negative} cross correlation spectrum, $C^{gT}_\ell$, on some scales, very much like what was found in the case of cubic Galileons\footnote{Another way of getting a negative ISW contribution is discussed in \cite{Das:2013sca}}. 

\begin{figure}
    \centering
    \includegraphics[width=\linewidth]{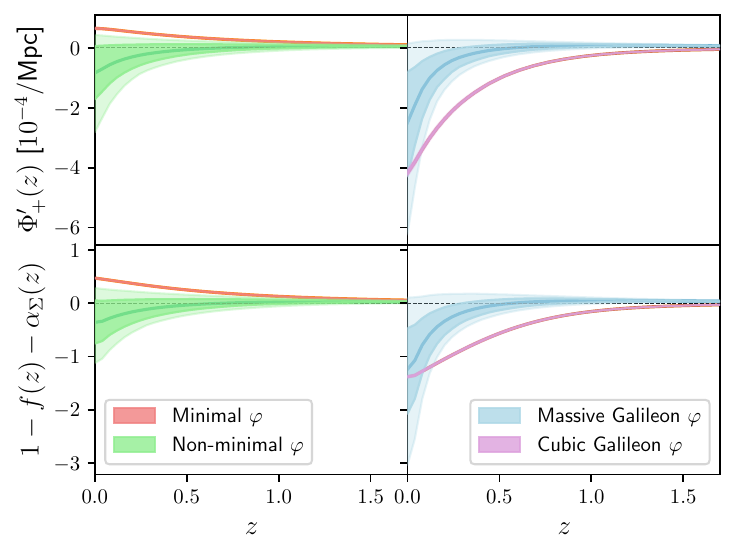}
    \caption{Posterior distributions of the time derivative of the potential $\Phi_+$ (top panel) and the quantity $(1-f(z)-\alpha_\Sigma(z))$ (bottom panel) for the four scalar dark energy models studied in this work, as given by current \desidesdovcmb data. Both the minimally coupled and the shift symmetric scalar fields have tight distributions, whereas the Non-minimal and Massive Galileon scalar fields have a much broader dynamical range.}
    \label{fig:isw_th_bands}
\end{figure}

\begin{figure*}[t]
    \centering
    \includegraphics[width=0.9\linewidth]{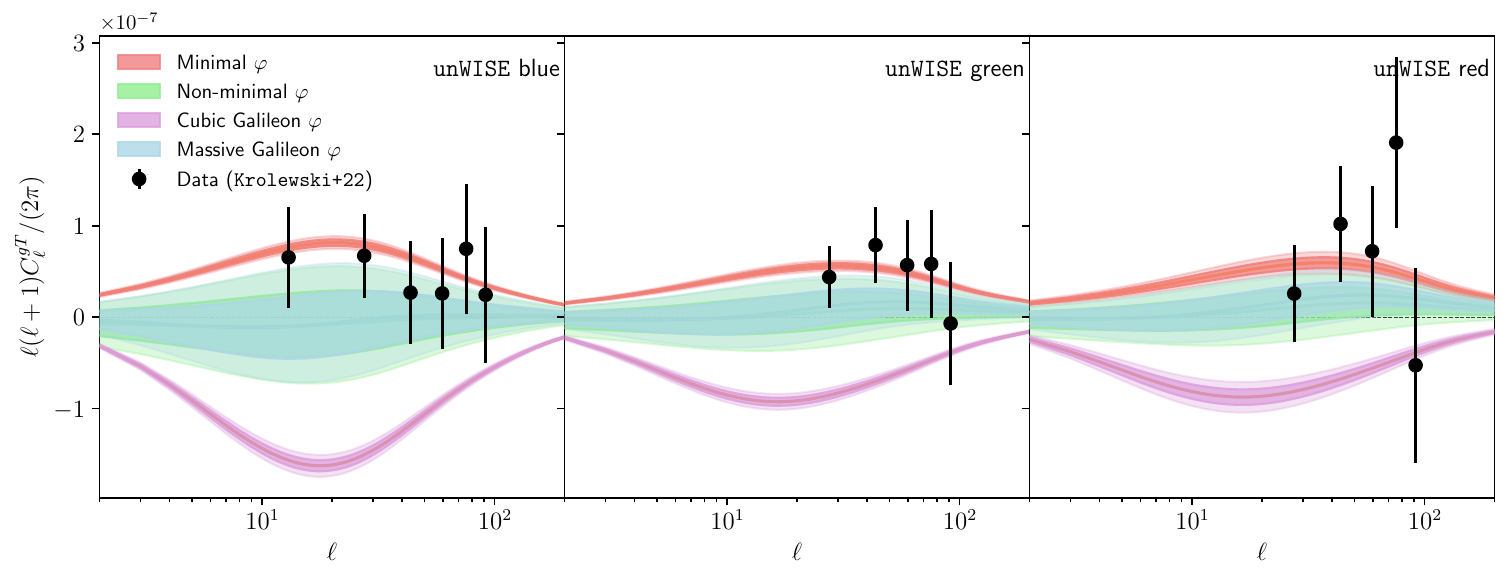}
    \caption{In black, the angular power spectra for the \unwise galaxy sample cross Planck 18 CMB temperature from \cite{Planck:2018yye}. The different bands are the projection of the \desidesdovcmb chains. We see a reasonably good agreement for the minimally coupled $\varphi$, NM and mG cases, taking into account that no ISW data has been used in the chains. The disagreement of cubic Galileon is a know problem of this theory \cite{Renk:2017rzu}. When computing the $A_{\rm ISW}$ and $b_{\rm g}$ fits, we do it per bin, since we do not have the cross-covariance.}
    \label{fig:Krolewski}
\end{figure*}

For both the NM and the mG models, we can have an $\alpha_\Sigma$ which is large enough to change the sign of the integrand in the ISW effect.  In Figure \ref{fig:isw_th_bands} we plot $1-f-\alpha_\Sigma$ which are consistent with current observations from the expansion history data (i.e. \desidesdovcmb) and we can clearly see that both can be positive across the redshift range of relevance for the ISW effect. We also show the effect on $\Phi_+$. Unlike the case of $\Lambda$CDM where $\Phi_+$ is negative but slowly evolves towards $0$, in the case of the evolving dark energy models, $\Phi_+$ diverges away from $0$, becoming more negative with time.

In order to assess the compatibility of NM and mG models with ISW data, we have collated three different public data sets that span a wide range of redshifts (for the details of the data sets and their analysis, we refer the reader to Appendix \ref{app:ISW-extra}).
The first data set, shown in Figure \ref{fig:Krolewski}, corresponds to the cross-correlation of \planck 2018 temperature fluctuations \cite{Planck:2018yye} with the full sky \unwise galaxy sample \cite{2019ApJS..240...30S}. We call this data set \krolewski and we refer the readers to \cite{Krolewski:2021znk} for a comprehensive description of the sample. 
A second data set, that has been widely used by the dark energy community, is the one compiled in \cite{Stolzner:2017ged} and contains cross-correlations of \planck 2015's CMB temperature fluctuations \cite{Planck:2015mis} with the NRAO VLA Sky Survey (\nvss) radio source catalog \cite{Condon:1998iy, DeZotti:2009an},  the 2MASS Photometric Redshift catalog (\tmpz) \cite{Bilicki:2013sza, Alonso:2014xca},  the Sloan Digital Sky Survey (\sdss) DR12 photo-z sample compiled by \cite{2016MNRAS.460.1371B},  the WISE × SuperCOSMOS (\wisc) galaxy sample \cite{Bilicki:2016irk} and  the SDSS DR6 photometric quasars (QSOs) from \cite{Richards:2008eq, 2009JCAP...09..003X}. We use the data publicly available in MontePython\footnote{Available at: \url{https://github.com/brinckmann/montepython_public/tree/3.6/data/ISW}. Note that there are some differences with respect to the $C_\ell$ plots in \cite{Stolzner:2017ged}.} \cite{Brinckmann:2018cvx, Audren:2012wb} and will refer to them as \stolzner and can be found in Figure~\ref{fig:Stolzner}, in the Appendix~\ref{app:ISW-gT}. 
The third and final data set we use comes from \cite{Reeves:2025xau} and comprises the cross-correlations between \planck 2018 temperature \cite{Planck:2018yye} and the Luminous Red Galaxy (LRG) catalogue from \desi Legacy Imaging survey DR9. It is the most recent ISW measurement available, we will call it \reeves and can be found in Figure~\ref{fig:Reeves_individual}, in the Appendix~\ref{app:ISW-gT} 
In order to calibrate the galaxy bias, we include the cross-correlation of the galaxy maps with \planck 18 CMB lensing \cite{Planck:2018lbu}, in the case of \krolewski, \act DR6 \cite{ACT:2023dou} lensing, in the case of \reeves, and the auto-correlation of the galaxy overdensities for \stolzner. These can be found in Figures~\ref{fig:Krolewski_bg}, \ref{fig:Stolzner_bg} and \ref{fig:Reeves_individual_bg}, in Appendix~\ref{sec:ISW-data}.

\begin{figure*}
    \centering
    \includegraphics[width=\linewidth]{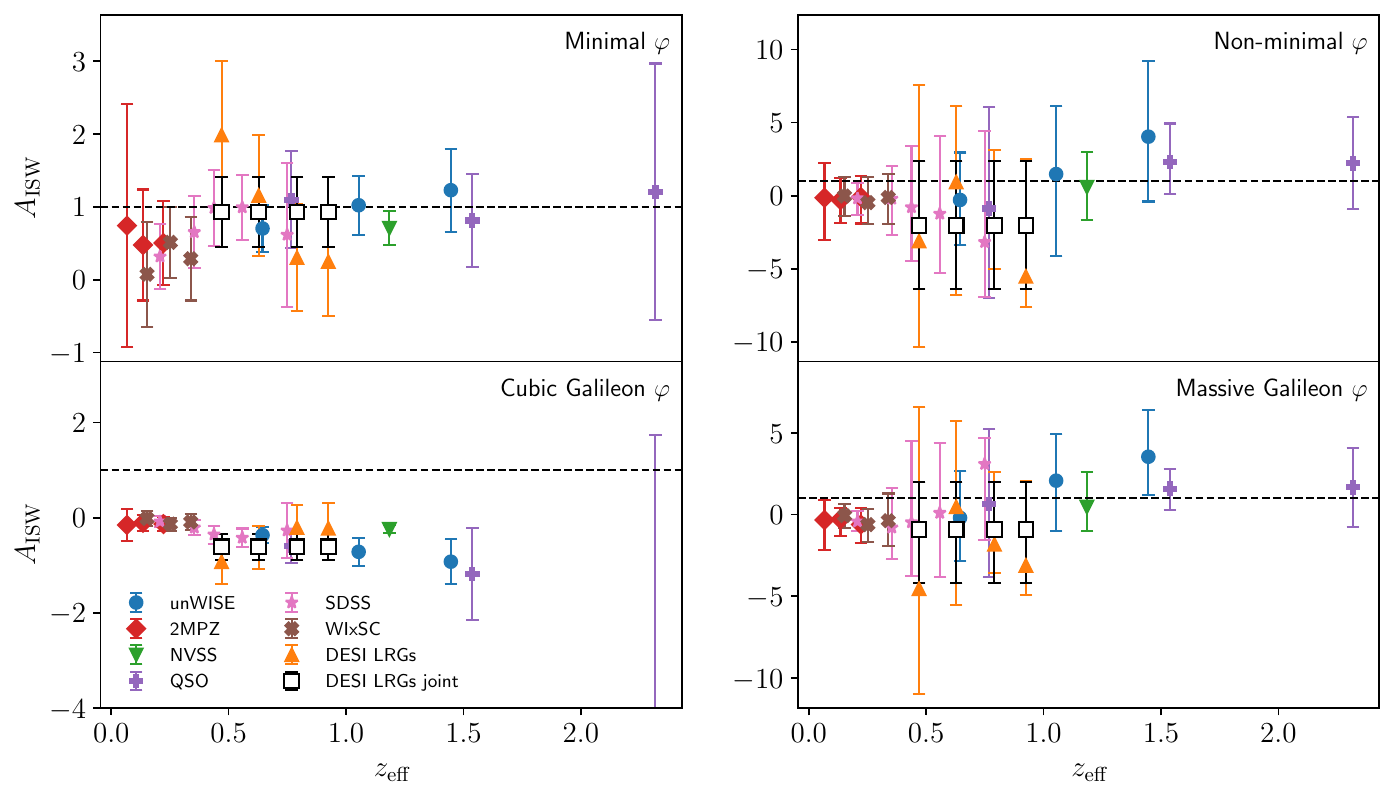}
    \caption{$A_{\rm ISW}$ results at the mean redshift of each galaxy sample. Each survey has its own distinct marker and color. The dashed line is $A_{\rm ISW} = 1$ that corresponds to perfect agreement with the data. Note that we have included the value from the joint analysis of \reeves, at each redshift bin location. Of course, it is the same in all of them as they were jointly fitted with a single $A_{\rm ISW}$ and cannot be interpreted as independent measurements.}
    \label{fig:A_ISW}
\end{figure*}

Given the low signal-to-noise of ISW measurements, it is customary to estimate the agreement with the data using the $A_{\rm ISW}$ parameter, defined as the best-fitting value for the modified power spectrum $\tilde C_\ell^{gT} = A_{\rm ISW} C_\ell^{gT}$. The agreement with the data is measured as 
$n_\sigma \equiv {|A_{\rm ISW} - 1|}/{\sigma_{A_{\rm ISW}}}$ where $n_\sigma$ is the number of $\sigma$ away the theory is from the data. In order to estimate the mean $A_{\rm ISW}$ and its uncertainty for the parameter space compatible with CMB, SNe and BAO data, we simultaneously fit $A_{\rm ISW}$ and the galaxy bias, $b_{\rm g}$, at each step of the chains\footnote{To speed up the process, instead of using the O(100k) MCMC steps, we used a subset of equally weighted $\sim500$ steps that matches the cosmological parameters posterior distribution (see Figure~\ref{fig:ISW_thinning}). The thinning of the chain is done closely following the \texttt{thin()} implementation in \texttt{GetDist}'s \texttt{chain.py} module \cite{Lewis:2019xzd}. This same subset was used to produce Figure~\ref{fig:isw_th_bands}.} by minimizing  $\chi^2 = (\tilde {\bf C} - \hat {\bf C})^T {\rm Cov}^{-1} (\tilde {\bf C} - \hat {\bf C})$, where $\tilde {\bf C}$ and $\hat {\bf C}$ are the concatenated theory angular power spectra and data estimator, respectively. In the case that we fit a single bin, $\tilde {\bf C}^T = \{ A_{\rm ISW} b_{\rm g}\, C_\ell^{gT}, b_{\rm g} C_\ell^{g\kappa} \}$ (or  $b_{\rm g}^2 C_\ell^{gg}$ for \stolzner which does not use the CMB lensing cross-correlation). The scales included in the fit are the same as in their respective analyses and can be found in Appendix~\ref{sec:ISW-data}. 

In the absence of magnification, the best fit and its associated uncertainty can be computed analytically in the basis $\{q = A_{\rm ISW} b_{\rm g},\, b_{\rm g}\}$ ($b_{\rm g}^2$ in the case of \stolzner) and propagating the error with a linear approximation around the minimum. However, when magnification cannot be neglected (as is the case for \krolewski and \reeves), the relation becomes non-linear and needs to be numerically minimized. In our case, only \stolzner data can be analyzed analytically. For convenience, we use \texttt{scipy}'s \texttt{curve\_fit} for all data sets to estimate both $b_{\rm g}$ and $A_{\rm ISW}$. We have checked the result against the analytical method for \stolzner, and found good agreement. Once fitted and with an estimate of their covariance, we then estimate the posterior distribution of the full chain by stacking all the individual step fits, assuming they are described by a Normal distribution. Numerically, for each fit we generate 200 realizations, that we stack together with all other steps in the chain and compute the 68\% and 95\% confidence levels. The results can be found in Figures \ref{fig:Krolewski}, and Figures~\ref{fig:Reeves_individual} and \ref{fig:Stolzner}, in the Appendix~\ref{app:ISW-gT}, and the $A_{\rm ISW}$ values on Table \ref{tab:AISW} and Figure~\ref{fig:A_ISW}. It is worth noticing that most measurements lie below $A_{\rm ISW} = 1$. If we were able to do a joint analysis, the tension with the data will most likely increase. In addition, some of the $A_{\rm ISW}$ values are large, such as \unwise \texttt{red} for the mG or NM. These are related to having angular power spectra very close to 0, requiring a larger amplitude correction to fit the data.

\begin{table*}
    \centering
    \begin{tabular}{|lc|rrrr|}
        \hline
        Survey & Bin &
        \multicolumn{1}{c}{Minimal $\varphi$} &
        \multicolumn{1}{c}{Non-minimal $\varphi$} &
        \multicolumn{1}{c}{Cubic Galileon} &
        \multicolumn{1}{c|}{Massive Galileon} \\ 
        \hline
        \multirow{3}{*}{\unwise}
            & 0 & $0.70 \pm 0.32$ ($0.93\sigma$) [$0.0\sigma$] & $-0.29^{+3.2}_{-3.1}$ ($0.41\sigma$) [$0.0\sigma$] & $-0.37 \pm 0.17$ ($8.2\sigma$) [$6.9\sigma$] & $-0.19^{+2.9}_{-2.6}$ ($0.44\sigma$) [$0.0\sigma$] \\
            & 1 & $1.02 \pm 0.40$ ($0.05\sigma$) [$0.0\sigma$] & $1.48^{+4.6}_{-5.6}$ ($0.10\sigma$) [$0.0\sigma$] & $-0.72 \pm 0.29$ ($5.9\sigma$) [$4.7\sigma$] & $2.08^{+2.8}_{-3.1}$ ($0.38\sigma$) [$0.0\sigma$] \\
            & 2 & $1.23 \pm 0.57$ ($0.40\sigma$) [$0.0\sigma$] & $4.04^{+5.2}_{-4.4}$ ($0.58\sigma$) [$0.5\sigma$] & $-0.92 \pm 0.47$ ($4.1\sigma$) [$3.2\sigma$] & $3.55^{+2.8}_{-2.4}$ ($0.90\sigma$) [$0.4\sigma$] \\
            
        \hline

        \multirow{4}{*}{\desi LRGs}
            & 0 & $2.0 \pm 1.0$ ($0.97\sigma$) [$0.0\sigma$] & $-3.1^{+11}_{-7.3}$ ($0.46\sigma$) [$0.4\sigma$] & $-0.92 \pm 0.48$ ($4.0\sigma$) [$2.9\sigma$] & $-4.6^{+11}_{-6.4}$ ($0.63\sigma$) [$0.4\sigma$] \\
            & 1 & $1.16 \pm 0.83$ ($0.19\sigma$) [$0.0\sigma$] & $0.93^{+5.2}_{-7.7}$ ($0.01\sigma$) [$0.0\sigma$] & $-0.62 \pm 0.45$ ($3.6\sigma$) [$2.3\sigma$] & $0.48^{+5.3}_{-6.0}$ ($0.09\sigma$) [$0.0\sigma$] \\
            & 2 & $0.30 \pm 0.73$ ($0.96\sigma$) [$0.7\sigma$] & $-2.09^{+5.2}_{-2.9}$ ($0.76\sigma$) [$0.5\sigma$] & $-0.21 \pm 0.47$ ($2.6\sigma$) [$2.0\sigma$] & $-1.8^{+4.4}_{-1.8}$ ($0.90\sigma$) [$0.5\sigma$] \\
            & 3 & $0.25 \pm 0.75$ ($1.0\sigma$) [$0.8\sigma$] & $-5.5^{+8.0}_{-2.1}$ ($1.3\sigma$) [$0.6\sigma$] & $-0.22^{+0.52}_{-0.53}$ ($2.3\sigma$) [$1.9\sigma$] & $-3.1^{+5.2}_{-1.8}$ ($1.2\sigma$) [$0.6\sigma$] \\
            & Joint & $0.93 \pm 0.48$ ($0.15\sigma$) [$0.0\sigma$] & $-2.0 \pm 4.4$ ($0.70\sigma$) [$0.0\sigma$] & $-0.61 \pm 0.28$ ($5.8\sigma$) [$3.4\sigma$] & $-0.90^{+2.9}_{-3.3}$ ($0.62\sigma$) [$0.0\sigma$] \\
        \hline

        \multirow{1}{*}{\nvss}
            & 0 & $0.71^{+0.24}_{-0.23}$ ($1.2\sigma$) [$0.9\sigma$] & $0.6^{+2.4}_{-2.2}$ ($0.18\sigma$) [$1.1\sigma$] & $-0.24 \pm 0.09$ ($14\sigma$) [$100.0\sigma$] & $0.5^{+2.2}_{-1.5}$ ($0.29\sigma$) [$1.1\sigma$] \\
        \hline
            
        \multirow{3}{*}{\tmpz}
            & 0 & $0.7 \pm 1.7$ ($0.16\sigma$) [$0.0\sigma$] & $-0.1^{+2.4}_{-2.9}$ ($0.43\sigma$) [$0.0\sigma$] & $-0.15 \pm 0.33$ ($3.5\sigma$) [$2.4\sigma$] & $-0.3^{+1.3}_{-1.8}$ ($0.87\sigma$) [$0.0\sigma$] \\
            & 1 & $0.48 \pm 0.76$ ($0.68\sigma$) [$0.0\sigma$] & $-0.3 \pm 1.5$ ($0.85\sigma$) [$0.0\sigma$] & $-0.11 \pm 0.17$ ($6.5\sigma$) [$5.5\sigma$] & $-0.3^{+0.70}_{-1.0}$ ($1.5\sigma$) [$0.0\sigma$] \\
            & 2 & $0.50 \pm 0.58$ ($0.86\sigma$) [$0.0\sigma$] & $-0.1^{+1.5}_{-1.8}$ ($0.67\sigma$) [$0.0\sigma$] & $-0.13 \pm 0.15$ ($7.5\sigma$) [$6.6\sigma$] & $-0.6 \pm 1.1$ ($1.5\sigma$) [$0.0\sigma$] \\
        \hline
            
        \multirow{3}{*}{QSO}
            & 0 & $1.10 \pm 0.67$ ($0.15\sigma$) [$0.8\sigma$] & $-0.9^{+6.9}_{-6.1}$ ($0.29\sigma$) [$1.3\sigma$] & $-0.59 \pm 0.37$ ($4.3\sigma$) [$4.0\sigma$] & $0.6 \pm 4.5$ ($0.08\sigma$) [$1.2\sigma$] \\
            & 1 & $0.81 \pm 0.64$ ($0.30\sigma$) [$0.0\sigma$] & $2.3^{+2.6}_{-2.2}$ ($0.54\sigma$) [$0.0\sigma$] & $-1.18 \pm 0.97$ ($2.2\sigma$) [$0.9\sigma$] & $1.6^{+1.2}_{-1.3}$ ($0.47\sigma$) [$0.0\sigma$] \\
            & 2 & $1.2 \pm 1.8$ ($0.12\sigma$) [$0.0\sigma$] & $2.2 \pm 3.1$ ($0.39\sigma$) [$0.0\sigma$] & $-7.1 \pm 8.9$ ($0.92\sigma$) [$0.2\sigma$] & $1.7 \pm 2.4$ ($0.28\sigma$) [$0.0\sigma$] \\
        \hline
            
        \multirow{3}{*}{\wisc}
            & 0 & $0.07 \pm 0.72$ ($1.3\sigma$) [$0.1\sigma$] & $0.0^{+1.3}_{-1.4}$ ($0.77\sigma$) [$0.0\sigma$] & $-0.02 \pm 0.17$ ($6.0\sigma$) [$5.3\sigma$] & $-0.06^{+0.73}_{-0.74}$ ($1.4\sigma$) [$0.0\sigma$] \\
            & 1 & $0.51 \pm 0.49$ ($1.0\sigma$) [$0.3\sigma$] & $-0.5^{+1.8}_{-1.5}$ ($0.92\sigma$) [$0.5\sigma$] & $-0.13 \pm 0.14$ ($8.1\sigma$) [$7.7\sigma$] & $-0.6^{+1.0}_{-1.1}$ ($1.6\sigma$) [$0.7\sigma$] \\
            & 2 & $0.29 \pm 0.57$ ($1.2\sigma$) [$0.1\sigma$] & $-0.1^{+1.6}_{-1.8}$ ($0.66\sigma$) [$0.0\sigma$] & $-0.09 \pm 0.18$ ($6.1\sigma$) [$5.4\sigma$] & $-0.4^{+1.7}_{-1.5}$ ($0.86\sigma$) [$0.0\sigma$] \\
        \hline
            
        \multirow{5}{*}{\sdss}
            & 0 & $0.32 \pm 0.45$ ($1.5\sigma$) [$0.7\sigma$] & $-0.2 \pm 1.1$ ($1.1\sigma$) [$0.3\sigma$] & $-0.08 \pm 0.11$ ($9.8\sigma$) [$100.0\sigma$] & $-0.42^{+0.65}_{-0.57}$ ($2.3\sigma$) [$0.6\sigma$] \\
            & 1 & $0.65 \pm 0.50$ ($0.70\sigma$) [$0.0\sigma$] & $-0.2^{+2.2}_{-2.5}$ ($0.52\sigma$) [$0.0\sigma$] & $-0.21 \pm 0.16$ ($7.6\sigma$) [$6.8\sigma$] & $-0.8^{+2.4}_{-1.9}$ ($0.82\sigma$) [$0.0\sigma$] \\
            & 2 & $0.98 \pm 0.52$ ($0.04\sigma$) [$0.0\sigma$] & $-0.8^{+4.2}_{-3.7}$ ($0.45\sigma$) [$0.3\sigma$] & $-0.35 \pm 0.19$ ($7.1\sigma$) [$6.3\sigma$] & $-0.5^{+5.0}_{-3.3}$ ($0.36\sigma$) [$0.2\sigma$] \\
            & 3 & $0.99 \pm 0.45$ ($0.02\sigma$) [$0.3\sigma$] & $-1.2^{+5.3}_{-4.1}$ ($0.48\sigma$) [$1.2\sigma$] & $-0.42 \pm 0.19$ ($7.5\sigma$) [$6.7\sigma$] & $0.1^{+4.3}_{-4.0}$ ($0.22\sigma$) [$1.0\sigma$] \\
            & 4 & $0.62 \pm 0.99$ ($0.38\sigma$) [$0.9\sigma$] & $-3.2^{+7.6}_{-3.8}$ ($0.74\sigma$) [$1.0\sigma$] & $-0.27 \pm 0.59$ ($2.2\sigma$) [$1.9\sigma$] & $3.1^{+1.6}_{-4.7}$ ($0.67\sigma$) [$0.9\sigma$] \\
        \hline
    \end{tabular}
    \caption{$A_{\rm ISW}$ results for each survey tomographic bin when fitted in isolation and together. This is only possible for the \desi LRGs as it is the only sample for which we have the full cross-covariance. In parenthesis, the tension with the data; i.e. with $A_{\rm ISW} = 1$. In square brackets, the tension computed from the $C_\ell^{gT}$ posterior distribution using Eq.~\ref{eq:suspicion} and setting it to $0\sigma$ for $p<0.5$. The \unwise sample is from \krolewski \cite{Krolewski:2021znk}, \desi LRGs from \reeves \cite{Reeves:2025xau} and the others from \stolzner \cite{Stolzner:2017ged}.}
    \label{tab:AISW}
\end{table*}

The ISW effect provides a useful consistency test of the scalar-field models considered above because it probes not only the background expansion history but also the time evolution of the Weyl potential. This is particularly important for the non-minimally coupled and massive Galileon models, for which the same interactions that improve the fit to BAO, SNe, and CMB distances also modify the relation between matter perturbations and the metric potentials. In $\Lambda$CDM the late-time decay of gravitational potentials gives a positive cross-correlation between CMB temperature anisotropies and low-redshift tracers of large-scale structure. By contrast, sufficiently rapid evolution of the effective lensing modification, encoded through $\alpha_\Sigma$ can change the sign of the ISW source term and lead to a suppressed or even negative $C_\ell^{gT}$.

This behaviour is visible in Figure  \ref{fig:isw_th_bands}. The minimally coupled scalar field remains close to the $\Lambda$CDM expectation, with a comparatively narrow range of allowed potential evolution. The non-minimally coupled and massive Galileon models, however, span a much wider range in both $\Phi_+'$ and $1-f-\alpha_\Sigma$. Thus, although these models are viable at the level of the expansion history, they predict a more varied phenomenology for the late-time gravitational potentials. The ISW effect is therefore a direct probe of the gravitational sector of these theories rather than merely another distance measurement.

The comparison with the ISW data in Figures \ref{fig:Krolewski}, \ref{fig:Reeves_individual} \ref{fig:Stolzner} 
shows that this additional phenomenology is not currently fatal for the models preferred by \desidesdovcmb, with all $A_{\rm ISW}$ fits within $\lesssim1\sigma$ agreement with the data. This is also true if we directly compute the tension between the $C_\ell^{gT}$ posterior distributions and the data using Eq.~\ref{eq:suspicion}, which can be seen in square brackets in Table~\ref{tab:AISW}. Note that we have defined $0\sigma$-tension for those models that agree with the data better than expected for a Gaussian realization; i.e. $p<0.5$. The minimally coupled model gives the closest match to the standard positive ISW signal, as expected from its proximity to $\Lambda$CDM. The cubic Galileon case, by contrast, tends to predict a negative or strongly suppressed cross-correlation over a range of tracers and redshift bins, reproducing the known tension of this class of models with ISW measurements \cite{Renk:2017rzu}. The non-minimally coupled and massive Galileon models occupy an intermediate position: they allow a broader range of ISW amplitudes, reflecting their fifth-force phenomenology, but their predicted bands remain broadly compatible with the existing measurements.

This result should be interpreted conservatively.  On the one hand, the current ISW data have low signal-to-noise, however the datasets used here have different sky coverage, tracer population and redshift distributions, providing complementary information. In the future, provided a consistent covariance can be estimated, it will be possible to do a fully consistent joint analysis, increasing the constraining power of these data and, possibly, the level of tension (or not) with the dark energy predictions. On the other hand, the datasets used here come with shortcomings (see Appendix~\ref{sec:ISW-data} for a more detailed discussion). First, for \krolewski and \stolzner, we do not have the cross-covariance between $C_\ell^{gT}$ and $C_\ell^{g\kappa}$ and $C_\ell^{gg}$, respectively, effectively assuming they are uncorrelated in our fits. Similarly, for \stolzner, the treatment of the redshifts distribution can be improved. As discussed in the Appendix~\ref{sec:ISW-data}, these approximations are probably good enough for our qualitative analysis but might need to be revisited for precision cosmology. In addition, \reeves data set has not been optimized for an ISW analysis and there may be room for improvements to maximize the signal-to-noise. For instance, looking at the error bars in Table~\ref{tab:AISW}, one can appreciate that they are systematically larger than those of \krolewski and \stolzner. Finally, as it is further discussed in Appendix~\ref{sec:ISW-prec}, in the case of the mG model, the theory vector computation fails for a significant fraction of models in some low-$z$ samples, which could potentially impact the ISW constraints. However, we have checked that even in these cases, we still recover the full sample PDF good enough for the purposes of this work (see Figure~\ref{fig:ISW_thinning}).
The agreement seen in Figures \ref{fig:Krolewski}, \ref{fig:Reeves_individual} and \ref{fig:Stolzner}  should therefore be regarded primarily as a qualitative consistency check rather than a precision model comparison. Nevertheless, the fact that the non-minimal and massive Galileon posteriors inferred from CMB, BAO, and SNe data do not obviously overshoot or reverse the observed ISW signal is non-trivial. It suggests that present ISW measurements do not yet exclude the fifth-force dynamics required by these models, while also highlighting the ISW effect as a promising route for testing them with a more homogeneous future analysis.

\section{The future}
\label{sec:future}

We have made some progress in assessing single scalar field dark energy. Yet we still find ourselves in a situation of profound underdetermination \cite{Ferreira:2025fpn}. To begin with,  there is not any clear observational evidence {\it against} $\Lambda$CDM. There are questions about the self-consistency of individual data sets, their interpretation and the cross consistency between them. Furthermore, any quantitative statistical statement for or against $\Lambda$CDM seems to be very dependent on assumptions such as what cosmological parameters are included, their range of values (i.e priors) and the data sets consider. This makes it difficult, if not impossible, to make definitive statements with current data. Thus, we need to look forward.

Fortunately we are in the middle of a new era in cosmological surveys -- dubbed the {Stage IV} era -- in which more and better data will be available in the foreseeable future. The \euclid satellite mission \cite{Laureijs:2011gra}, the Rubin Observatory \cite{LSSTScienceBook:2009jmu}, the Nancy Roman Telescope \cite{Spergel:2015sza} will all, along with the next iterations of the \desi collaboration \cite{DESI:2016fyo}, contribute to a better understanding of the expansion of the Universe. All of these missions will improve measurements of the expansion rate of the Universe as well as of the large scale structure of the matter distribution and, in particular, the growth rate (or any other measure of growth through, for example, constraints from weak lensing and galaxy clustering, in the so-called 3x2pt analyses). 

We already have an idea of how much better our constraints will improve as there have been multiple attempts at forecasting the outcomes of these various experiments. For example we can adopt the assumptions made in the \texttt{LSST} Dark Energy Science Colalboration (\texttt{DESC}) Science Readiness Document (SRD) \cite{thelsstdarkenergysciencecollaboration2021lsstdarkenergyscience}, noting that forecasts for other combinations of Stage IV surveys will be equivalent, within factors of a few. Following the \srd, and restricting ourselves to the CPL parameters, $(w_0,w_a)$, we can use the Figure of Merit, FOM, defined to be $ {\rm FOM}=\left[{\rm det}\left({\rm  Cov}\right)\right]^{-1/2}$ as our measure of how well we will perform \cite{Albrecht:2006um}.  In this expression, ${\rm Cov}$ is the parameters covariance matrix, i.e.\ a measure of their predicted uncertainty and correlations. Focusing on $(w_0,w_a)$ alone, from the \srd, we expect the ${\rm FOM}\simeq 500-1000$, an improvement over current constraints (${\rm FOM} =  187.5$ for \desidesdovcmb, and ${\rm FOM} = 222$ when including {\tt DES} weak lensing and galaxy clustering 3x2pt data \cite{DES:2026jmi}) by a factor of few . Clearly, this is not a step change in our knowledge and merely attests to the fact that we are already in the Stage IV era. 

Let us now use these forecasts to imagine what the future holds. As mentioned above, there are still (systematic) uncertainties about the current data and one must consider three possible outcomes:  
\begin{itemize}
    \item ${\bf \Lambda}$: a re-analysis of the current data consolidates $\Lambda$CDM as actually fully consistent with current measurements and future measurements just tighten this conclusion.
    \item {\bf Q}: as above, a reanalysis pushes the current constraints upwards but not leftwards in the $(w_0,w_a)$ plane. As a result, minimally coupled (quintessence) scalar field models (but not $\Lambda$) are consistent with current and future data.
    \item {\bf X}: as a result of further scrutiny of current data and reinforced with future data, the constraints remain stubbornly where they are or migrate downwards in the $(w_0,w_a)$ plane in region favoured by extended models.
\end{itemize}

\begin{figure}[t]
    \centering
    \includegraphics[width=\columnwidth]{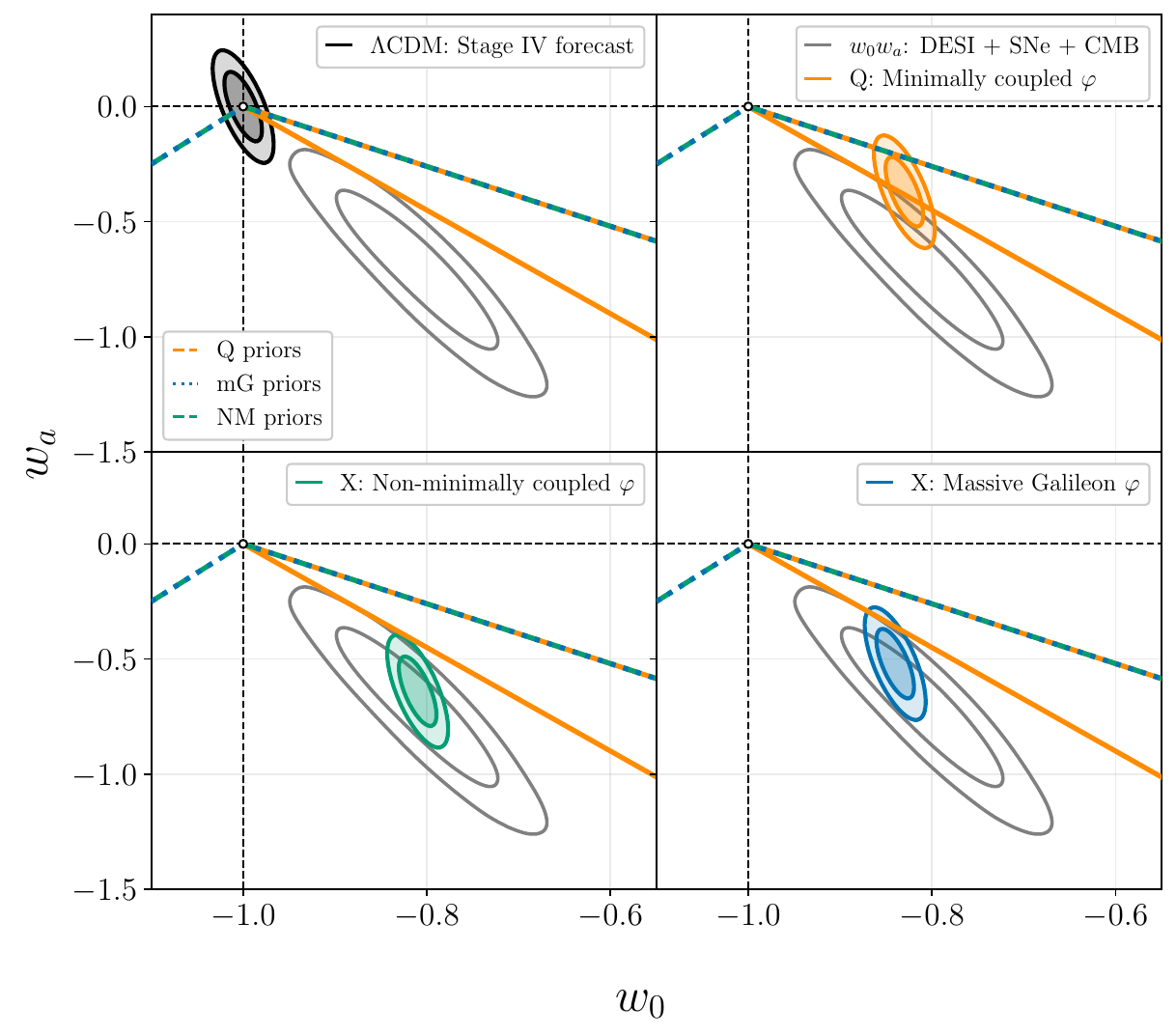}
    \caption{Qualitative depiction of the possible different scenarios in the near future. The ellipses show the 68\% and 95\% C.L. posterior distribution of the CPL parameters assuming the forecast errors from the \srd \cite{thelsstdarkenergysciencecollaboration2021lsstdarkenergyscience}, under the scenarios mentioned in the text: either we go back to $\Lambda$CDM (\textbf{$\Lambda$}, they move to agree with quintessence (\textbf{Q}) or they stay compatible with the exotic scenarios (\textbf{X}). We have centered the ellipses on the best fits obtained with \desidesdovcmb in Section~\ref{sec:cosmoconstraints}, and compared with current constraints, in gray. The improvement will be, in terms of the FOM, of a few factors (${\rm FOM} \sim 1000$, in the \srd very optimistic case, vs. $187.5$ with current data).} 
    \label{fig:expansion_forecast}
\end{figure}
Figure~\ref{fig:expansion_forecast} illustrates these scenarios and compares them with current \desidesdovcmb constraints from Section~\ref{sec:cosmoconstraints}. We use the {\it very} optimistic prediction for $w_0-w_a$ ${\rm FOM}=1000$ ($\sigma(w_0) \simeq 0.014$ and $\sigma(w_a) \simeq 0.10$) from the \srd \cite{thelsstdarkenergysciencecollaboration2021lsstdarkenergyscience}, centered at the models' best fit $w_0$-$w_a$. Although the \srd considers that a more realistic target is ${\rm FOM}=500$ ($\sigma(w_0) \simeq 0.02$ and $\sigma(w_a) \simeq 0.14$) we prefer to use the overly optimistic case to emphasize our point and to make room for possible improvements due to e.g. combinations with newer external probes.

If we assume that future cosmological constraints continue converging on the region of parameter space favored by current data with the expected precision of Stage IV, the statistical evidence for dynamical dark energy would go from moderate ($\simeq 3\sigma$) to very strong/overwhelming. For {\bf Q} or {\bf X}, the optimistic forecast shown in Figure \ref{fig:expansion_forecast} (${\rm FOM} \sim 1000$) would yield a $\simeq 12\sigma-14\sigma$ detection of dynamical dark energy over $\Lambda$ ($w_0 = -1, w_a = 0$). For the more realistic target (${\rm FOM} \sim 500$), these would be reduced to the still very strong result of $\simeq 8\sigma -10\sigma$ significance level. This does leave open the possibility of eliminating $\Lambda$ as an explanation for dark energy, where the statistical evidence would be sufficiently strong to overcome many of the current issues that preclude a clear evidential preference for $\Lambda$ or dynamical dark energy (priors, systematics, choice of data, differing analysis techniques, etc.). The question then would be whether we will be able to distinguish between different kinds of single scalar field dark energy scenarios.

In this sense, the ${\bf \Lambda}$ and {\bf Q} possibilities place us firmly in the regime where quintessence suffices to explain the observation. Furthermore, as emphasized above, we will need at most an overall constant ($\Lambda$ or $V_0$ depending on how one wants to view it) and $m^2$. This will give us a complete description of dark energy apart from (significant) theoretical concerns about naturalness and UV-completeness. Note that extensions are still viable (such as the NM or mG model) but nothing more can be learnt from background\footnote{Recall, we are calling \desisnecmb loosely background data as the background is the most dominant information in the dataset combination.} cosmological data. Any deviations from quintessence or $\Lambda$ must be found in exotic effects in gravitational clustering which may indicate new physics at play. 

The {\bf X} hypothesis is more intriguing. As we have seen there are a number of possibilities. If we assume that the EFT we are considering is valid all the way down to solar system scales we have established that the NM and mG models have severe problems with local, astrophysical constraints and that, for example, \llr measurements introduce moderate tensions between large scale and small scale constraints. Screening may protect these theories from these constraints (although there are theoretical concerns about the viability of these mechanisms). Possibilities are theories in which the weak equivalence principle is violated or the possibility that there is a hidden, as yet undetected, other component of the Universe which is masquerading as matter or dark energy \cite{Caldwell:2025inn}. Or the (not unreasonable) possibility that the EFT we are considering is not valid on Solar System scales.

\begin{figure}[t]
    \centering
    \includegraphics[width=\columnwidth]{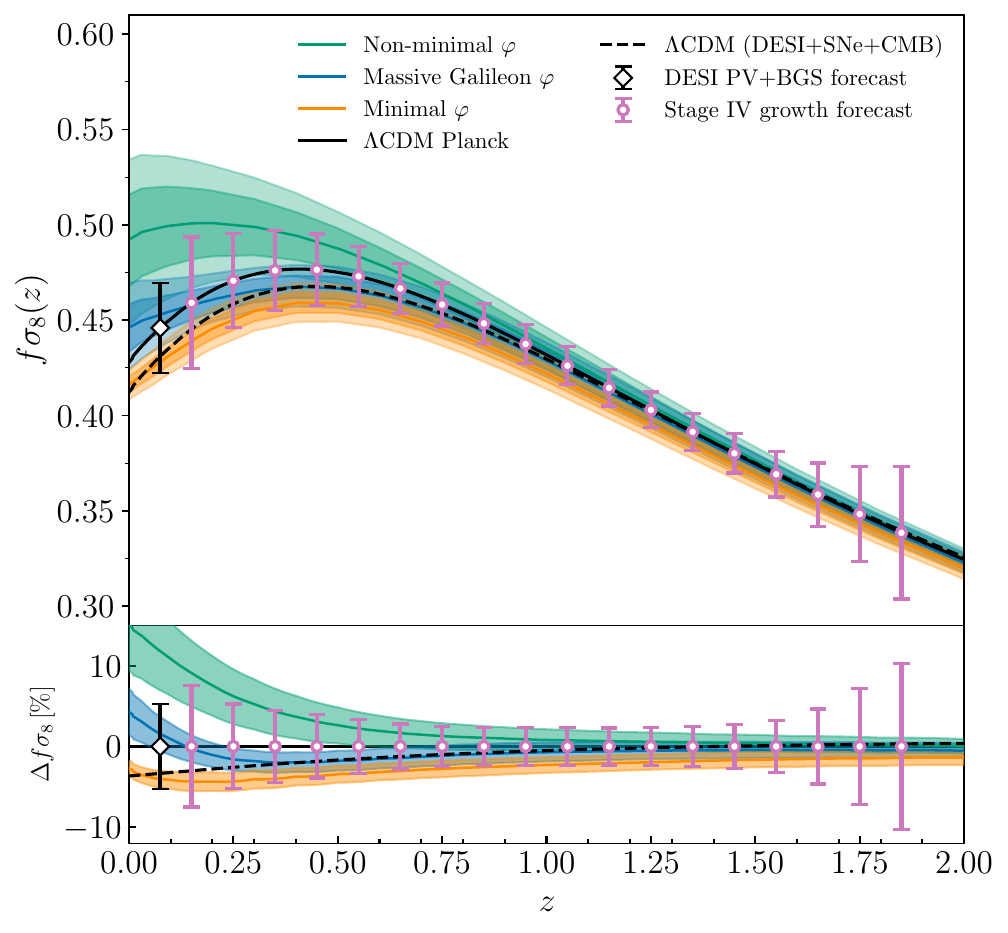}
    \includegraphics[width=\columnwidth]{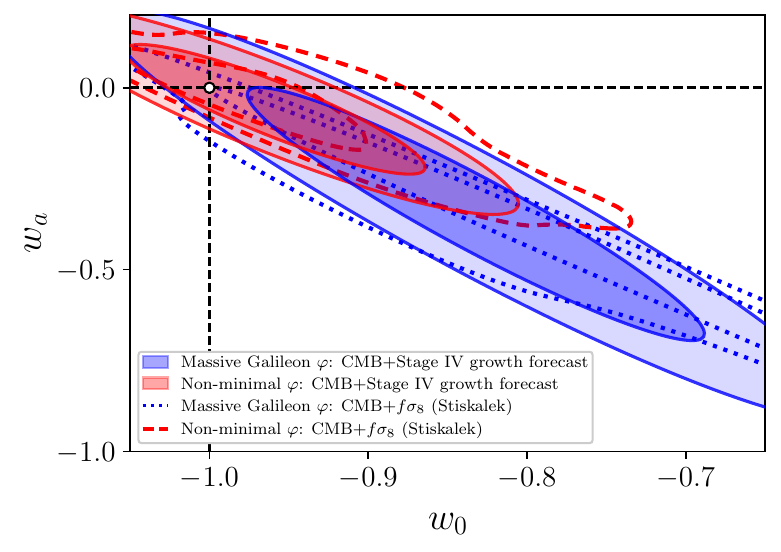}
    \caption{
    {\it Top}: Forecast uncertainties on $f\sigma_8$ using Table V from \cite{Font-Ribera:2013rwa} and the fiducial predictions from the \planck 2018 TTTEEE+lowE+lensing $\Lambda$CDM cosmology (Table 2 in \cite{Planck:2018vyg}), compared with the constraints on $f\sigma_8$ using our baseline expansion history dataset as in Figure~\ref{fig:fsigma8data}.
    {\it Bottom}: 68\% and 95\% C.L. constraints for the mG and NM dark energy models from CMB and $f\sigma_8$ data in the $(w_0, w_a)$ plane. The dashed contours are actual constraints with CMB data and the low-$z$ data of \cite{Stiskalek:2025oht}. The solid contours are obtained with mock CMB and growth Stage IV data (shown in the top panel), with $\Lambda$CDM as fiducial model. They are approximated as ellipses to highlight that they are a forecast. The constraints are projected onto the $(w_0, w_a)$ plane following Section~\ref{sec:prior_procedure} with \desidesdovcmb data properties.} \label{fig:fsigma8_forecast}
\end{figure}

To further pin down {\bf X} theories with ancillary gravitational effects beyond their impact on the cosmological background, we will need improved measurements of the ISW effect and the late time growth rate of structure, both of which are, currently, inconclusive. Future measurements will sharpen constraints on gravitational growth, with e.g. \desi \cite{Font-Ribera:2013rwa} or \euclid \cite{Majerotto:2012mf} expected to produce constraints at the percent level on $f\sigma_8$ over a range of redshifts \footnote{As noted above, \desi is not measuring $f\sigma_8$. However, for our qualitative assessment of the near future, the forecast values of \cite{Font-Ribera:2013rwa} suffice.}. But the focus is on redshifts of $z\simeq 0.5$ and greater where, as we can see in Figure \ref{fig:fsigma8data} or \ref{fig:fsigma8_forecast}, what we really need are tighter constraints at very {\it low} redshifts, $z\sim 0.1$ and lower. 

The most direct method for doing so is with cosmic flows measurements and this has been done to remarkable effect in \stiskalek \cite{Stiskalek:2025oht}. DESI has also predicted a similar low-$z$ $\lesssim 5\%$-accurate growth measurements by combining the clustering of galaxies and PV from the BGS and PV surveys \cite{Saulder:2023oqm}. It may also be possible to obtain independent measurements from, for example, weak lensing or redshift space distortions. These methods will be hampered by severe limitations: cosmic variance due to the fact that we can only probe a narrow redshift range around us, the severe effect of non-linear growth which may be difficult to disentangle from the linear growth rate which we are trying to constrain, and, of course, the modern bugbear of cosmology, baryonic effects which has, until now proven to be difficult to model with enough accuracy.

In order to emphasize the importance of low-$z$ growth measurements, we show in the Figure \ref{fig:fsigma8_forecast} top panel, as an example, the forecast errors for \desi measurements from \cite{Font-Ribera:2013rwa}.  We then consider the following scenario: measurements of the expansion rate are consistent with one of the three scalar field dark energy scenarios -- Q, NM or mG -- but measurements of growth are firmly consistent with $\Lambda$CDM (motivated by a number of measurements from, for example, \boss \cite{BOSS:2016wmc} and \desi). In this situation, will it be possible to definitely rule out any of the scalar field scenarios?

As can be seen in Figure \ref{fig:fsigma8_forecast} top panel, one finds that, despite the substantial departures from the fiducial $\Lambda$CDM values at low redshift, the extended models remain reasonably compatible with the forecast growth data in the intermediate redshift range where most of the constraining power is. It is however important to note that, given the expected precision, and assuming uncorrelated measurements, a small constant shift between the different models, as the one seen in Figure~\ref{fig:fsigma8_forecast}, might bump the $\chi^2$ up and become detectable. However, given that the largest deviations occur at very low redshifts, low-$z$ growth measurements will offer the most important discriminatory power between $\Lambda$, a minimally coupled scalar field and more exotic extended models. 

To illustrate it further, we will compare the constraining power of future growth data with the current constraints with the low-$z$ measurement from \stiskalek \cite{Stiskalek:2025oht}, that we discussed in Section~\ref{sec:growth}. Note that \stiskalek has a similar constraining power to that predicted for \desi BGS+PV at $z<0.15$ \cite{Saulder:2023oqm} so the conclusion will hold even if \stiskalek error were underestimated. In Section~\ref{sec:growth}, we combined the $f\sigma_8$ measurement of \stiskalek \cite{Stiskalek:2025oht} with \planck PR3 CMB measurements in order to constrain the cosmological parameters that would otherwise remain unconstrained. Similarly, for the forecast mock data, we combine the \planck PR3 compressed CMB likelihood from \cite{Wolf:2024eph, Chen:2018dbv} with the $f\sigma_8$ predictions from \cite{Font-Ribera:2013rwa} (shown in Figure~\ref{fig:fsigma8_forecast} top panel), assuming a fiducial \planck 2018 $\Lambda$CDM model for both data sets. The results can be seen in the bottom panel of Figure~\ref{fig:fsigma8_forecast}.

To facilitate the comparison, the results shown in Figure~\ref{fig:fsigma8_forecast} bottom panel are the projection of the dark energy model posteriors onto the $(w_0, w_a)$ plane as described in Section~\ref{sec:prior_procedure}, using the fiducial \desidesdovcmb data combination (despite the MCMC having used CMB+$f\sigma_8$ measurements). In addition, to distinguish the constraints from actual and mock data, we represent the forecast $(w_0, w_a)$ posteriors as ellipses centered at the posterior means, with their corresponding covariances. Remarkably, the single low-$z$ growth data point has comparable constraining power to the complete Stage IV growth data. This has to do with the fact that, even though the Stage IV forecast has significantly more data and similarly tight error bars, the differences from the predicted $\Lambda$CDM values manifest most strongly at very low redshifts. Consequently, a single data point targeting the cosmological epoch where observable predictions are most divergent from each other yields comparable constraining power. This highlights the important role of high quality, low redshift growth measurements in revealing the microphysical nature of dark energy. 

\begin{figure}[t]
    \centering
    \includegraphics[width=0.95\columnwidth]{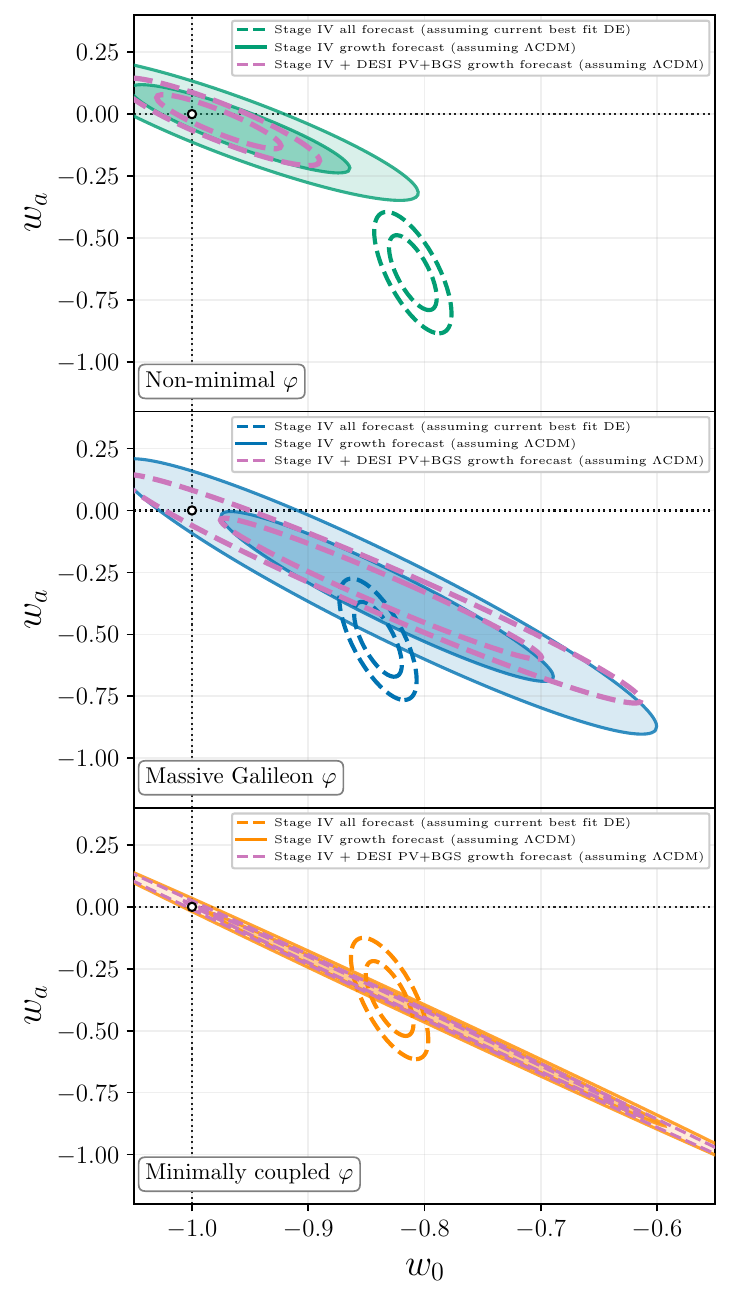}
    \caption{Comparison of Stage IV growth forecasts for the dark energy models assuming $\Lambda$CDM growth data from Figure \ref{fig:fsigma8_forecast} and the full Stage IV forecasts assuming the data is centered on the best fit dark energy model from the current expansion history data from Figure \ref{fig:expansion_forecast}. Only the NM model exhibits (mild) tension between the full and growth forecasts. This suggests that intermediate-$z$ growth data alone will not be highly discriminatory. However, in combination with low-$z$ data (\desipv), the NM growth constraints significantly improve, reaching a $4.8\sigma$ tension with the background constraints ($4\sigma$ in the \srd less optimistic scenario), hinting for possible discriminatory power in the near future. We do not show the \desipv results alone for clarity, but the constraining power is similar to that of Stage IV intermediate-$z$ growth data. To project the forecast expansion histories, we follow the procedure explained in Section~\ref{sec:prior_procedure}.}
    \label{fig:forecast_3pannel}
\end{figure}

Along these lines, it is also worth noting that, even if one assumes that Stage IV achieves the goals that have been set out in terms of their ability to measure the growth of cosmic structures in the intermediate redshift range, these results will likely have only a small impact on the overall Stage IV results. 

For example, we can compare the ``Stage IV all'' and Stage IV growth forecast results, which are depicted in Figure \ref{fig:forecast_3pannel}.  As discussed earlier, the ``Stage IV all'' forecast assumes that the data continues to favour the current \desidesdovcmb best fits from Section~\ref{sec:cosmoconstraints} for the respective dark energy models, while the growth forecast explores the constraints on these dark energy models assuming that the growth data is centered on $\Lambda$CDM. Even in this somewhat extreme scenario, we can see that for both the mG model and for quintessence, the $(w_0, w_a)$ constraints on these dark energy models coming from the growth data are consistent with the $(w_0, w_a)$ constraints on these dark energy models coming from the ``Stage IV all'' case (corresponding to $n_\sigma \simeq 1.2$ and $n_\sigma \simeq 0.1$ respectively). It is only the constraints on $(w_0, w_a)$ that are obtained when assuming the NM model that display an interesting tension ($n_\sigma \simeq 3.4$--although, this decreases to $\simeq 2.7$ with the more realistic Stage IV FOM of 500) between the $\Lambda$CDM-centered Stage IV growth forecast and the ``Stage IV all'' forecast. This is due to the extreme departures from $\Lambda$CDM growth predictions seen in this model, but even here, one would hope that these departures would be more discriminatory than they are, as even moderate shifts of current \desidesdovcmb constraints would make them consistent. Furthermore, and as mentioned earlier, the extreme departures from $\Lambda$CDM growth are only realized at the very lowest redshifts, while in the redshift window considered here (representative of RSDs in DESI-type surveys) the departures are only moderate. Whether the expansion history data shifts or not, a small shift in dark energy model parameters like $\xi$ can easily satisfy both this type of expansion history preference and $\Lambda$CDM growth data within the relevant uncertainties and find the middle ground between the two $(w_0, w_a)$ projections. Consequently, high quality low-$z$ growth measurements have the potential to provide this discriminatory power.

To further illustrate this, in Figure~\ref{fig:forecast_3pannel} we also show the combination of the intermediate-$z$ Stage IV growth forecast with a low-$z$ \desipv forecast data point at $z\sim0.075$, generated with the same \planck18 fiducial $\Lambda$CDM cosmology, and relative error of 5.3\%, as predicted by \cite{Saulder:2023oqm}. We assume no cross-covariance since the intermediate-$z$ data forecast did not include the BGS survey \cite{Font-Ribera:2013rwa}. We find only a moderate improvement on the constraints for mG and quintessence, whose growth does not significantly deviate from the \lcdm prediction. However, for the case of NM, which predict larger deviations from \lcdm at low redshift,  the impact is significant. This single addition substantially shrinks the $(w_0, w_a)$ contours and creates a $n_\sigma \simeq 4.8 $ tension with respect to the expansion history when the ${\rm FOM} \sim 1000$ ($n_\sigma \simeq 4.0$ with ${\rm FOM} \sim 500$), in contrast to $n_\sigma \simeq 3.4 $ ($n_\sigma \simeq 2.7$) with the Stage IV growth alone. These constraints are not completely driven by the single low-$z$ data point but from the combination of both low- and intermediate-$z$ growth measurements. In all three cases, the \desipv point alone has a similar constraining power as the Stage IV intermediate-$z$ forecast. We have omitted it from Figure~\ref{fig:forecast_3pannel} for clarity. These results provide more concrete hope that future growth data could possibly offer some measure of discriminatory power. If the low-$z$ regime can be properly leveraged (perhaps with multiple well-constrained data points), we may be able to make firm statements concerning the cosmological viability of dark energy models directly coupled to gravity in the future.

The ISW effect may also, conceivably, be used to place constraints on these theories. Indeed, it has been successfully used to place significant constraints on the shift symmetric cubic Galileon model \cite{Renk:2017rzu}. Nevertheless, looking back at the last decade, we have not witnessed significant improvements in using the ISW to constrain cosmology. While there is a forecast of how an ideal measurement of the ISW effect might constrain $\Lambda$ \cite{Hu:2004yd} it is unclear, given the persistent systematic limitations of surveys of structure on the largest scales, whether it will be powerful enough, as a method, to constrain extended scalar field models, such as the NM or mG models.

In this paper we have not discussed the role that projected large scale structure measurements will play in constraining dark energy. In particular, the role of galaxy clustering and weak lensing and their cross-correlations (the so-called 3x2pt measurements) that are the main observational targets of \euclid and \rubin. Although we have used the forecast FOM from \rubin's \srd, which includes these observables, we have not used nor discussed the impact of these kind data. For instance, weak lensing will in principle be a powerful probe, capable of measuring $\Phi_+$ directly (e.g. \cite{Tutusaus:2023aux}), and place constraints on the amplitude and growth rate of structure without confounding properties such as galaxy bias. However, while weak lensing is a powerful probe, it is very sensitive to the non-linear regime of structure growth as well as very dependent on our understanding of baryonic physics and feedback. If we are to use projected data, we need to develop accurate models for the galaxy, matter and gas distributions, beyond the quasi-linear regime. Work on this is ongoing (e.g. \cite{Winther:2014cia,Winther:2017jof, Wright:2022krq,Fiorini:2023fjl,Srinivasan:2023qsu,Atayde:2024tnr,Bose:2024qbw,Davies:2024nlc,Euclid:2025yud,BeltranJimenez:2025yad,Brando:2026pff,Pantiri:2026iaq}), with a range of methods being developed and there is some optimism that some aspects of modelling will be well understood by the end of the Stage IV era (indeed, this is assumed in the forecasts presented in Figure \ref{fig:expansion_forecast}). However, the vast landscape of dark energy spreads the effort too thin, with a main focus on specific models. Given the general approach of our EFT framework, their (moderate) preference by current data, and the gap we have left untested at cosmological non-linear scales, we believe (and we will work on this in upcoming papers) that there is a convincing argument to model such scales for the NM and mG cases and test the compatibility with current and future galaxy clustering and weak lensing data.

\section{Conclusions}\label{sec:conclusions}

In this paper we have laid out the status and prospects of single scalar field dark energy as an explanation for late time accelerated expansion. Unsurprisingly, there is no clear answer to whether we have detected evidence for such a form of dark energy as much of our interpretation of the data, or of the validity of the data, is dictated by our preconceptions. Nevertheless, we have been able to establish what are the possible scenarios which should be considered and how they will play out as the quality and quantity of the data improves. 

We now go through the main take away messages of this paper:
\begin{enumerate}
    \item There is very little evidence that whatever form of energy -- dark energy -- which is responsible for the accelerated expansion of the Universe is {\it not} the cosmological constant, $\Lambda$. While the strongest, credible evidence for it comes from a combination of the CMB, \desi measurements of the BAO, and the original calibration of \desyf measurements of apparent magnitudes of SNe Ia, when using the recent recalibrated \desyf data -- \desdov -- the evidence is greatly reduced and, we emphasize, ``marginal''.
    \item If one's prejudice is firmly in favour of $\Lambda$, it is possible to make such a case. While there are concerns about possible systematics with parts of either the \desi data or the \desyf compilation, one can argue that the \desdov analysis on its own as well as discarding measurements of the low-$\ell$ E mode from the \planck 2018 data set (and thus allowing for a higher value of the optical depth, $\tau$) allow current measurements to be compatible with $\Lambda$. One is, of course, allowing one's prior for $\Lambda$ dictate the choices about which data sets to consider. And, it should be noted that recent constraints on $\tau$, which do not rely on the CMB \cite{2508.21069, Kageura:2026ryq}, favour a low $\tau$ as well (although concerns can also be raised about these data sets)
    \item Dark energy, in the form of a single scalar field, is a natural extension to $\Lambda$. Given the very narrow range of time and length scales accessible to cosmological measurements, one is very limited to what one might say about the fundamental properties of these scalar fields. In particular, at most we will be able to constrain a few, effective, fundamental constants such as $V_0$ (or $\Lambda$), $\beta$, $m^2$, $\xi$, $\alpha$ and $\gamma$. It is, thus, pointless to consider more complex constructions as they will be completely degenerate with this, the simplest, formulation arising from EFT.
    \item An interesting consequence of this perspective is that $V_0$ (or $\Lambda$) is {\it always} present. Indeed, from the analysis we have undertaken, we have always found uncontrovertible evidence for $\Lambda$ which means that, from the EFT perspective, one can view dark energy as ``$\Lambda$ + something dynamical''. Alternatively, one can think of $\Lambda$ as simply the lowest order of a local Taylor expansion of a potential which might have, globally, a ground state which is $0$ (i.e.~{\it no} cosmological constant).
    \item The simplest single scalar field form of dark energy is quintessence, completely characterized in terms of a quadratic potential. Current measurements are only marginally consistent  with quintessence. Furthermore, the data favours $m^2<0$, which means from cosmological data alone, we cannot determine the full form of the potential. 
    \item There is slightly stronger evidence for non-trivial modifications to quintessence either in the form of a non-minimal coupling (NM) or in the form of a massive Galileon (mG), both with a quintessence like potential. These modifications comprehensively cover the EFT of single scalar field dark energy models on cosmological scales. The evidence weakens if one considers selections of the data (as described above) that are more consistent with $\Lambda$.
    \item If one assumes that the scope of the EFT extends all the way down to Solar System scales, then both NM and mG models inevitably lead to fifth forces at late times which, bar any other mechanism, should be detectable and constrainable on astrophysical scales, effectively ruling out these modification to quintessence. That is a substantial extrapolation for the EFT which might not, in practice, be valid. 
    \item Both NM and mG have internal theoretical resources that can generate gravitational screening -- either of the Vainshtein or chameleon form -- which would shield them from being detected and ruled out with astrophysical constraints. Nevertheless, in both cases, there are severe theoretical obstructions to implementing these screening mechanisms. Although there may be possibilities of circumventing these obstructions, it is still an open question of whether screening is viable or if it involves extra, added corrections to enable them.
    \item To circumvent the issues of whether there is a viable screening mechanism or the cosmological-scales EFT validity at astrophysical scales, one should focus on larger scales and, in particular on the effect on the growth rate of structure. There one clearly sees that the largest deviations from $\Lambda$CDM occur at very late times. The current, tightest, constraint on the density-weighted growth rate, $f\sigma_8$ at $z\sim 0.02$ already leads to a tension between what the growth rate constraints imply for $w_0$ and $w_a$ and what one obtains from  constraints from the expansion rate.
    \item A corollary of this, is that one should not expect current endeavours to constrain $f\sigma_8$ on intermediate to high redshift ranges to deliver a killer blow to these theories -- the uncertainties will remain too large in a regime where the deviations are small. Quite clearly the focus should be on improving constraints at very low redshifts where, unfortunately, a host of complications -- cosmic variance, non-linear growth, baryonic effects -- can wreak havoc.
    \item The ISW effect has been used to rule out the simplest shift symmetric models -- cubic Galileons. The key effect is that the time derivative of the gravitational potential, $\Phi_+'$ has the opposite sign to what one finds in a $\Lambda$ dominated cosmology. In both NM and mG, one finds that, indeed, $\Phi_+'$ is pulled away from the $\Lambda$ prediction but not as much as for the cubic Galileon. Thus measurements of the ISW effect are not yet sufficient to rule out these possibilities. 
    \item While we have focused on a universally coupled single scalar field, it has been argued that a scalar field coupled solely to the dark matter will have the right behaviour to explain current data. Such a theory naturally evades astrophysical constraints but will have a similar phenomenology to that of the NM model in the growth rate, albeit with a slightly suppressed effect. Thus it should be amenable to the same tests discussed above.
    \item We have a good idea of how the quality and quantity of data will improve over the next decade. In rough terms, we expect our constraints on $w_0$ and $w_a$ to improve by a factor of a few but not more. If future constraints converge on the same location as they are today, future data will be able to rule the cosmological constant out as source of dark energy, with more than $8\sigma$ significance level. However, if dark energy is dynamical, we should not expect a step change in our understanding of its microphysical properties.
    \item From the point of view of scalar field dark energy, there are a few possible outcomes. Contours may drift back to $\Lambda$CDM although quintessence and its extensions remain a possibility.  Contours may drift upwards in the ($w_0$,$w_a$) plane, favouring quintessence and its extensions but increasing the evidence against $\Lambda$CDM. Finally,  constraints remain resolutely where they are or drift downwards in the ($w_0$,$w_a$). These will make $\Lambda$CDM and quintessence exceedingly unlikely and will point towards extensions such as the NM and mG case. If, in addition, constraints in $f\sigma_8$ can be improved and are shown to be inconsistent with both NM and mG, one then has to consider the possibility that single scalar field dark energy is not a viable explanation for the accelerated expansion of the Universe.

\end{enumerate}

These are remarkable times in which cosmological data is opening new vistas on the Universe. One would have hoped that, by now, we would be in a position in which we could make definitive statements about one of the main open questions in cosmology -- ``what is dark energy?''. Unfortunately that is not yet the case.  While we have had to roll back our ambitions of coming up with a data driven microphysical answer, in this paper we have attempted to make clear statements about one of the main hypothesis -- single scalar field dark energy -- and map out possible different outcomes for how this question may be answered. Even in the face of this persisting underdetermination, it is still worth reminding ourselves of the lofty goal we have undertaken: to learn about particle physics with galaxies and the distribution of matter and light in the Universe; i.e. to learn about the microphysics of the Universe from its largest scales. It is then not a surprise that we are limited, and we should be amazed that we can even learn a few properties under a very sensible assumption -- that dark energy is given by a single scalar field -- such as its mass, its possible coupling to Gravity and/or whether its kinetic sector is canonical or not. We look forward to how new data will further refine these answers.

\begin{acknowledgments}
We thank David Alonso, Sofia Chiarenza, Harry Desmond, Kazuya Koyama, Johannes Noller, Eusebio S\'anchez, Nacho Sevilla, Richard Stiskalek and Tariq Yasin for many valuable conversations. We thank Alex Krolewski, Alexander Reeves and Benjamin Stölzner for sharing their ISW data and useful insights. CGG was supported by the Beecroft Trust and the C\'esar Nombela  Research Talent Attraction grant from the Community of Madrid  (Ref. 2025-T1/TEC-36302). PGF is supported by STFC and the Beecroft Trust. For the purposes of open access, the authors have applied a Creative Commons Attribution (CC BY) license to any Author Accepted Manuscript version arising.

{\it Software}:  We made extensive use of {\tt hi\_class} \cite{hi_class_1,hi_class_2,Blas:2011rf}, {\tt Cobaya} \cite{Torrado:2020dgo} and the {\tt numpy} \cite{oliphant2006guide, van2011numpy}, {\tt scipy} \cite{2020SciPy-NMeth}, and {\tt matplotlib} \cite{Hunter:2007} python packages.

\end{acknowledgments}

\appendix

\section{Dark energy models priors and posteriors}\label{sec:appendixA}
As Bayesian evidence depends on the choice of priors, we report the priors on the cosmological and dark energy model parameters used in the computation in Table.~\ref{tab:priors}. We also note that for the mG model, due to the shape of the $m^2$ posteriors, we find it more efficient to sample in the variable $A\equiv \phi_i |m^2|^{1.41}$ \cite{Wolf:2025acj}.

\begin{figure}[t]
   \centering
    \includegraphics[width=\columnwidth]{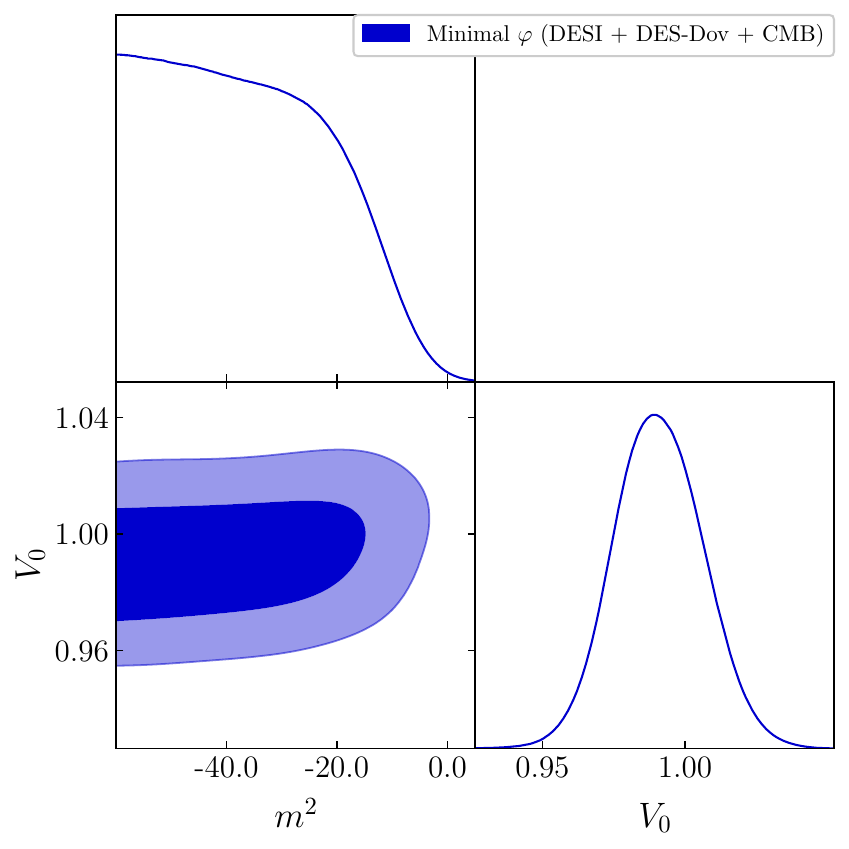}
   \vskip -0.1in
   \caption{Constraints (68\% and 95\% C.L.) on thawing quintessence parameters using our baseline dataset combination.}
   \label{fig:quint_fullpost}
\end{figure}

\begin{figure}[t]
   \centering
    \includegraphics[width=\columnwidth]{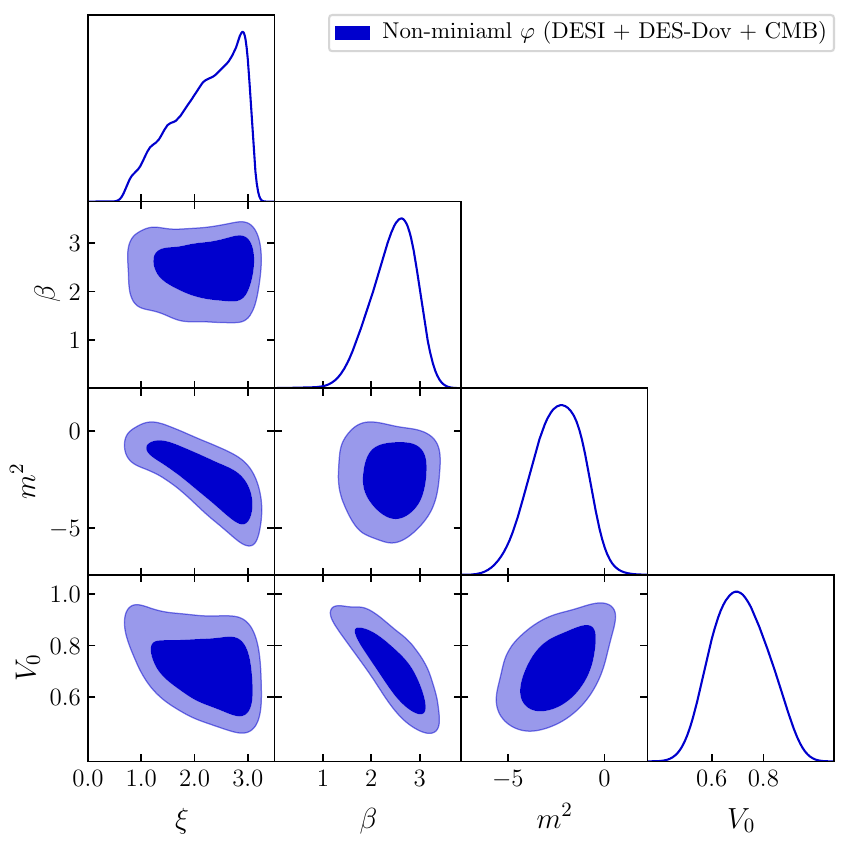}
   \vskip -0.1in
   \caption{Constraints (68\% and 95\% C.L.) on the non-minimally coupled scalar field parameters using our baseline dataset combination.}
   \label{fig:nm_fullpost}
\end{figure}

\begin{figure}[t]
   \centering
    \includegraphics[width=\columnwidth]{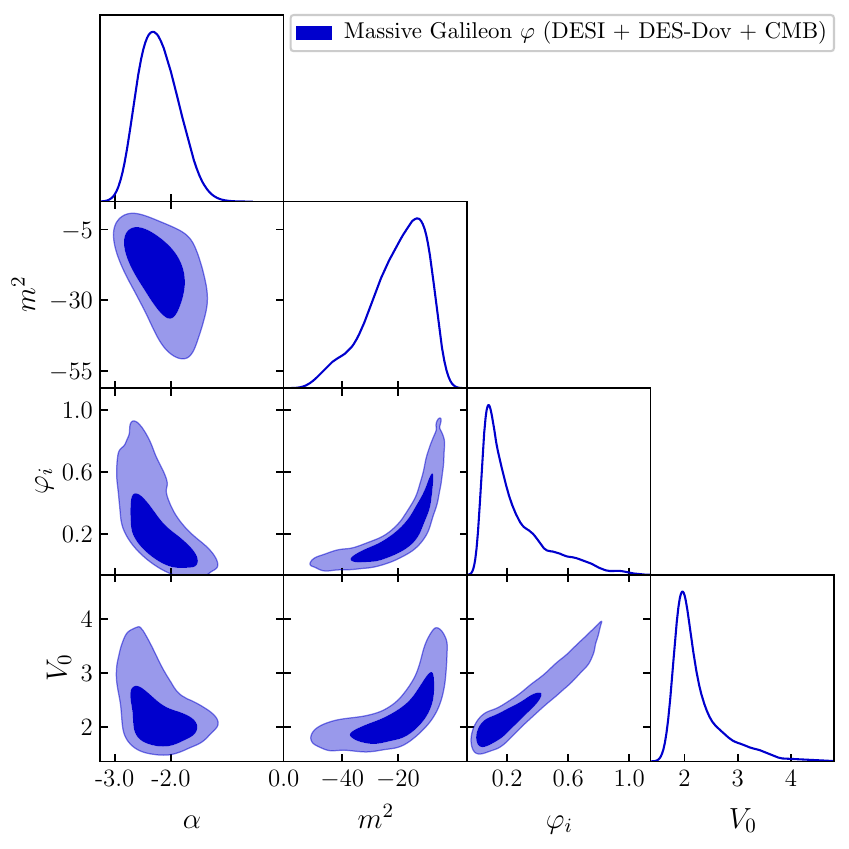}
   \vskip -0.1in
   \caption{Constraints (68\% and 95\% C.L.) on the massive Galileon parameters using our baseline dataset combination.}
   \label{fig:mG_fullpost}
\end{figure}

We also provide the posterior distributions for the various models considered in this work in Figures~\ref{fig:quint_fullpost}, \ref{fig:nm_fullpost} and \ref{fig:mG_fullpost} for quintessence, NM and mG, respectively. In particular, it is worth noting that across the board the cosmological data displays a preference for negative cosmological masses, which can lead to more rapid evolution in the scalar field and dark energy equation of state \cite{Wolf:2023uno, Dutta:2008qn, Shlivko:2024llw, Wolf:2024eph}. The quintessence mass is essentially unconstrained in this direction because the field can be tuned to rest on top of its potential for much of cosmic history, before only beginning to roll quite recently. Consequently, much of the scalar field evolution occurs at very recent redshifts, outside the regimes where the cosmological data has most of its constraining power. However, the presence of the terms seen in the NM and mG models changes the field dynamics and allows for tight constraints on the cosmological masses.

\begin{table}[t]
\centering
\begin{tabular}{lc}
\hline
Parameter & Prior \\
\hline

\multicolumn{2}{l}{\textbf{CDM parameters}} \\
\hline
$\Omega_{\rm m}$ & $[0.25,\ 0.40]$ \\
$\omega_{\rm b}$ & $[0.015,\ 0.030]$ \\
$h$ & $[0.6,\ 0.8]$ \\
$n_s$ & $[0.85,\ 1.10]$ \\
$\tau_{\rm reio}$ & $[0.01,\ 0.15]$ \\
$\ln(10^{10}A_s)$ & $[2.9,\ 3.2]$ \\

\hline
\multicolumn{2}{l}{\textbf{CPL model (DESI Priors)}} \\
\hline
$w_0$ & [-3.0, 1.0] \\
$w_a$ & [-3.0, 2.0] \\

\hline
\multicolumn{2}{l}{\textbf{CPL model (Restricted Priors)}} \\
\hline
$w_0$ & [-1.5, 0.0] \\
$w_a$ & [-2.0, 0.5] \\

\hline
\multicolumn{2}{l}{\textbf{Minimal $\varphi$}} \\
\hline
$m^2$ & [-100, 20] \\
$V_0$ & [-0.5, 1.5] \\

\hline
\multicolumn{2}{l}{\textbf{Non-minimal $\varphi$}} \\
\hline
$\xi$ & [0.0, 10.0] \\
$\beta$ & [0.0, 10.0] \\
$m^2$ & [-10.0, 10.0] \\

\hline
\multicolumn{2}{l}{\textbf{Massive Galileon $\varphi$}} \\
\hline
$\alpha$ & [-10.0, 2.5] \\
$m^2$ & [-80.0, 0.0] \\
$A \equiv \phi_i |m^2|^{1.41}$ & [0.0, 10.0] \\

\hline
\end{tabular}
\caption{Parameter priors for $\Lambda$CDM and extended dark energy  models.}
\label{tab:priors}
\end{table}

\section{Violating the Weak Equivalence Principle: Interacting dark matter--dark energy models}\label{sec:appendixB}

The focus in this paper is on single scalar field dark energy which is universally coupled to the rest of the world. By this, we mean that it couples with the same strength to all the other constituents of the Universe, via gravity, i.e. it satisfies the Weak Equivalence principle. This can be made explicit in the case of the non-minimally coupled theories by transforming to the Einstein frame -- there the coupling of the scalar field to the matter components is through a conformal factor multiplying the Einstein frame metric and is the same for all species. As a result, baryon and dark matter (as well as everything else) is affected in the same way.

An interesting alternative can also be considered: theories in which the dark energy (in the form of a single scalar field) couples only to the dark matter. In this case one can automatically avoid laboratory and astrophysical fifth force constraints as these arise from observation of test bodies made of baryons.  While this is not the focus of this paper, we comment on these proposals for completeness.

Theories of interacting dark matter and dark energy have been studied for a number of decades. One can frame these theories at a level in which one does not have to specify the actual form of the dark energy but simply assume that energy momentum tensors of the dark energy, $T^{\rm DE}_{\mu\nu}$, and dark matter, $T^{\rm M}_{\mu\nu}$ are intertwined through the conservation equations in the form
\begin{eqnarray}
\nabla^\mu T^{\rm DE}_{\mu\nu} &=&Q \nonumber \\
\nabla^\mu T^{\rm M}_{\mu\nu} &=&-Q
\end{eqnarray}
where $Q$ then dictates the particular nature of the interaction \cite{Amendola:1999er,Chimento:2003iea}. There is a wide range of phenomenological forms for $Q$ which have been studied in detail, and a more complete perspective of these theories was developed in \cite{Pourtsidou:2013nha}. More recently, such approaches have been constrained with current data \cite{Figueruelo:2026eis, deCruzPerez:2025dni, Silva:2025hxw}.

We are interested here in a more microphysical view of theories as advocated in \cite{Farrar:2003uw} and further developed in \cite{Boehmer:2008av, Das:2005yj}. The simplest idea is that the masses of the dark matter particles are dependent on the dark energy scalar field, $m(\varphi)$, which in practice means that the dark matter becomes non-minimally coupled, unlike the baryons or the rest of the Universe. More elaborate scenarios can be constructed in which a rich dark matter section (including ``dark radiation'') will interact directly with the dark energy \cite{Berghaus:2023ypi}.

It should be clear that such theories can be seen as a deformed version of the non-minimal theory described in this paper as emerging in the EFT construction of dark energy -- assuming now, in the Einstein frame, that different constituents will have different coupling strengths to the dark energy scalar field. And thus, while it may lead to specific predictions which, in detail differ from those of the NM model, qualitatively they should be similar. At the background level, such theories are already known to produce similar phenomenology for the evolution of the equation of state (or equivalently the dark energy density) \cite{Khoury:2025txd, Chakraborty:2025syu, Gomez-Valent:2026ept}. Furthermore, the fifth force -- the dark energy --  will affect the gravitational collapse of the dark matter. If one notes that the growth rate of structure on large scales is mostly driven by that of dark matter, then we expect to see similar effects on the time evolution of $f\sigma_8$ and on the ISW that we see in the NM model.

A more detailed analysis of this WEP violating single scalar field dark energy, using the tools of EFT advocated here, is required if we are to make more quantitative statements about our ability to constraint it and to distinguish it from the class of models being considered here.

\begin{figure*}[t]
   \centering
   \includegraphics[width=0.48\textwidth]{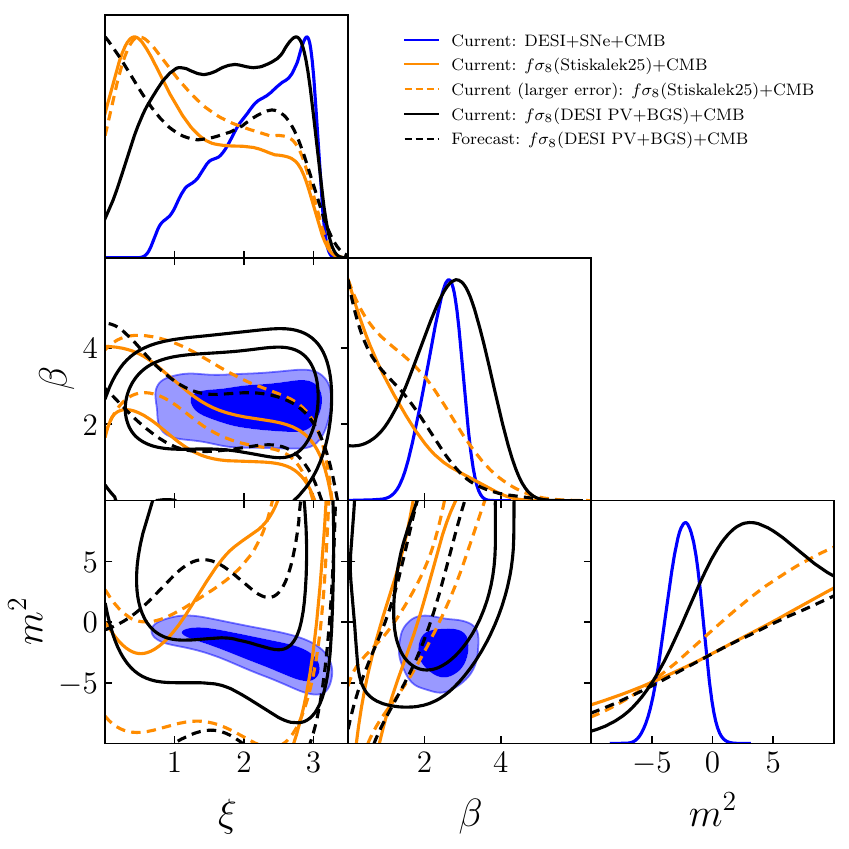}
   \hfill
   \includegraphics[width=0.48\textwidth]{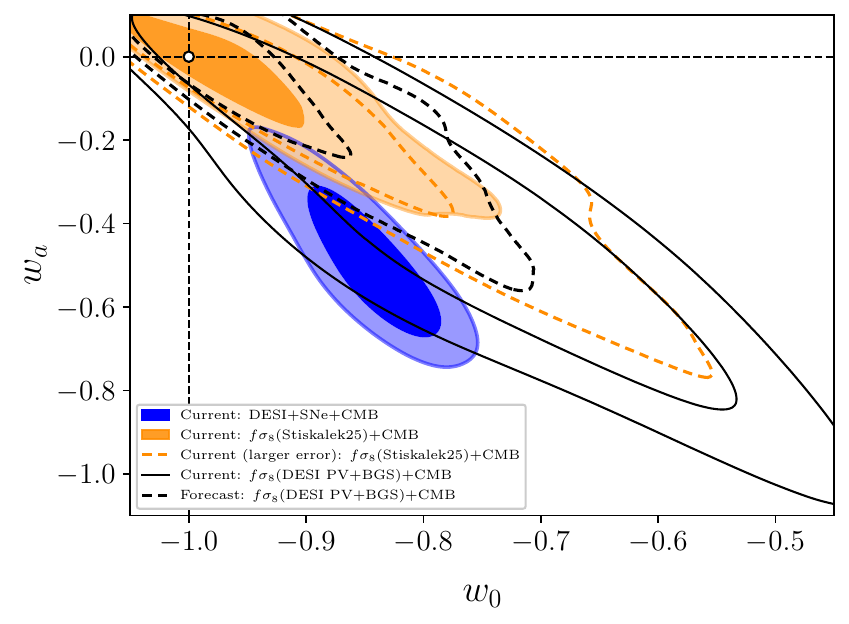}
   \vskip -0.1in
   \caption{Marginalized posterior distributions (68\% and 95\% C.L.) on the non-minimally coupled dark energy parameters (left) and its projection into the $(w_0, w_a)$ plane (right) for: the current \desi PV data point ($\simeq 12\%$ error, black solid), the forecast \desipv data point centered at the fiducial $\Lambda$CDM prediction ($\simeq 5.3\%$ error, black dotted), the \stiskalek PV data point ($\simeq 4.5\%$ error, red solid), the \stiskalek PV data point with a larger error ($\simeq 6.5\%$, red dotted), and finally the results from the current expansion history for reference (in blue solid).}
   \label{fig:growth_appendix}
\end{figure*}

\section{Further Analysis on Cosmological Growth Measurements}\label{sec:appendix_growth}

The analysis in this appendix has two purposes: (1) to present the constraints using \desipv measurement and forecast its constraining power at the end of the survey program using the expected $\simeq 5.3\%$-error measurement at $0<z<0.15$ (see Table 5 in \cite{Saulder:2023oqm}), and (2) to assess the impact of the \stiskalek more conservative error budget, as motivated by \cite{Hollinger:2023mrp}, which translates into a change on the $f\sigma_8$ relative error from $\simeq 4.5\%$ to $6.5\%$, assuming the same relative change as predicted for in $S_8$ in \cite{Stiskalek:2025oht}. To do so, we focus on the non-minimally coupled model, as the most departing case from $\Lambda$CDM growth. As in the main text, we carry out the inference on the model parameter space and then project the resulting posteriors into the $(w_0, w_a)$ plane, following Section~\ref{sec:prior_procedure}. Both parameter and CPL constraints can be found in Figure \ref{fig:growth_appendix} for the following scenarios: the current \desi PV data point ($\simeq 12\%$ error, black solid), the forecast \desipv data point centered at the fiducial $\Lambda$CDM prediction ($\simeq 5.3\%$ error, black dotted), the \stiskalek PV data point ($\simeq 4.5\%$ error, red solid), the \stiskalek PV data point with a larger error ($\simeq 6.5\%$, red dotted), and finally the results from the current \desidesdovcmb expansion history for reference (in blue solid).

Both the parameter constraints and $(w_0, w_a)$ projections tell a consistent story. The current \desi PV measurement, despite having a mean value much more favorable to enhanced growth, has a large uncertainty and provides loose constraints, both at the scalar field parameter level as for $(w_0, w_a)$. In contrast, the other more precise $f\sigma_8$ scenarios clearly pull the NM model back towards an effective $\Lambda$/quintessence in comparison to the expansion history data. This justifies our focus on \stiskalek to illustrate the potential discriminative power of low-$z$ growth measurements. This is still true in the conservative scenario where we used the larger error for \stiskalek, which still provides tighter constraints than \desi PV.  Finally, assuming that the future \desipv measurement remains centered at \lcdm growth, and noting that the precision will be similar to \stiskalek's, we can expect similar results to \stiskalek's in terms of the constraining power and pulling of the parameters towards an effective $\Lambda$/quintessence. As a consequence, we conclude \stiskalek is a good representative of both current and future low-$z$ measurements.

\section{Integrated Sachs-Wolfe  -- Additional Information} \label{app:ISW-extra}
\subsection{Extra $C_\ell^{gT}$ plots}\label{app:ISW-gT}
\begin{figure*}
    \centering
    \includegraphics[width=0.9\linewidth]{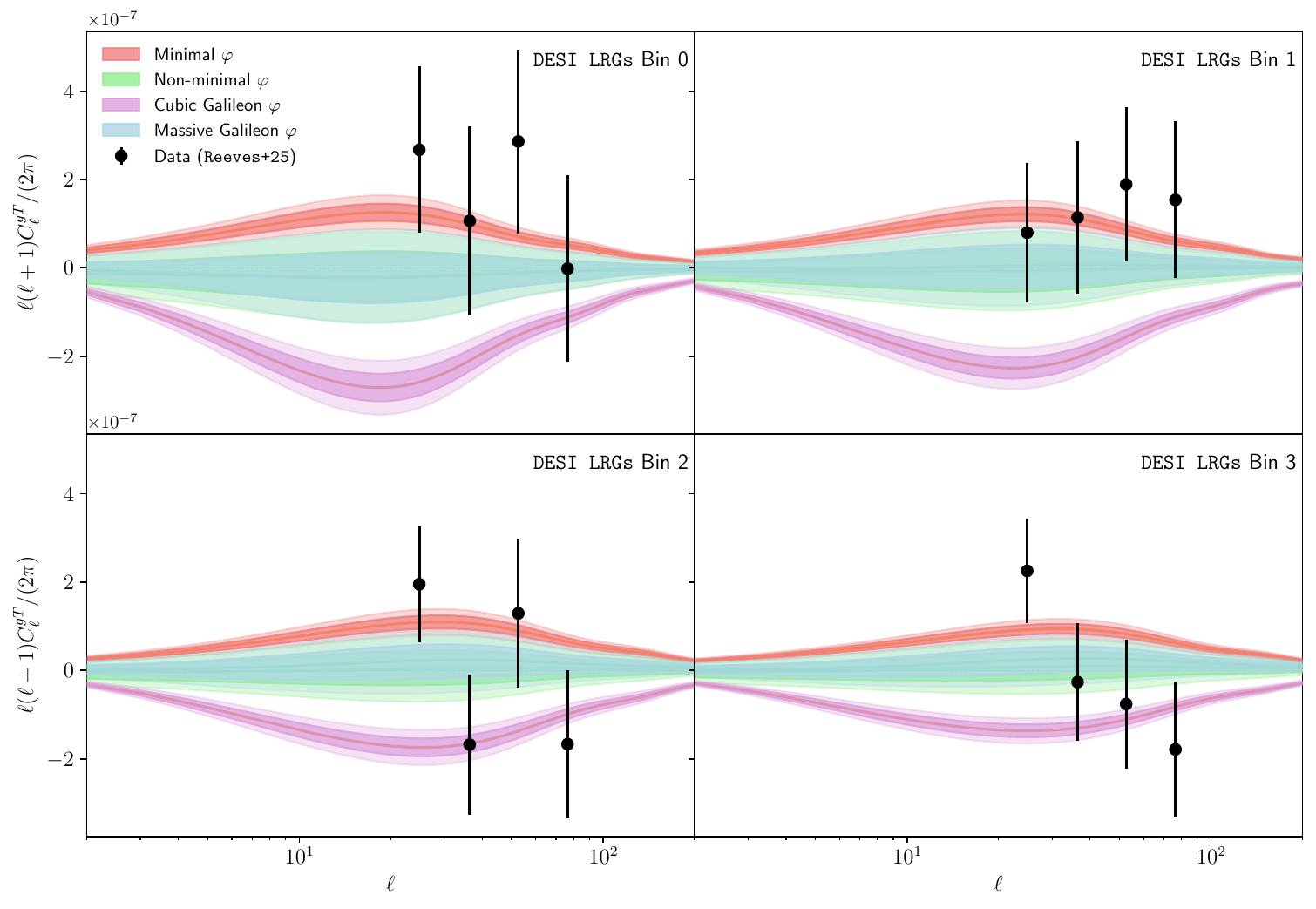} 
    \caption{Same as Figure \ref{fig:Krolewski} but for Reeves+25 \cite{Reeves:2025xau}. In this case, the $A_{\rm ISW}$ and $b_{\rm g}$ have been fitted per bin. The plot remains the same for the joint fit.}
    \label{fig:Reeves_individual}
\end{figure*}

\begin{figure*}
    \centering
    \includegraphics[width=\linewidth]{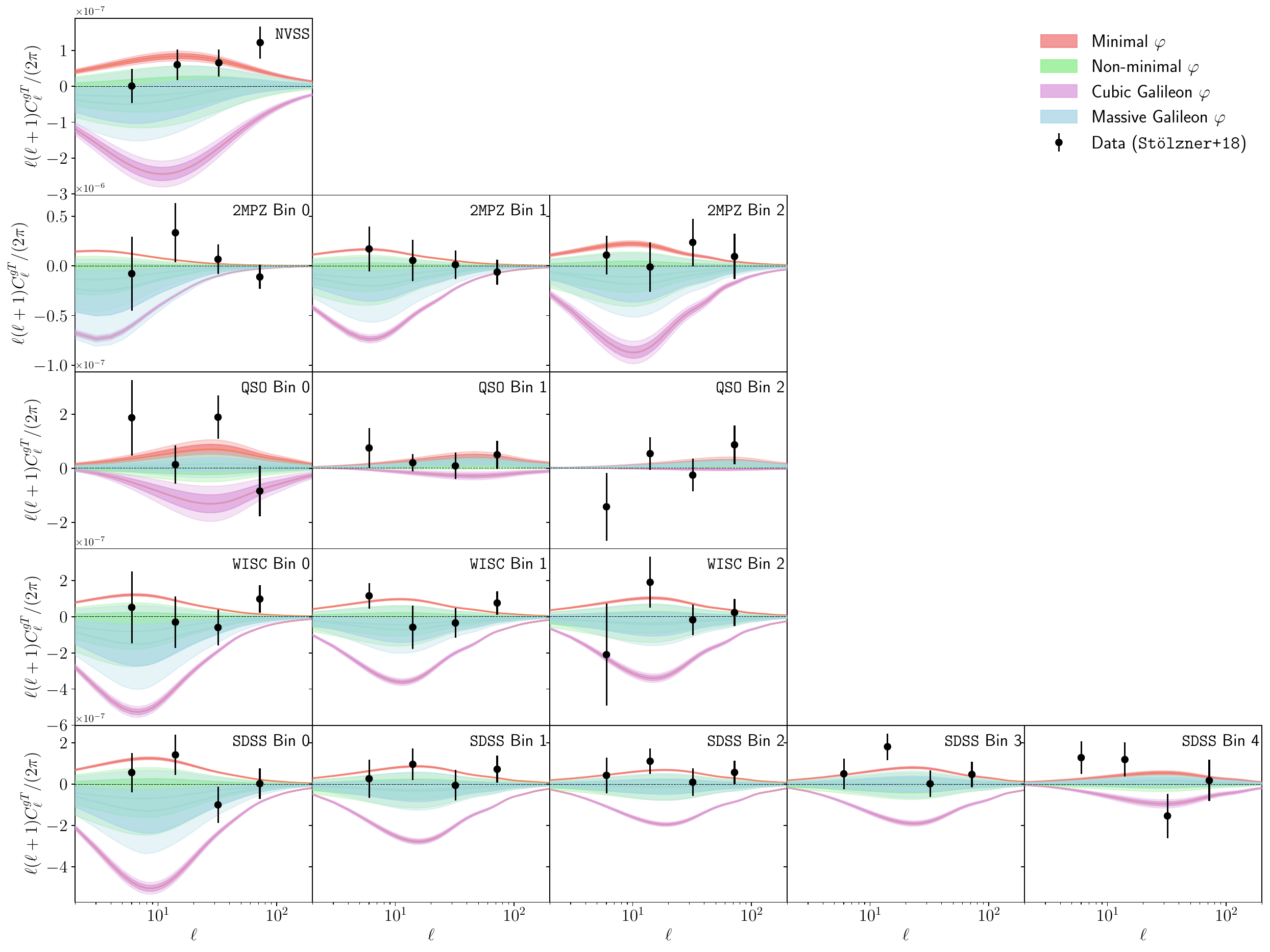}        
    \caption{Same as Figure \ref{fig:Krolewski} but for \stolzner \cite{Stolzner:2017ged}.}
    \label{fig:Stolzner}
\end{figure*}

In this section we present the ISW angular power spectra for the galaxy and CMB temperature cross-correlation for \reeves (Figure~\ref{fig:Reeves_individual}) and \stolzner (Figure~\ref{fig:Stolzner}).

\subsection{Data sets}\label{sec:ISW-data}
We now describe the three data sets considered for the ISW analysis in more detail. The main take-away is that the three data sets have undergone different data analysis choices, that we are missing the cross-covariance between most of the redshift bins and surveys, preventing us from doing a joint analysis, and that the underlying approximations in some of them may not be good enough for precision cosmology, although they will most likely not change the qualitative picture given the low signal-to-noise ratio of the sample. 

{\it Krolewski+22} \cite{Krolewski:2021znk}: 
This data set uses the cross correlation of \planck 2018 temperature fluctuations \cite{Planck:2018yye} with the full sky \unwise galaxy sample \cite{2019ApJS..240...30S}, probing the range of redshifts $z\in [0, 2]$ in three different tomographic bins that peak at $z\sim 0.5, 1.1$ and $1.5$. In addition, they use the cross-correlation with \planck 2018 CMB lensing \cite{Planck:2018lbu} to calibrate the linear galaxy biases. Given the broad tomographic bins they need to account for their redshift evolution, which they do by calibrating the galaxy redshift distribution cross-correlating the \unwise samples with the spectroscopic galaxies of \sdss, split in thin $z$-bins \cite{Krolewski:2019yrv}. This method directly provides an estimate for $b_g(z) dN/dz$, leaving the overall effective galaxy bias factor to be marginalized over. However, the \unwise sample has a non-negligible magnification contribution, which is independent of the galaxy bias and requires a calibration of the $dN/dz$. This is obtained cross-matching the galaxy sample with {\tt COSMOS} \cite{Laigle:2013tsa}. Following \cite{Krolewski:2021znk}, we use each estimate when computing the theory vector, computing the clustering and lensing parts separately with \texttt{hi\_class}, multiplying the effective $b_{\rm g}$ (and $A_{\rm ISW}$), and then adding them together. For the magnification parameter, we use the value in Table 1 of \cite{Krolewski:2021znk}. When fitting $b_{\rm g}$ and $A_{\rm ISW}$, we bin the angular power spectra in the following way: for $C_\ell^{gT}$, we use the bandpower window function, which accounts for the survey geometry; however, for the $C_\ell^{g\kappa}$, we evaluate it at the effective bandpower. This is because although we have the window functions for the former, we do not have it for the latter. In addition, since we do not have the cross-covariance between $C_\ell^{gT}$ and $C_\ell^{g\kappa}$, the fit assumes them uncorrelated. Although this is not valid for a precision cosmology analysis, for our qualitative study, it is good enough: first, as we have checked in \reeves, the correlation between $C_\ell^{gT}$ and $C_\ell^{g\kappa}$, is non-negligible but small; moreover, we use $C_\ell^{g\kappa}$ only to calibrate the galaxy bias but the $A_{\rm ISW}$ uncertainties are dominated by the low signal-to-noise of $C_\ell^{gT}$. Finally, we also lack the cross-covariance between the different tomographic bins, impeding us to do a joint fit. 

{\it St\"olzner+18} \cite{Stolzner:2017ged}: This data set is the largest compilation of ISW cross-correlations to date and can be found in {\tt MontePython}'s repository \footnote{Available at: \url{https://github.com/brinckmann/montepython_public/tree/3.6/data/ISW}} \cite{Brinckmann:2018cvx, Audren:2012wb}. It contains cross-correlations of \planck 2015's CMB temperature fluctuations with \cite{Planck:2015mis} with \nvss \cite{Condon:1998iy, DeZotti:2009an},  \tmpz \cite{Bilicki:2013sza, Alonso:2014xca},  \sdss DR12 photo-$z$ sample from \cite{2016MNRAS.460.1371B},  the \wisc galaxy sample \cite{Bilicki:2016irk} and the \sdss DR6 photometric QSOs from \cite{Richards:2008eq, 2009JCAP...09..003X}. The range of redshifts covered is $z\in[0, 5]$. These samples were subdivided in different number of tomographic bins, which resulted in 15 angular cross-power spectra $C_\ell^{gT}$ and 15 auto-power spectra $C_\ell^{gg}$, which were used to calibrate the galaxy bias. In the building of the samples, \stolzner paid particular attention to reducing the cross-correlation between different samples by masking overlapping areas and using tomographic bins with little or no overlap. In particular, they masked the \sdss area from the {\tt 2MASS} and \wisc, and used the {\tt 2MASS} catalog only at low redshifts. In addition, although {\tt 2MASS} and \wisc cover the same area, \wisc was built so that it excludes galaxies in {\tt 2MASS}. This leaves a remaining correlation due to both samples tracing the same underlying matter field that they minimized by not using the lowest redshift \wisc bin.  Another two approximations were made. First, the galaxy redshift distribution for each subsample is not recalibrated but assumes a tophat selection function on the range of redshifts covered. They explored the impact of this approximation by convolving their distributions with the photometric redshift error of each survey and found that the overall effect is negligible for the full sample and at most $\sim 10\%$ for the highest redshift bin galaxy bias (and corresponding $A_{\rm ISW}$), which is good enough for our qualitative assessment here. Second, the authors neglected the cross-covariance between $C_\ell^{gT}$ and $C_\ell^{gg}$, when doing the joint fit. For the same arguments as in the case of \krolewski, this approximation is likely to be good enough for our goals in this paper. However, in contrast to \cite{Stolzner:2017ged}, and given we do not have the cross-covariance between different tomographic bins and samples, nor a straightforward way of testing the impact of the approximations made in the data analysis, we decide not to do a joint analysis of all probes. We worry that with the increased sensitivity we might introduce some bias or artificially increase the precision of the $A_{\rm ISW}$ measurement, yielding a misleading conclusion. In addition, we would not be able, anyway, to combine them with the \krolewski and \reeves data sets. Finally, in contrast to the other two data sets \stolzner used the \texttt{PolSpice} estimator, instead of {\tt NaMaster} \cite{Hivon:2001jp, Alonso:2018jzx}, correcting for the mask effects in the configuration space, before transforming into harmonic space. In this case, \stolzner angular power spectra are given at all $\ell$, and then binned with equal weights to remove the mode coupling. We use the same binning scheme as in {\tt MontePython}'s \texttt{Likelihood\_isw} class\footnote{\url{https://github.com/brinckmann/montepython_public/blob/3.6/montepython/likelihood_class.py\#L2971}}.

{\it Reeves+25} \cite{Reeves:2025xau}: This data set  comprises the cross-correlations between \planck 2018 temperature \cite{Planck:2018yye} and the \desi Legacy Survey DR9 LRGs. They are the newest ISW measurement available, covering $z\in[0.2, 1.4]$. Although they have not been optimized for ISW analysis, they provide the cross-covariance between the different tomographic bins and angular power spectra, letting us do a joint fit. In particular, we have the $C_\ell^{gT}$ and the \act DR6 CMB convergence \cite{ACT:2023dou} cross-correlation with the galaxy density $C_\ell^{g\kappa}$ to calibrate the galaxy bias. For the CMB convergence, we account for the Monte-Carlo correction \cite{ACT:2023oei} to capture the noise properties of the CMB convergence map due to the reduced sky coverage of the overlap with \desi LRGs. Finally, as in the case of \unwise, we also need to account for the magnification effect. We use a fixed value for each bin, taken from the prior distribution means in Table 1 in \cite{Reeves:2025xau}. 

When fitting the $A_{\rm ISW}$ and $b_{\rm g}$ parameters we use the scale cuts from the original analyses. For \krolewski $\ell < 100$ and $\ell < 300$ for the $C_\ell^{gT}$ and $C_\ell^{g\kappa}$ terms, respectively. The latter correspond to the "conservative" scale cuts in \cite{Krolewski:2021znk}. For \stolzner, we used the same cuts as in {\tt MontePython}'s likelihood; i.e. for $C_\ell^{gT}$ we use $4 \leq \ell \leq 99$ for all samples, except for \nvss, for which it extends to $\ell_{\rm max}=100$, and for $C_\ell^{gg}$, the scale cuts depend on the sample, so that we use $\ell_{\rm min} = 10$ for all samples and $\ell_{\rm max} = 60,\, 100,\, 70,\, 70,\, 50$ for \tmpz, \nvss, QSO, \sdss and \wisc, respectively. Finally, for \reeves, we use $20\leq \ell \leq 90$ for $C_\ell^{gT}$ and $40 \leq \ell \leq 185,\, 215,\, 235,\, 245$ for each galaxy tomographic bin of $C_\ell^{g\kappa}$, respectively.

\begin{figure*}
    \centering
\includegraphics[width=0.9\linewidth]{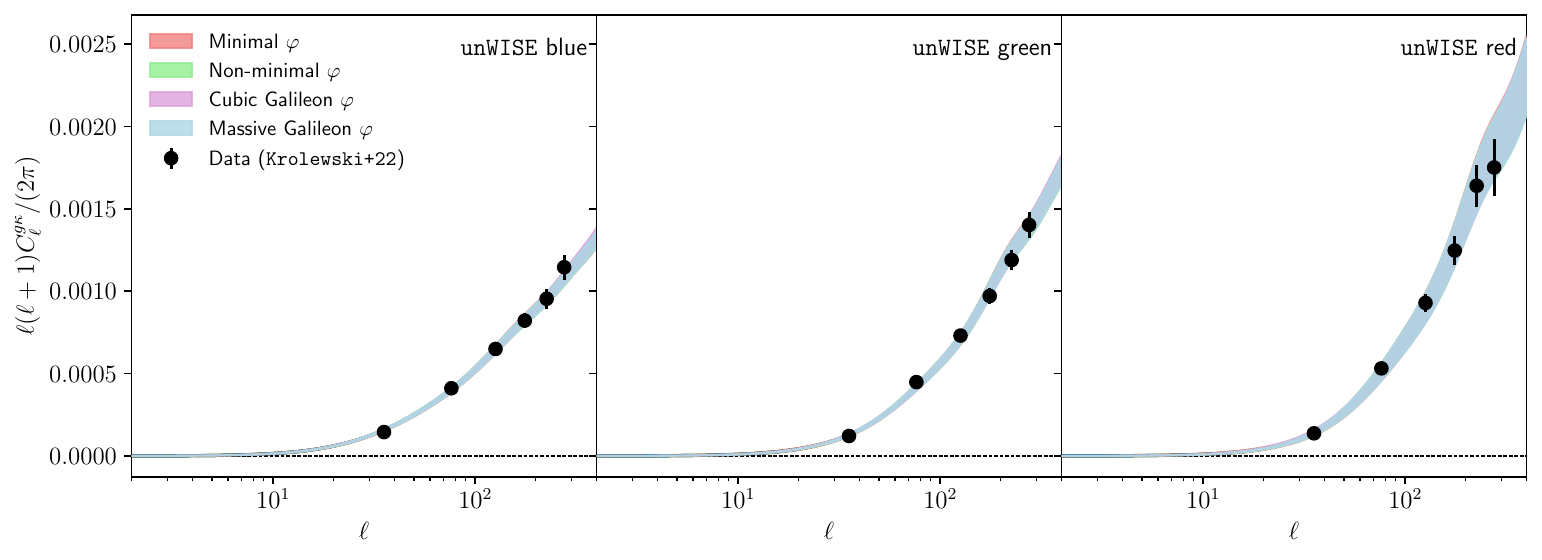}
    \caption{Angular power spectra $C_\ell^{g\kappa}$ used to calibrate the galaxy bias and fitted simultaneously with $C_\ell^{gT}$, shown in Figure~\ref{fig:Krolewski}. The color bands are the posterior distributions obtained as described in Section~\ref{sec:isw}, and includes the galaxy bias factor. As can be seen in Table~\ref{tab:bg_ISW}, the $b_{\rm g}$ uncertainties are similar for all scalar field cases, yielding posteriors that can be hardly distinguished.}
    \label{fig:Krolewski_bg}
\end{figure*}

\begin{figure*}
    \centering
    \includegraphics[width=0.9\linewidth]{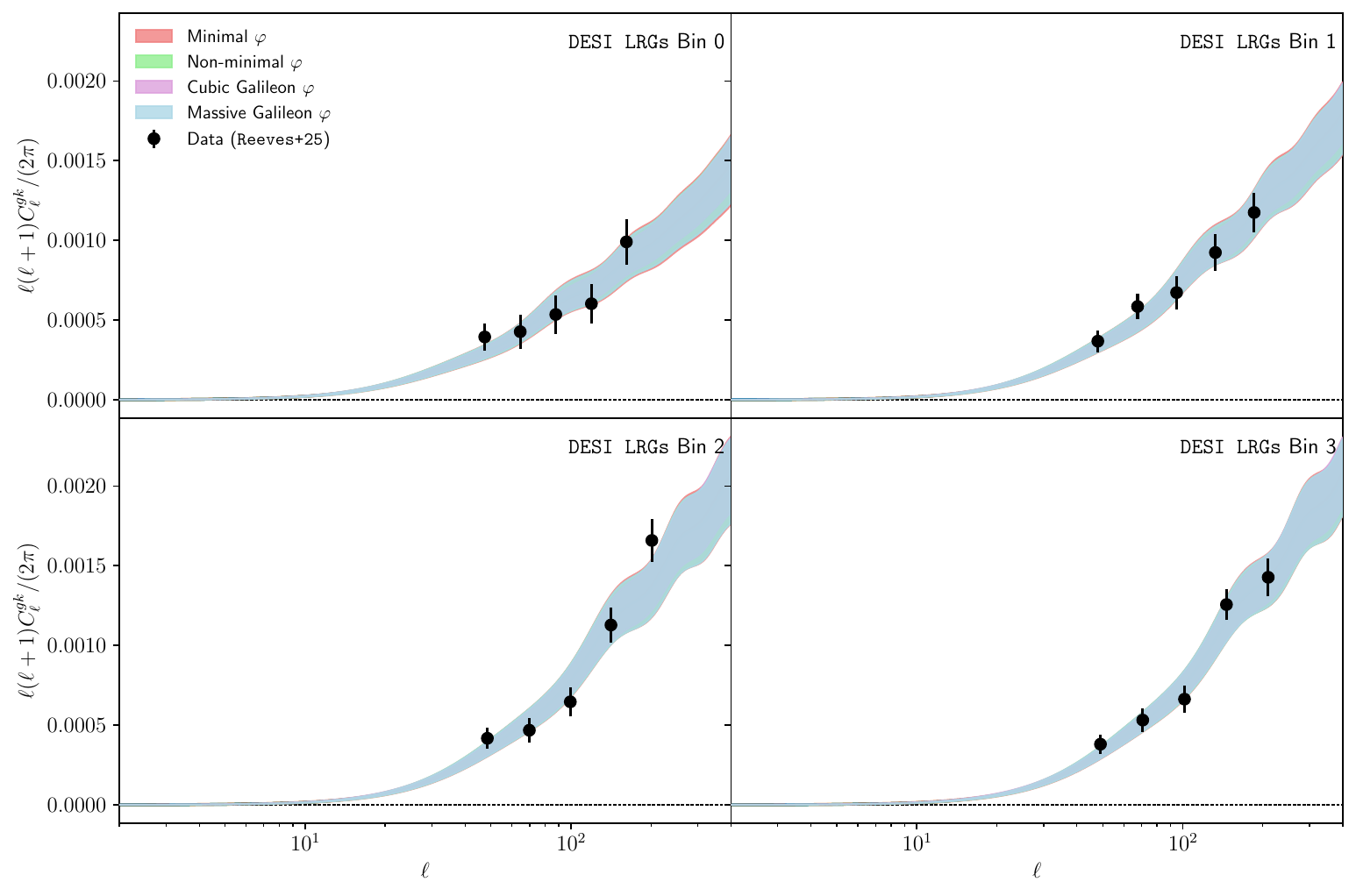}
    \caption{Same as Figure~\ref{fig:Krolewski_bg} but for \reeves. In this case, each tomographic bin has been fitted individually. The case when all bins are fitted simultaneously is virtually indistinguishable.}
    \label{fig:Reeves_individual_bg}
\end{figure*}

\begin{figure*}
    \centering
    \includegraphics[width=\linewidth]{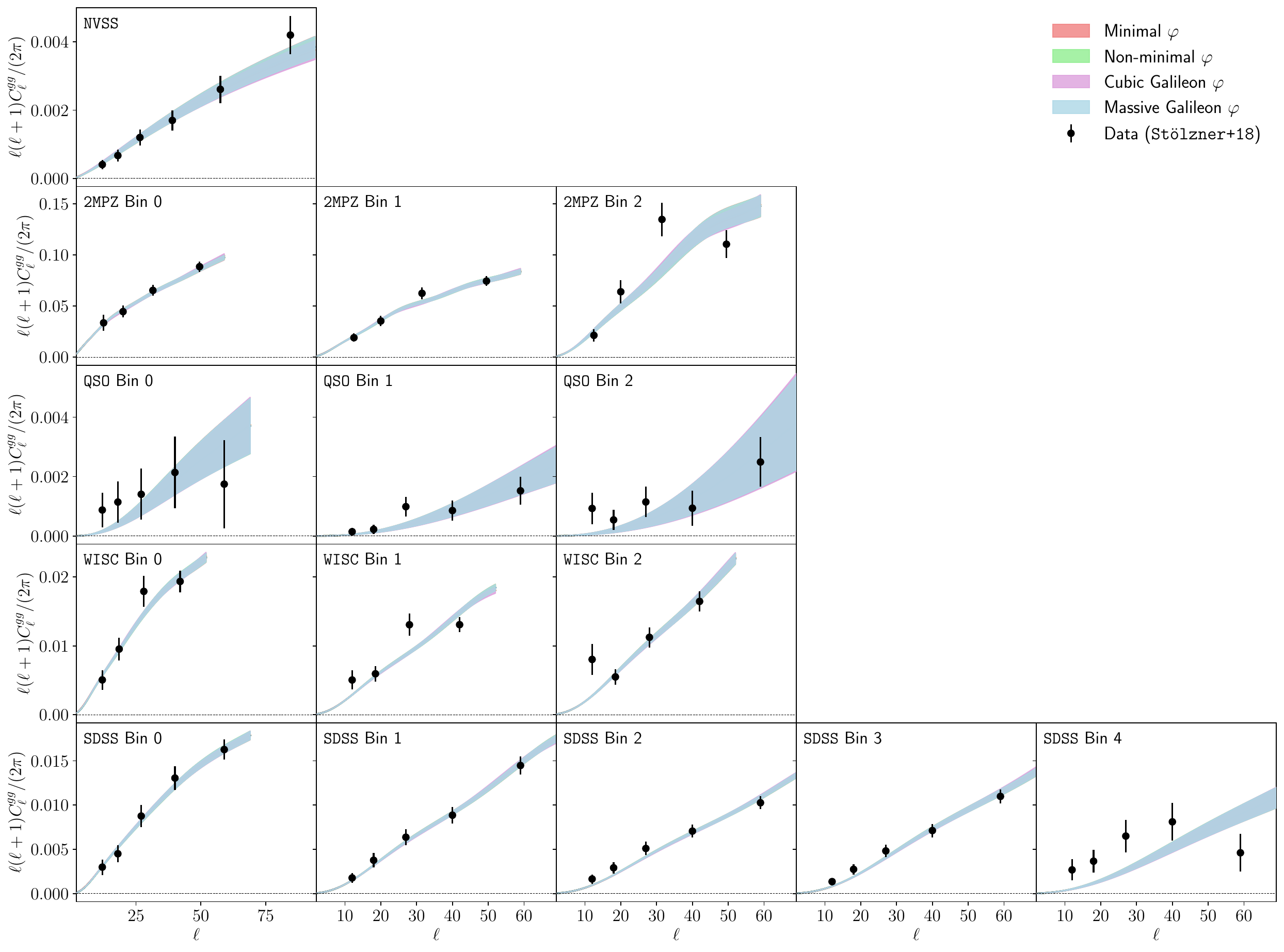}
    \caption{Same as Figure~\ref{fig:Krolewski_bg} but for \stolzner. Instead of $C_\ell^{g\kappa}$, \stolzner used $C_\ell^{gg}$ to calibrate the galaxy bias. The bands cover the range of scales used in the fit, and since we are using the same limit for each column, that results in some white space in some panels.}
    \label{fig:Stolzner_bg}
\end{figure*}

For completeness in Figures~\ref{fig:Krolewski_bg}, \ref{fig:Reeves_individual_bg} and \ref{fig:Stolzner_bg} we provide the angular power spectra that were used to calibrate the galaxy bias, as well as the their fitted values. These plots show how our fit is matching the data and our error estimation compatible with the data uncertainties. In general, fit to these calibration angular power spectra are good. The worst fit is for \stolzner \sdss\xspace \texttt{Bin 4} and \tmpz\xspace \texttt{Bin 2} with $p=0.02$, for all models, followed by \wisc\xspace \texttt{Bin 1} with $p=0.014$ and \wisc\xspace \texttt{Bin 2} already with a reasonable $p=0.1$. We leave to future work the exploration if these cases are due to modeling or data issues (e.g. magnification having been neglected). 

\begin{table*}
    \centering
    \begin{tabular}{|lc|rrrr|}
        \hline
        Survey & Bin &
        \multicolumn{1}{c}{Minimal $\varphi$} &
        \multicolumn{1}{c}{Non-minimal $\varphi$} &
        \multicolumn{1}{c}{Cubic Galileon} &
        \multicolumn{1}{c|}{Massive Galileon} \\
        \hline
        \multirow{3}{*}{\unwise}
        & 0 & $1.554 \pm 0.045$ [$0.0\sigma$] & $1.448^{+0.055}_{-0.054}$ [$0.0\sigma$] & $1.209 \pm 0.033$ [$0.0\sigma$] & $1.471 \pm 0.051$ [$0.0\sigma$] \\
        & 1 & $2.270 \pm 0.059$ [$0.1\sigma$] & $2.187 \pm 0.063$ [$0.1\sigma$] & $1.935 \pm 0.048$ [$0.3\sigma$] & $2.205 \pm 0.061$ [$0.1\sigma$] \\
        & 2 & $3.14 \pm 0.11$ [$0.0\sigma$] & $3.04 \pm 0.11$ [$0.0\sigma$] & $2.69 \pm 0.10$ [$0.0\sigma$] & $3.06 \pm 0.11$ [$0.0\sigma$] \\
        \hline
        
        \multirow{4}{*}{\desi LRGs}
        & 0 & $1.60 \pm 0.16$ [$0.0\sigma$] & $1.44 \pm 0.15$ [$0.0\sigma$] & $1.13 \pm 0.11$ [$0.0\sigma$] & $1.47 \pm 0.15$ [$0.0\sigma$] \\
        & 1 & $1.97 \pm 0.13$ [$0.0\sigma$] & $1.83 \pm 0.13$ [$0.0\sigma$] & $1.49 \pm 0.10$ [$0.0\sigma$] & $1.86 \pm 0.13$ [$0.0\sigma$] \\
        & 2 & $2.29 \pm 0.14$ [$1.4\sigma$] & $2.16 \pm 0.14$ [$1.4\sigma$] & $1.83 \pm 0.12$ [$1.3\sigma$] & $2.19 \pm 0.14$ [$1.4\sigma$] \\
        & 3 & $2.34 \pm 0.12$ [$0.0\sigma$] & $2.22 \pm 0.13$ [$0.0\sigma$] & $1.90 \pm 0.11$ [$0.0\sigma$] & $2.25 \pm 0.13$ [$0.0\sigma$] \\
        \hline
        
        \multirow{4}{*}{\desi LRGs joint}
        & 0 & $1.54 \pm 0.15$ \phantom{[$0.1\sigma$]} & $1.39 \pm 0.14$ \phantom{[$0.1\sigma$]} & $1.08 \pm 0.11$ \phantom{[$0.1\sigma$]} & $1.42 \pm 0.15$ \phantom{[$0.1\sigma$]} \\
        & 1 & $1.92 \pm 0.13$ [$0.1\sigma$] & $1.77 \pm 0.13$ [$0.1\sigma$] & $1.44 \pm 0.10$ [$0.1\sigma$] & $1.81 \pm 0.13$ [$0.1\sigma$] \\
        & 2 & $2.29 \pm 0.13$ \phantom{[$0.1\sigma$]} & $2.16 \pm 0.13$ \phantom{[$0.1\sigma$]} & $1.83 \pm 0.11$ \phantom{[$0.1\sigma$]} & $2.20 \pm 0.14$ \phantom{[$0.1\sigma$]} \\
        & 3 & $2.34 \pm 0.12$ \phantom{[$0.1\sigma$]} & $2.23 \pm 0.12$ \phantom{[$0.1\sigma$]} & $1.91 \pm 0.10$ \phantom{[$0.1\sigma$]} & $2.26 \pm 0.12$ \phantom{[$0.1\sigma$]} \\
        \hline
        
        \multirow{1}{*}{\nvss}
        & 0 & $2.173 \pm 0.082$ [$0.0\sigma$] & $2.126 \pm 0.082$ [$0.0\sigma$] & $1.985 \pm 0.074$ [$0.0\sigma$] & $2.154 \pm 0.082$ [$0.0\sigma$] \\
        \hline
        
        \multirow{3}{*}{\tmpz}
        & 0 & $1.115 \pm 0.025$ [$0.0\sigma$] & $1.078 \pm 0.027$ [$0.0\sigma$] & $0.961 \pm 0.020$ [$0.0\sigma$] & $1.104 \pm 0.025$ [$0.0\sigma$] \\
        & 1 & $1.234 \pm 0.030$ [$0.0\sigma$] & $1.204 \pm 0.031$ [$0.0\sigma$] & $1.093 \pm 0.026$ [$0.0\sigma$] & $1.223 \pm 0.030$ [$0.0\sigma$] \\
        & 2 & $1.890 \pm 0.072$ [$2.8\sigma$] & $1.851 \pm 0.072$ [$2.8\sigma$] & $1.716 \pm 0.065$ [$2.7\sigma$] & $1.872 \pm 0.072$ [$2.8\sigma$] \\
        \hline
        
        \multirow{3}{*}{QSO}
        & 0 & $1.49 \pm 0.25$ [$0.2\sigma$] & $1.48 \pm 0.25$ [$0.2\sigma$] & $1.47 \pm 0.25$ [$0.2\sigma$] & $1.49 \pm 0.25$ [$0.2\sigma$] \\
        & 1 & $2.52 \pm 0.25$ [$0.5\sigma$] & $2.53 \pm 0.25$ [$0.5\sigma$] & $2.61 \pm 0.26$ [$0.5\sigma$] & $2.53 \pm 0.25$ [$0.5\sigma$] \\
        & 2 & $3.49 \pm 0.41$ [$1.2\sigma$] & $3.52 \pm 0.42$ [$1.2\sigma$] & $3.66 \pm 0.43$ [$1.2\sigma$] & $3.52 \pm 0.42$ [$1.2\sigma$] \\
        \hline
        
        \multirow{3}{*}{\wisc}
        & 0 & $0.940 \pm 0.029$ [$0.1\sigma$] & $0.918 \pm 0.030$ [$0.1\sigma$] & $0.840 \pm 0.026$ [$0.1\sigma$] & $0.931 \pm 0.029$ [$0.1\sigma$] \\
        & 1 & $0.866 \pm 0.026$ [$2.2\sigma$] & $0.849 \pm 0.026$ [$2.2\sigma$] & $0.787 \pm 0.023$ [$2.3\sigma$] & $0.857 \pm 0.026$ [$2.2\sigma$] \\
        & 2 & $1.078 \pm 0.036$ [$1.2\sigma$] & $1.058 \pm 0.036$ [$1.2\sigma$] & $0.994 \pm 0.033$ [$1.3\sigma$] & $1.068 \pm 0.036$ [$1.2\sigma$] \\
        \hline
        
        \multirow{5}{*}{\sdss}
        & 0 & $1.079 \pm 0.028$ [$0.0\sigma$] & $1.055 \pm 0.029$ [$0.0\sigma$] & $0.974 \pm 0.025$ [$0.0\sigma$] & $1.068 \pm 0.028$ [$0.0\sigma$] \\
        & 1 & $0.923 \pm 0.024$ [$0.0\sigma$] & $0.907 \pm 0.024$ [$0.0\sigma$] & $0.858 \pm 0.022$ [$0.0\sigma$] & $0.915 \pm 0.024$ [$0.0\sigma$] \\
        & 2 & $0.852^{+0.022}_{-0.023}$ [$0.9\sigma$] & $0.838 \pm 0.023$ [$0.9\sigma$] & $0.801 \pm 0.021$ [$1.0\sigma$] & $0.845 \pm 0.022$ [$0.9\sigma$] \\
        & 3 & $1.222 \pm 0.033$ [$0.7\sigma$] & $1.206 \pm 0.033$ [$0.7\sigma$] & $1.173 \pm 0.031$ [$0.8\sigma$] & $1.214 \pm 0.033$ [$0.7\sigma$] \\
        & 4 & $1.13 \pm 0.11$ [$2.9\sigma$] & $1.12 \pm 0.11$ [$2.9\sigma$] & $1.11 \pm 0.11$ [$2.9\sigma$] & $1.13 \pm 0.11$ [$2.9\sigma$] \\
        \hline
    \end{tabular}
   \caption{Same as Table~\ref{tab:AISW} but for the galaxy bias, $b_{\rm g}$. As in Table~\ref{tab:AISW}, we show in square brackets the tension between the $b_{\rm g}^2C_\ell^{gg}$ or $b_{\rm g} C_\ell^{g\kappa}$ posterior distributions and the data. In this case, the joint fit to \reeves \desi LRGs has four entries because the common parameter is $A_{\rm ISW}$, not $b_{\rm g}$, but a single tension to show the global fit to $C_\ell^{g\kappa}$. In general, the agreement with the data is good (i.e. $p > 0.05$, which corresponds to $1.6\sigma$).}
    \label{tab:bg_ISW}
\end{table*}

\subsection{Precision parameters}\label{sec:ISW-prec}

\begin{figure*}
    \centering
    \includegraphics[width=\linewidth]{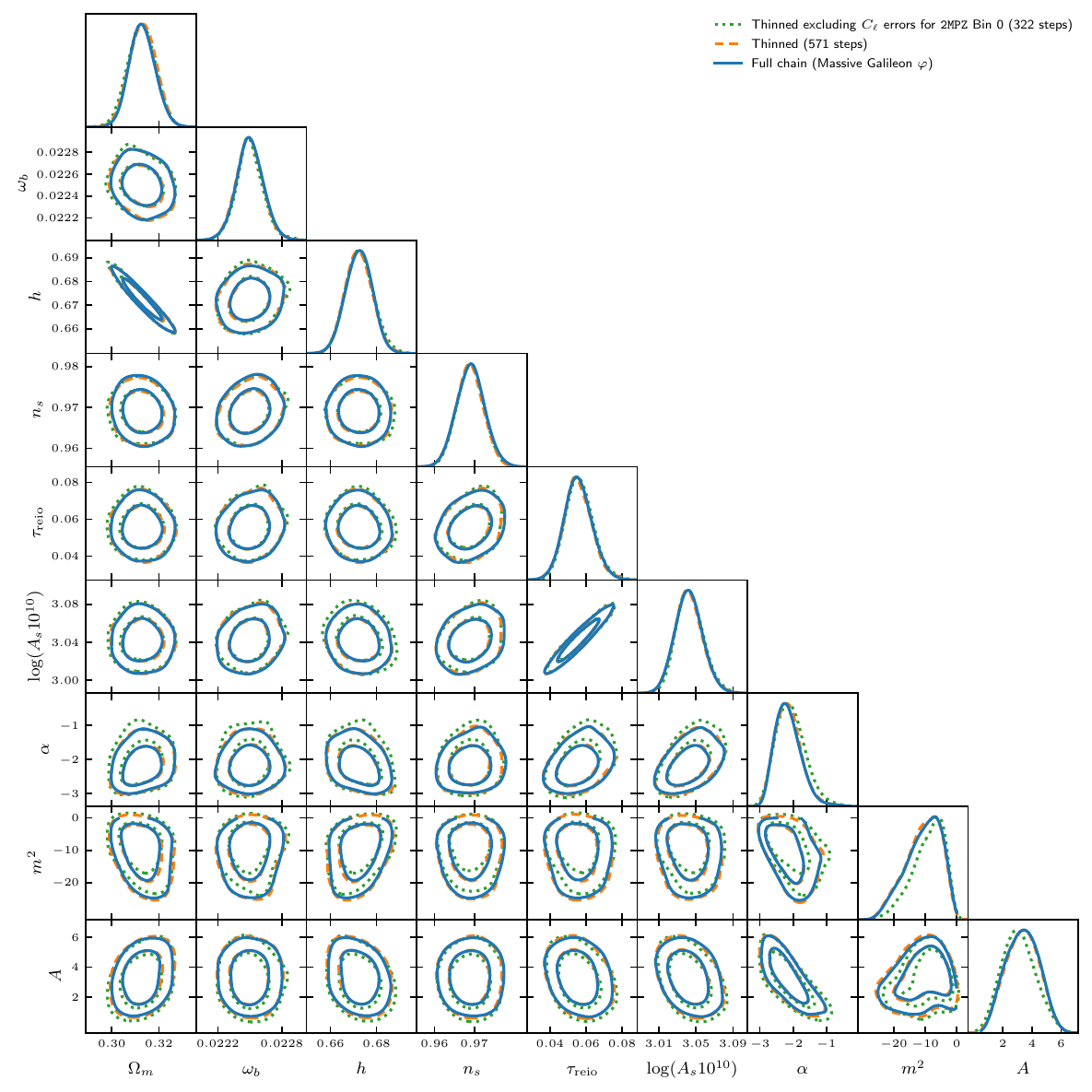}
    \caption{68\% and 95\% C.L. cosmological parameters posterior distribution with \desidesdovcmb data and the mG model. In solid blue line, the posterior distribution using the full chain; in orange dashed line, the subset used in the ISW analysis; and in green dotted line, the same subset excluding the steps that had integration issues for the worst case scenario: mG and \stolzner \tmpz {\tt Bin 0}. Even in this case, with a rejection rate of 44\%, the posterior distribution is well recovered.}
    \label{fig:ISW_thinning}
\end{figure*}

Despite our best efforts to produce reliable angular power spectra, we found that there were still models for which the perturbations, the non-Limber integral or other computational step failed. In addition, apart from those that crashed, there were some that produced unreliable results, with unphysical oscillations or values, a clear sign of integration issues. We rejected those steps before computing the angular power spectra posterior distribution and fitting $A_{\rm ISW}$ and $b_{\rm g}$. The most problematic case is mG, especially with low redshift samples. However, as it can be seen in Figure~\ref{fig:ISW_thinning}, even in the worst case scenario, with 44\% of the samples rejected, the cosmological constraints posterior distribution is well captured. The subsampled (thinned) chains have the following lengths that depend on the dark energy case: 559 steps for Quintessence, 555 for NM, 589 for Cubic Galileon and 571 for mG. Out of these, when computing the angular spectra, we additionally removed the failing steps, as described above.

For \krolewski, we found that, for the NM model, we only had to remove 6 samples out of 555 (1.1\%
), just for the \unwise\xspace {\tt Blue} sample. In comparison, for the mG model, we removed  58 (10\%), 44 (7.7\%), and 44 (7.7\%) for the {\tt Blue}, {\tt Green} and {\tt Red} samples, respectively, out of 571 total steps.

In the case of \reeves, when fitting each tomographic bin independently, the NM model has less than 1.5\% of the steps rejected across all of them: 7 (1.26\%), 7 (1.26\%), 1 (0.18\%), 1 (0.18\%), for each tomographic bin, respectively, out of 555 steps. In the case of mG, these values are larger: 80 (14\%), 103 (18\%), 93 (16\%), and 78 (14\%) out of 571 steps. When fitting all bins together, the numbers grow to 16 (2.88\%) and 217 (38\%) for NM and mG, respectively.

Finally, for the \stolzner data set we find that for NM, only one case has rejection rates above 2\% (QSO-2). In detail, we removed 7 (\(1.26\%\)), 3 (\(0.54\%\)), and 1 (\(0.18\%\)) samples for \tmpz bins 0, 1, and 2, respectively; 6 (\(1.08\%\)), 9 (\(1.62\%\)), and 26 (\(4.68\%\)) samples for QSO bins 0, 1, and 2; 5 (\(0.90\%\)), 5 (\(0.90\%\)), and 6 (\(1.08\%\)) samples for \wisc bins 0, 1, and 2; and 2 (\(0.36\%\)), 8 (\(1.44\%\)), 9 (\(1.62\%\)), 15 (\(2.70\%\)), and 10 (\(1.80\%\)) samples for \sdss bins 0--4. For mG, once again, the rejection rate is substantially larger. We removed 61 samples (\(10.68\%\)) for \nvss; 249 (\(43.61\%\)), 136 (\(23.82\%\)), and 89 (\(15.59\%\)) samples for \tmpz bins 0, 1, and 2; 98 (\(17.16\%\)), 31 (\(5.43\%\)), and 45 (\(7.88\%\)) samples for QSO bins 0, 1, and 2; 81 (\(14.19\%\)), 62 (\(10.86\%\)), and 141 (\(24.69\%\)) samples for \wisc bins 0, 1, and 2; and 50 (\(8.76\%\)), 96 (\(16.81\%\)), 170 (\(29.77\%\)), 179 (\(31.35\%\)), and 197 (\(34.50\%\)) samples for \sdss bins 0--4. The highest rejection rate was found for \tmpz bin 0, where 249 samples (\(43.61\%\)) were discarded, followed by \sdss bins 4 and 3 with rejection rates of \(34.50\%\) and \(31.35\%\), respectively.

These, sometimes very large, rejection rates of mG, are another reason to treat the results presented here with caution and motivates a more detailed study of the ISW computations in a future work. However, we have checked the impact on the cosmological parameters posterior distribution and can be neglected for the purposes of this work (see Figure~\ref{fig:ISW_thinning}).

The \texttt{hi\_class} parameters we used to compute the perturbations were:
\begin{verbatim}
# Common CLASS settings
non_linear = halofit
output = mTk,mPk
extra_metric_transfer_functions = yes
z_max_pk = 4.1
k_output_values = 0.01, 0.1, 1.0

nonlinear_min_k_max = 10
P_k_max_h/Mpc = 100
\end{verbatim}
and for the angular power spectra:
\begin{verbatim}
# Angular power spectrum settings
output = tCl, nCl, mTk, mPk, lCl
selection = zeros
dNdz_selection =  file_path
selection_mean = 1
selection_magnification_bias = magnification_bias

number_count_contributions = dens, lensing, gr
selection = zeros
temperature contributions = lisw
lensing = yes

l_max_scalars = 400
l_max_lss = 400

delta_l_max = 300
l_switch_limber = 1000
l_switch_limber_for_nc_local_over_z = 10000
l_switch_limber_for_nc_los_over_z = 10000

k_max_tau0_over_l_max = 5

transfer_neglect_delta_k_S_t0 = 100
transfer_neglect_delta_k_S_t1 = 100
transfer_neglect_delta_k_S_t2 = 100
neglect_CMB_sources_below_visibility = 1e-30

perturbations_sampling_stepsize = 0.01
tol_perturbations_integration = 1e-6

start_small_k_at_tau_c_over_tau_h = 0.0004
start_large_k_at_tau_h_over_tau_k = 0.05
tight_coupling_trigger_tau_c_over_tau_h = 0.005
tight_coupling_trigger_tau_c_over_tau_k = 0.008
start_sources_at_tau_c_over_tau_h = 0.006
\end{verbatim}
where \texttt{selection = zeros} corresponds to an option we have implemented in \texttt{hi\_class} to use the input $dN/dz$ without convolving it with any other distribution. As described in the previous section, for all the samples that have a non-negligible magnification bias, we compute the ``density'' and ``lensing + GR'' parts separately and then combine them after applying the galaxy bias only to the ``density'' part. 

\newpage
\bibliography{refs}

\end{document}